\documentclass[prd,preprint,eqsecnum,amsfonts,amsmath,amssymb,aps]{revtex4}

\allowdisplaybreaks

\newcommand{\sous}[2]{\stackrel{\phantom{(n,p)}}{#2} \! \! \! \! \! \!
			\! \! \! 
			{{} \atop \scriptstyle #1} {}}

\begin{document}

\title{Gravitational waves from inspiralling compact binaries: \\ 
Energy flux to third post-Newtonian order}

\author{Luc Blanchet}
\affiliation{D\'epartement d'Astrophysique Relativiste et de Cosmologie (UMR 8629 du CNRS),\\
Observatoire de Paris, 92195 Meudon Cedex, France,\\
and Institut d'Astrophysique de Paris, 98bis boulevard Arago, 75014 Paris, France}
\author{Bala R. Iyer} 
\affiliation{Raman Research Institute, Bangalore 560 080, India}
\author{Benoit Joguet} 
\affiliation{Institut d'Astrophysique de Paris, 98bis boulevard Arago, 75014 Paris, France}  

\date{\today}

\begin{abstract}
The multipolar-post-Minkowskian approach to gravitational radiation is
applied to the problem of the generation of waves by the compact
binary inspiral. We investigate specifically the third post-Newtonian
(3PN) approximation in the total energy flux. The new results are the
computation of the mass  quadrupole moment of the binary to the 3PN
order, and the current quadrupole and mass octupole to the 2PN order.
Wave tails and tails of tails in the far zone are included up
to the 3.5PN order. The recently derived 3PN equations of binary
motion are used to compute the time-derivatives of the moments. We
find perfect agreement to the 3.5PN order with perturbation
calculations of black holes in the test-mass limit for one
body. Technical inputs in our computation include a model of point
particles for describing the compact objects, and the Hadamard
self-field regularization. Because of a physical incompleteness of the
Hadamard regularization at the 3PN order, the energy flux depends on
one unknown physical parameter, which is a combination of a parameter
$\lambda$ in the equations of motion, and a new parameter $\theta$
coming from the quadrupole moment.
\end{abstract}

\maketitle

\section{Introduction}

Inspiralling compact binaries are systems of two neutron stars and/or
black holes undergoing an adiabatic orbital decay by gravitational
radiation emission. These systems constitute an important target for
the gravitational-wave detectors like LIGO and VIRGO. The currently
favoured theory for describing the binary inspiral is the
post-Newtonian approximation.  Since inspiralling compact binaries are
very relativistic the Newtonian description (corresponding to the
quadrupole approximation) is grossly inadequate for constructing the
theoretical templates to be used in the signal analysis of
detectors. In fact, from several measurement-accuracy analyses
\cite{3mn,CFPS93,FCh93,CF94,TNaka94,P95,PW95,KKS95,DIS98} it follows
that the third post-Newtonian (3PN) approximation, corresponding to
the order $1/c^6$ when the speed of light $c\to +\infty$, constitutes
a necessary achievement in this field.  Note that the 3PN
approximation is needed to compute the time evolution of the binary's
orbital phase, that depends {\it via} an energy balance equation on
the total gravitational-wave energy flux. The energy flux is therefore
a crucial quantity to predict.

Following earliest computations at the 1PN level \cite{WagW76,BS89}
(at a time where post-Newtonian corrections were of purely academic
interest), the energy flux generated by compact binaries was
determined to the 2PN order
\cite{BDIWWi95,BDI95,WWi96,BIWWi96,GopuI97}, by means of a formalism
based on multipolar and post-Minkowskian approximations
\cite{BD89,DI91a,BD92,B95,B98mult}, and independently using a direct
integration of the relaxed Einstein equations \cite{WWi96,W99,PW00}
(see also Refs. \cite{EW75,Th80}).  Since then the calculations have
been extended to include the non-linear effects of tails at higher
post-Newtonian orders. The tails at the 2.5PN and 3.5PN orders were
computed in Refs. \cite{B96,B98tail} (this extended the computation of
tails at the dominant 1.5PN order \cite{P93,Wi93,BS93}), and the
contribution of tails generated by the tails themselves (so-called
``tails of tails'') at the 3PN order were obtained in
Ref. \cite{B98tail}. However, unlike the 1.5PN, 2.5PN and 3.5PN orders
that are entirely composed of tail terms, the 3PN approximation
involves also, besides the tails of tails, many non-tail contributions
coming from the relativistic corrections in the multipole moments of
the binary.

The present paper is devoted to the computation of the multipole
moments, chiefly the quadrupole moment at the 3PN order, in the case
where the binary's orbit is circular (the relevant case for most
inspiralling binaries). We reduce some general expressions for the
multipole moments of a slowly-moving extended system \cite{B98mult} to
the case of a point-particle binary at the 3PN order. The self-field
of point-particles is systematically regularized by means of
Hadamard's concept of ``partie finie''
\cite{Hadamard,Schwartz,Sellier}. The time-derivatives of the 3PN
quadrupole moment are computed with the help of the equations of
binary motion at the 3PN order in harmonic coordinates (the coordinate
system chosen for this computation).  The 3PN equations of motion have
been derived recently by two groups working independently with
different methods: ADM-Hamiltonian formulation of general relativity
\cite{JaraS98,JaraS99,JaraS00,DJS00,DJS01}, and direct post-Newtonian
iteration of the field equations in harmonic coordinates
\cite{BF00,BFreg,BFregM,BFeom,ABF}. There is complete physical
equivalence between the results given by the two approaches
\cite{DJS01,ABF}. We shall find that our end result for the energy
flux at the 3.5PN order is in perfect agreement, in the test-body
limit for one body, with the result of black-hole perturbation theory,
which is currently known up to the higher 5.5PN approximation
\cite{Sasa94,TSasa94,TTS96} (see Ref. \cite{MSSTT} for a review). In a
separate work \cite{BFIJ01} we report the computation of the
3.5PN-accurate orbital phase which constitutes the crucial component
of the theoretical template of inspiralling binaries.

One conclusion of the investigation of the equations of motion of
compact binaries is that from the 3PN order the model of
point-particles (described by Dirac distributions) might become
physically incomplete, in the sense that the equations involve one
undetermined coefficient, $\omega_{\rm static}$ in the ADM-Hamiltonian
formalism \cite{JaraS98,JaraS99,JaraS00,DJS00,DJS01} (see, however,
\cite{DJS01b}) and $\lambda$ in the harmonic-coordinate approach
\cite{BF00,BFreg,BFregM,BFeom,ABF}. Technically this is due to some
subtle features of the self-field regularization {\it \`a la}
Hadamard. In the present paper, we shall be lead to introduce a second
undetermined coefficient, called $\theta$, coming from our computation
of the 3PN quadrupole moment. However, we shall find that the total
energy flux contains only one unknown parameter, which is a certain
linear combination of $\theta$ and $\lambda$ entering the 3PN
coefficient. All other terms in the flux up to the 3.5PN order are
completely specified.

The plan of this paper is as follows. Sections II to IV are devoted to
the basic expressions of the moments we shall apply. Section V
presents the needed information concerning our point-particle model,
and Sections VI to IX deal with the computation of all the different
types of terms in the required multipole moments. Section X explains
our introduction of the $\theta$-ambiguity. Finally we present our
results for the moments and energy flux in Sections XI and XII. The
intermediate values for all the terms composing the moments in the
case of circular orbits are relegated to Appendix A.

\section{Expressions of the multipole moments}

In this section we give a short summary on the expressions of
multipole moments in the post-Newtonian approximation. The moments
describe some general isolated sources that are weakly
self-gravitating and slowly-moving, i.e. whose internal velocities are
much smaller than the speed of light: $v \ll c$. In this paper we
order all expressions according to the formal order in $1/c$, and we
pose ${\cal O}(n)\equiv {\cal O}(1/c^n)$. In addition, the moments are
{\it a priori} valid only in the case where the source is continuous
(for instance a hydrodynamical fluid); however, we shall apply these
moments to the case of point-particles by supplementing the above
expressions with a certain regularization ansatz based on Hadamard's
concept of ``partie finie'' \cite{Hadamard,Schwartz,Sellier}. We adopt
a system of harmonic coordinates, which means

\begin{subequations}\label{1}\begin{eqnarray}
 \partial_\nu h^{\mu\nu} &=& 0 \label{1a}\;,\\ h^{\mu\nu} &\equiv&
|g|^{1/2} g^{\mu\nu}-\eta^{\mu\nu}\label{1b}
\;,\end{eqnarray}\end{subequations}$\!\!$ where $g^{\mu\nu}$ and $g$
denote respectively the inverse and the determinant of the covariant
metric $g_{\mu\nu}$, and where $\eta^{\mu\nu}$ denotes the Minkowski
metric with signature $+2$. The Einstein field equations, relaxed by
the harmonic-coordinate condition, take the form of d'Alembertian
equations for all the components of the field variable,

\begin{subequations}\label{2}\begin{eqnarray}
\Box h^{\mu\nu}
 &=& {16\pi G\over c^4} \tau^{\mu\nu} \label{2a}\;,\\
\tau^{\mu\nu} &\equiv& |g| T^{\mu\nu} + {c^4 \over 16\pi G}
   \Lambda^{\mu\nu} \label{2b}
\;,\end{eqnarray}\end{subequations}$\!\!$
where $\Box=\eta^{\mu\nu}\partial_\mu\partial_\nu$ and where we have
introduced the effective stress-energy (pseudo-)tensor $\tau^{\mu\nu}$
of the matter and gravitational fields in harmonic coordinates. The
matter stress-energy is described by $T^{\mu\nu}$ and the
gravitational stress-energy by the non-linear interaction term
$\Lambda^{\mu\nu}$. The latter is given in terms of the metric by the
exact expression

\begin{eqnarray}\label{3}
\Lambda^{\mu\nu} = &-& h^{\rho\sigma}
\partial^2_{\rho\sigma} h^{\mu\nu}+\partial_\rho h^{\mu\sigma} 
\partial_\sigma h^{\nu\rho} 
+{1\over 2}g^{\mu\nu}g_{\rho\sigma}\partial_\lambda h^{\rho\tau} 
\partial_\tau h^{\sigma\lambda} \nonumber\\
&-&g^{\mu\rho}g_{\sigma\tau}\partial_\lambda h^{\nu\tau} 
\partial_\rho h^{\sigma\lambda} 
-g^{\nu\rho}g_{\sigma\tau}\partial_\lambda h^{\mu\tau} 
\partial_\rho h^{\sigma\lambda} 
+g_{\rho\sigma}g^{\lambda\tau}\partial_\lambda h^{\mu\rho} 
\partial_\tau h^{\nu\sigma}\nonumber\\
&+&{1\over 8}(2g^{\mu\rho}g^{\nu\sigma}-g^{\mu\nu}g^{\rho\sigma})
(2g_{\lambda\tau}g_{\epsilon\pi}-g_{\tau\epsilon}g_{\lambda\pi})
\partial_\rho h^{\lambda\pi} 
\partial_\sigma h^{\tau\epsilon}
\;.\end{eqnarray}
Both the matter and gravitational contributions in $\tau^{\mu\nu}$
depend on the field $h$, with the gravitational term
$\Lambda^{\mu\nu}$ being at least quadratic in $h$ and its space-time
derivatives.

The multipole moments of slowly-moving sources are in the form of some
functionals of the (formal) {\it post-Newtonian} expansion of the
pseudo-tensor $\tau^{\mu\nu}$; we denote the formal post-Newtonian
expansion with an overbar, so ${\overline \tau}^{\mu\nu}={\rm
PN}(\tau^{\mu\nu})$. It is convenient to introduce the auxiliary
notation

\begin{equation}\label{4}
\Sigma = {\overline\tau^{00} +\overline\tau^{ii}\over c^2} \;;\quad
\Sigma_i = {\overline\tau^{0i}\over c} \;;\quad \Sigma_{ij} =
\overline{\tau}^{ij} \;.\end{equation} From a general study
\cite{B95,B98mult} of the matching between the exterior gravitational
field of the source and the inner post-Newtonian field, we obtain some
``natural'' definitions for the $l$th order mass-type ($I_L$) and
current-type ($J_L$) multipole moments of the source. The physics of
the isolated source, as seen in its exterior, is extracted from these
multipole moments when they are connected, in a consistent way, to the
observables of the radiative field at (Minkowskian) future null
infinity, given in this formalism by the so-called radiative multipole
moments. The connection between $I_L$ and $J_L$ and the mass-type
($U_L$) and current-type ($V_L$) radiative moments at infinity
involves up to say the 3.5PN order many tail effects and even a
particular ``tail-of-tail'' effect arising specifically at 3PN. All
these effects are known \cite{B98tail} and therefore will not be
investigated here but simply added at the end of our computation in
Section XII. Here we focus our attention on the reduction to
point-particle binaries of the general {\it source} multipole moments
(in symmetric-tracefree form), whose complete expressions are given by

\begin{subequations}\label{5}\begin{eqnarray}
 I_L(t)&=& \sous{\!\!\!\!B= 0}{\rm FP} \int d^3{\bf x}~|\tilde{\bf
  x}|^B \int^1_{-1} dz\left\{ \delta_l(z)\hat x_L\Sigma -{4(2l+1)\over
  c^2(l+1)(2l+3)} \delta_{l+1}(z) \hat x_{iL} \dot{\Sigma}_i\right.
  \nonumber\\ &&\qquad\qquad \left. +{2(2l+1)\over
  c^4(l+1)(l+2)(2l+5)} \delta_{l+2}(z) \hat x_{ijL} \ddot{\Sigma}_{ij}
  \right\} ({\bf x},t+z |{\bf x}|/c)\label{5a}\;,\\ J_L(t)&=&
  \sous{\!\!\!\!B= 0}{\rm FP} ~\varepsilon_{ab<i_l} \int d^3{\bf
  x}~|\tilde {\bf x}|^B \int^1_{-1} dz\biggl\{ \delta_l(z)\hat
  x_{L-1>a} \Sigma_b \nonumber\\ &&\qquad\qquad -{2l+1\over
  c^2(l+2)(2l+3)} \delta_{l+1}(z) \hat x_{L-1>ac} \dot{\Sigma}_{bc}
  \biggr\} ({\bf x},t+z |{\bf x}|/c)\label{5b}
  \;.\end{eqnarray}\end{subequations}$\!\!$ 
Our notation is as follows. $L=i_1i_2\cdots i_l$ is a multi-index
composed of $l$ indices; a product of $l$ spatial vectors $x^i\equiv
x_i$ is denoted $x_L=x_{i_1}x_{i_2}\cdots x_{i_l}$; the
symmetric-tracefree (STF) part of that product is denoted using a
hat: ${\hat x}_L={\rm STF}(x_L)$, for instance ${\hat x}_{ij}=x_i
x_j-\frac{1}{3}\delta_{ij}$, ${\hat x}_{ijk}=x_i x_j
x_k-\frac{1}{5}(x_i \delta_{jk}+x_j \delta_{ki}+x_k \delta_{ij})$;
the STF projection is also denoted using brackets surrounding the
indices, e.g. ${\hat x}_{ij}\equiv x_{<ij>}$, 
$x_{<i}v_{j>}=\frac{1}{2}(x_iv_j+x_jv_i)-\frac{1}{3}\delta_{ij}x_kv_k$;
$\varepsilon_{ijk}$ denotes the usual Levi-Civita symbol
($\varepsilon_{000}=+1$); the dots refer to the time
differentiation. The matter densities $\Sigma$, $\Sigma_i$ and
$\Sigma_{ij}$ in (\ref{5}) are evaluated at the position ${\bf x}$
and at time $t+z |{\bf x}|/c$. The function $\delta_l(z)$ is given
by

\begin{equation}\label{6}
  \delta_l (z) = {(2l+1)!!\over 2^{l+1} l!} (1-z^2)^l
\ ; \quad\int^1_{-1} dz\delta_l (z) = 1
\;.\end{equation} 
This function tends to the Dirac distribution when $l\to +\infty$.
Each of the terms composing $I_L$ and $J_L$ is to be understood in the
sense of post-Newtonian expansion, and computed using the (infinite)
post-Newtonian series

\begin{equation}\label{7}
\int^1_{-1} dz \delta_l(z) S({\bf x},t+z |{\bf
x}|/c)=\sum^\infty_{j=0}{(2l+1)!! \over 2^jj!(2l+2j+1)!!}|{\bf
x}|^{2j}\left({\partial\over c\partial t}\right)^{2j}S({\bf x},t)
\;.\end{equation}
Finally the symbol ${\rm FP}_{B=0}$ in front of the integrals in
(\ref{5}) refers to a specific finite part operation defined by
analytic continuation (see Ref. \cite{B98mult} for the details). Such
a finite part is crucial because the integrals have a non-compact
support due to the gravitational contribution in the pseudo-tensor,
and would be otherwise divergent at infinity (when $|{\bf x}|\to
+\infty$). The integral involves the regularization factor
$|\tilde{\bf x}|^B = |{\bf x}/r_0|^B$, where $B$ is a complex number
and $r_0$ denotes an arbitrary length scale. It is defined by complex
analytic continuation for any $B\in{\mathbb C}$ except at isolated
poles in ${\mathbb Z}$, including in general the value of interest
$B=0$. We expand the integral as a Laurent expansion when $B\to 0$ and
pick up the finite part (in short ${\rm FP}_{B=0}$), or coefficient
of the zeroth power of $B$ in that expansion. This finite part is in
fact equivalent to the Hadamard partie finie \cite{Hadamard}.

Thus, the moments depend {\it a priori} on the constant $r_0$
introduced in this analytic continuation process. This constant can be
thought of as due to the ``regularization'' of the field at infinity;
the moments will depend explicitly on $r_0$ when the integral develops
a polar part at $B=0$ due to the behaviour of the integrand when
$|{\bf x}| \to +\infty$. As we shall see the source moments start
depending explicitly on $r_0$ from the 3PN order. However, we know
that the metric is actually independent of $r_0$ (more precisely,
$r_0$ cancels out between the two terms of the multipole expansion
given by Eq. (3.11) in Ref. \cite{B98mult}). Indeed, as a good check
of the calculation, we shall see that because of non-linear tail
effects in the wave zone the constant $r_0$ is cancelled out, so the
physical energy flux does not depend on it.

To the 1PN order the expressions (\ref{5a}) and (\ref{5b}) are
equivalent to some alternative forms obtained earlier in
Refs. \cite{BD89} and \cite{DI91a}, respectively. The multipole
moments in the form (\ref{5}) were derived in \cite{B95} up to the 2PN
order, and shown subsequently in \cite{B98mult} to be in fact valid up
to any post-Newtonian order (formally). On the other hand, both
(\ref{5a}) and (\ref{5b}) reduce to the expressions obtained in
Ref. \cite{DI91b} in the limit of linearized gravity, where we can
replace $\tau^{\mu\nu}$ by the compact-support matter tensor
$T^{\mu\nu}$ (hence there is no need in this limit to consider a
finite part). Note that the source multipole moments $I_L$ and $J_L$
parametrize, by definition, the linearized approximation to the vacuum
metric outside the source \cite{B98mult}, but take into account all
the non-linearities due to the inner (near-zone) field of the
source. The non-linearities in the exterior field can be obtained by  some
specific post-Minkowskian algorithm (see Ref. \cite{B98mult} for proof
and details).  The inclusion of  these non-linearities permits one
to  relate the
source moments $I_L$ and $J_L$ to the radiative ones $U_L$ and
$V_L$. Some other source moments $W_L$, $X_L$, $Y_L$ and $Z_L$ should
also be taken into account (see Ref. \cite{B98mult} for discussion),
but these parametrize a (linearized) gauge transformation and do not
contribute to the radiation field up to a high post-Newtonian
order. We shall check that these moments do not affect the present
calculation.

\section{Definitions of potentials}

Our first task is to work out the expressions (\ref{5}) to the 3PN
order in the case of $I_L$ and 2PN order in the case of $J_L$. In this
paper we shall use some convenient retarded potentials, and then, from
these, the corresponding ``instantaneous'' potentials. For insertion
into the pseudo-tensor ${\overline \tau}^{\mu\nu}$ (and, most
importantly, its gravitational part ${\overline \Lambda}^{\mu\nu}$) we
need the components of the metric ${\overline h}^{\mu\nu}$ developed
to post-Newtonian order ${\cal O}(8,7,8)$. By this we mean ${\cal
O}(8)\equiv {\cal O}(1/c^8)$ in the $00$ and $ij$ components of the
metric, and ${\cal O}(7)\equiv {\cal O}(1/c^7)$ in the $0i$
components. With this precision the metric reads

\begin{subequations}\label{8}\begin{eqnarray}
{\overline h}^{00} &=& -{4\over c^2}V - {2\over c^4}\left(\hat{W}+4V^2
\right)- {8\over c^6} \left (\hat{Z}+2\hat{X}
+ V \hat{W}+ {4\over 3}V^3\right) +{\cal O}(8)\label{8a}\;,\\
{\overline h}^{0i} &=& -{4\over c^3} V_i - {8\over c^5} \left(\hat{R}_i
+V V_i\right)
+{\cal O}(7)\label{8b}\;,\\
{\overline h}^{ij} &=& -{4\over c^4}\left( \hat{W}_{ij}-{1\over 2}
\delta_{ij}\hat{W}\right) - {16\over c^6} \left(\hat{Z}_{ij} -{1\over 2}
\delta_{ij}\hat{Z}\right)+{\cal O}(8)\label{8c}  
\;.\end{eqnarray}\end{subequations}$\!\!$
The potentials are generated by the components of the matter tensor
$T^{\mu\nu}$ or, rather, using a notation similar to (\ref{4}), by

\begin{equation}\label{9}
 \sigma = {T^{00}+T^{ii}\over c^2}\;;\quad
 \sigma_i = {T^{0i}\over c}\;;\quad
 \sigma_{ij} = T^{ij} 
\;.\end{equation}
The potential $V$ is a retarded version of the Newtonian potential
and is defined by the retarded integral $\Box^{-1}_R$ acting
on the source $\sigma$,

\begin{equation}\label{10}
V({\bf x},t)=\Box^{-1}_R\{-4\pi G\sigma \} \equiv G \int {d^3{\bf
y}\over |{\bf x}-{\bf y}|} \sigma( {\bf y}, t - |{\bf x} -{\bf y}|/c)
\;.\end{equation} To the 1PN order we have the potentials $V_i$
and $\hat{W}_{ij}$ (together with the spatial trace
$\hat{W}=\hat{W}_{ii}$), which are generated by the current and stress
$\sigma_i$ and $\sigma_{ij}$ respectively,

\begin{subequations}\label{11}\begin{eqnarray}
V_i &=& \Box^{-1}_R\{-4\pi G \sigma_i\} \;,\\ \hat{W}_{ij} &=&
\Box^{-1}_R\{-4 \pi G (\sigma_{ij} - \delta_{ij} \sigma_{kk}) -
\partial_i V \partial_j V\} \;.\end{eqnarray}\end{subequations}$\!\!$
To the 2PN order there are the potentials $\hat{R}_i$, $\hat{Z}_{ij}$,
$\hat{X}$ (and also $\hat{Z}=\hat{Z}_{ii}$) whose expressions read

\begin{subequations}\label{12}\begin{eqnarray}
{\hat X} = \Box^{-1}_R\Bigl\{ &-&4\pi G \sigma_{ii}V + \hat{W}_{ij}
\partial^2_{ij} V +2 V_i \partial_t \partial_i V \nonumber\\ &+& V
\partial_t^2 V + {3\over 2} (\partial_t V)^2 - 2 \partial_i V_j
\partial_j V_i\Bigr\} \label{12a} \;,\\ {\hat R}_i = \Box^{-1}_R\Bigl\{
&-& 4\pi G (\sigma_i V - \sigma V_i) - 2 \partial_k V\partial_i V_k -
{3\over 2} \partial_t V \partial_i V \Bigr\} \;,\\ {\hat Z}_{ij} =
\Box^{-1}_R\Bigl\{ &-& 4\pi G (\sigma_{ij}-\delta_{ij} \sigma_{kk})V -
2\partial_{(i} V \partial_t V_{j)} \nonumber\\ &+& \partial_i V_k
\partial_j V_k + \partial_k V_i \partial_k V_j - 2 \partial_{(i} V_k
\partial_k V_{j)} \nonumber\\ &-& \delta_{ij} \partial_k V_m
(\partial_k V_m - \partial_m V_k) - {3\over 4} \delta_{ij} (\partial_t
V)^2\Bigr\}
\;.\end{eqnarray}\end{subequations}$\!\!$
Next we expand the retardations and define some associated
instantaneous potentials. The highest-order expansion is needed for
the $V$-potential, up to ${\cal O}(5)$, while ${\cal O}(3)$ is
sufficient for $V_i$ and $\hat{W}_{ij}$. We write these expansions in
the form

\begin{subequations}\label{13}\begin{eqnarray}
V &=& U +{1\over 2c^2} \partial_t^2 \chi -{2G\over 3c^3} {d^3 Q\over
dt^3} +{1\over 24c^4} \partial_t^4 P + {\cal O}(5)\;,\\ V_i &=& U_i
+{1\over 2c^2} \partial_t^2 \chi_i + {\cal O}(3)\;,\\ \hat{W}_{ij} &=&
U_{ij} - {G\over 2c} {d^3 Q_{ij} \over dt^3} +{1\over 2c^2}
\partial_t^2 \chi_{ij} -{1\over c^2} K_{ij} + {\cal O}(3)
\;,\end{eqnarray}\end{subequations}$\!\!$
where the instantaneous potentials are given by the Poisson-type
integrals

\begin{subequations}\label{14}\begin{eqnarray}
U &=& \Delta^{-1}\{ -4\pi G \sigma \} \equiv G \int {d^3{\bf y}\over
|{\bf x}-{\bf y}|} \sigma( {\bf y}, t)\label{14a}\;,\\ U_i &=&
\Delta^{-1}\{ -4\pi G \sigma_i\} \label{14b}\;,\\ U_{ij} &=&
\Delta^{-1}\{ -4 \pi G (\sigma_{ij} - \delta_{ij} \sigma_{kk}) -
\partial_i U \partial_j U\}\label{14c}\;,\\ \chi &=& 2\Delta^{-1}U = G
\int d^3{\bf y}~ |{\bf x}-{\bf y}| \sigma( {\bf y}, t)\label{14d}\;,\\
\chi_i &=& 2\Delta^{-1}U_i\label{14e}\;,\\ \chi_{ij} &=&
2\Delta^{-1}U_{ij}\label{14f}\;,\\ P&=& 24~\!\Delta^{-2}U = G \int
d^3{\bf y}~ |{\bf x}-{\bf y}|^3 \sigma( {\bf y}, t) \label{14g}\;,\\
K_{ij} &=& \Delta^{-1}\{ \partial_{(i}U \partial_{j)} \partial_t^2
\chi\}\label{14h}
\;.\end{eqnarray}\end{subequations}$\!\!$
In addition, the Newtonian precision ${\cal O}(1)$ is required
for the other potentials ${\hat X}$, ${\hat R}_i$ and ${\hat
Z}_{ij}$. For simplicity in the notation, we
shall keep the same names for the Newtonian approximations to these
potentials, henceforth re-defined as

\begin{subequations}\label{15}\begin{eqnarray}
{\hat X} = \Delta^{-1}\Bigl\{ &-&4\pi G \sigma_{ii}U + U_{ij}
\partial^2_{ij}U + 2 U_i \partial_t \partial_i U \nonumber\\ &+& U
\partial_t^2 U + {3\over 2} (\partial_t U)^2 - 2 \partial_i U_j
\partial_j U_i\Bigr\} \;,\\ {\hat R}_i = \Delta^{-1}\Bigl\{ &-& 4\pi G
(\sigma_i U - \sigma U_i) - 2 \partial_k U \partial_i U_k - {3\over 2}
\partial_t U \partial_i U\Bigr\} \;,\\ {\hat Z}_{ij} = \Delta^{-1}\Bigl\{
&-& 4\pi G (\sigma_{ij}-\delta_{ij} \sigma_{kk})U - 2\partial_{(i} U
\partial_t U_{j)} \nonumber\\ &+& \partial_i U_k \partial_j U_k +
\partial_k U_i \partial_k U_j - 2 \partial_{(i} U_k \partial_k U_{j)}
\nonumber\\ &-& \delta_{ij} \partial_k U_m (\partial_k U_m -
\partial_m U_k) - {3\over 4} \delta_{ij} (\partial_t U)^2\Bigr\}
\;.\end{eqnarray}\end{subequations}$\!\!$
Finally the ``odd'' terms in (\ref{13}) (having an odd power of $1/c$
in factor) are simple functions of time parametrized by

\begin{subequations}\label{16}\begin{eqnarray}
Q_{ij}(t) &=& \int d^3{\bf x}~\! (x_ix_j-{\bf x}^2\delta_{ij}) \sigma(
{\bf x}, t)\;,\\
Q(t) &=& \int d^3{\bf x}~\! {\bf x}^2 \sigma(
{\bf x}, t)
\;.\end{eqnarray}\end{subequations}$\!\!$
(Beware that $Q\not= Q_{ii}$.)

\section{Nomenclature of terms}

The post-Newtonian metric (\ref{8}) is inserted into the pseudo-tensor
(\ref{2b}), in which notably the term $\Lambda^{\mu\nu}$, given by
Eq. (\ref{3}), is developed up to quartic order $h^4$. Making use of
the formula (\ref{7}) we obtain the source moments $I_L(t)$ and
$J_L(t)$ as some functionals of all the retarded potentials, and,
then, of all the ``instantaneous'' potentials defined by
(\ref{13})-(\ref{16}). We transform some of the terms by integration
by parts, being careful to  take into account the presence of the
analytic continuation factor $|\tilde{\bf x}|^B$. The surface terms
are always zero by analytic continuation (starting from the case where
the real part of $B$ is a large negative number). Notice that we use
the Leibniz rule, which is surely valid in the case of potentials
corresponding to smooth (``fluid'') sources. However, when we shall
insert for the potentials some singular expressions corresponding to
point-like particles, and shall replace the derivatives by some
appropriate distributional derivatives, the Leibniz rule will no
longer be satisfied in general. This will be a source of some
indeterminacy discussed in Section X.

We find that the moments are quite complicated, so it is useful to
devise a good nomenclature of terms. First, we distinguish in $I_L$
and $J_L$ the contributions which are due to the source densities
$\Sigma$, $\Sigma_i$ and $\Sigma_{ij}$ [see Eq. (\ref{5})], and we
refer to them as scalar (S), vector (V) and tensor (T)
respectively. Furthermore, we split each of these contributions
according to the value of the summation index $j$ in Eq. (\ref{7}):
for instance the S-type term denoted SI is defined by the set of terms
in $I_L$ coming from the ``scalar'' $\Sigma$ in which we have used the
formula (\ref{7}) with only the contribution of the index $j=0$ (there
are no S-type terms in $J_L$); similarly we denote by SII, using Roman
letters, the S-terms corresponding to $j=1$ (these terms involve a
factor ${\bf x}^2$ and a second time-derivative); and for instance VII
denotes the set of terms in both $I_L$ and $J_L$ coming from the
``vector'' $\Sigma_i$ and which have $j=2$. With this notation the
mass moment to the 3PN order can be written as:

\begin{eqnarray}\label{17}
&&I_L = \hbox{SI+SII+SIII+SIV} \nonumber\\
&&\qquad       +\hbox{VI+VII+VIII+TI+TII}+{\cal O}(7)
\;.\end{eqnarray}
For simplicity's sake we omit writing the multi-index $L$ on each of
these separate pieces (there can be no confusion from the
context). Second, the numerous terms are numbered according to their
order of appearance in the following formulas. For instance the piece
SI which is part of the mass moment (\ref{17}) will be composed of the
terms SI(1), SI(2), etc; similarly VII is made of terms VII(1) and so
on. The numbering of terms is indicated in round brackets at the top
of each term. The explicit expressions of all the separate pieces
forming $I_L$ is as follows.
 
\begin{subequations}\label{18}\begin{eqnarray}
{\rm SI} &=& \sous{\!\!\!\!B= 0}{\rm FP} \int d^3{\bf x}~\!|\tilde{\bf
x}|^B \hat{x}_L \Biggl\{ \stackrel{(1)}{\sigma} -
\stackrel{(2)}{{1\over 2\pi Gc^2} \Delta (U^2)} +
\stackrel{(3)}{{4U\over c^4} \sigma ^{aa}} - \stackrel{(4)}{{2\over
\pi Gc^4} U_i \partial_t \partial_i U} \nonumber \\ &-&
\stackrel{(5)}{{1\over \pi Gc^4} (\partial^2_{ij}U) U_{ij}} -
\stackrel{(6)}{ {1\over 2\pi Gc^4} (\partial_t U)^2} + \stackrel{(7)}{
{2\over \pi Gc^4} \partial_i U_j \partial_j U_i} - \stackrel{(8)}{
{2\over 3\pi Gc^4} \Delta (U^3) } \nonumber\\ &-& \stackrel{(9)} {
{1\over 2\pi Gc^4} \Delta (U \partial ^2_t \chi)} - \stackrel{(10)} {
{1\over 2\pi Gc^4} \Delta (U U_{aa})} - \stackrel{(11)}{ {2\over
3}{G\over c^5} \sigma {d^3Q \over dt^3}} + \stackrel{(12)} { {1\over
2\pi c^5} (\partial^2_{ij} U) {d^3Q_{ij} \over dt^3}} \nonumber \\ &+&
\stackrel{(13)}{ {16\over c^6} \sigma U_a U_a} + \stackrel{(14)} {
{8\over c^6} \sigma ^{aa} U^2 }+ \stackrel{(15)}{ {2\over c^6}
\sigma^{aa} \partial^2_t \chi }+ \stackrel{(16)}{ {4\over c^6} U_{jk}
\sigma _{jk}} - \stackrel{(17)} { {1\over \pi Gc^6} U_i \partial_i
\partial^3_t \chi} \nonumber \\ &-& \stackrel{(18)} { {1\over \pi Gc^6}
(\partial^2_t \chi_i) (\partial_t \partial_i U)} - \stackrel{(19)}
{{1\over 2\pi Gc^6} (\partial^2_{ij} U) (\partial^2_t \chi_{ij})} +
\stackrel{(20)} { {1\over \pi G c^6} (\partial_{ij}^2 U) K^{ij}}
\nonumber\\ &-& \stackrel{(21)} { {1\over 2\pi Gc^6} (\partial^2_{ij}
\partial^2_t \chi) U_{ij}} + \stackrel{(22)} { {1\over 2\pi Gc^6}
U_{aa} \partial^2_t U }+ \stackrel{(23)} { {1\over 2\pi Gc^6} U
\partial^2_t U_{aa}} -\stackrel{(24)} {{1\over 2\pi Gc^6} \partial_t U
\partial^3_t \chi} \nonumber \\ &+& \stackrel{(25)}{ {2\over \pi Gc^6}
\partial_i U_j \partial_t^2 \partial_j \chi_i} - \stackrel{(26)}{
{2\over \pi Gc^6} U (\partial_t U)^2} -\stackrel{(27)}{ {6\over \pi
Gc^6} U_a \partial_t U \partial_a U} -\stackrel{(28)} {{12\over \pi
Gc^6} U U_a \partial_t \partial_a U } \nonumber\\ &+& \stackrel{(29)}
{ {4\over \pi Gc^6} U \partial_a U \partial_t U_a} - \stackrel{(30)} {
{8\over \pi Gc^6} U_a (\partial_b U_a) \partial_b U }+
\stackrel{(31)}{ {2\over \pi Gc^6} (\partial_t U_a)^2
}-\stackrel{(32)} {{1\over \pi Gc^6} (\partial_t U_{aa}) \partial_b
U_b } \nonumber\\ &+& \stackrel{(33)}{ {4\over \pi Gc^6} \partial_i
U_j \partial_t U_{ij}} +\stackrel{(34)} { {8\over \pi Gc^6} U U_i
\partial_t \partial_i U} - \stackrel{(35)}{ {4\over \pi Gc^6} {\hat
Z}_{ij} \partial^2_{ij} U} - \stackrel{(36)} { {4\over \pi Gc^6} U
\partial_i U \partial_t U_i } \nonumber\\ &-& \stackrel{(37)} {
{4\over \pi Gc^6}(\partial_t \partial_i U) {\hat R}_i} +
\stackrel{(38)} { {8\over \pi Gc^6} \partial_i U_j \partial_j {\hat
R}_i} - \stackrel{(39)}{ {2\over 3\pi Gc^6} \Delta (U^4)} -
\stackrel{(40)}{ {1\over \pi Gc^6} \Delta (U^2 U_{bb})} \nonumber\\
&-& \stackrel{(41)} {{1\over \pi Gc^6} \Delta (U^2 \partial_t^2 \chi)}
-\stackrel{(42)}{ {1\over 8 \pi Gc^6} \Delta (\partial_t^2 \chi
\partial_t^2 \chi)} - \stackrel{(43)}{ {1\over 24\pi Gc^6} \Delta (U
\partial^4_t P)} \nonumber\\ &-& \stackrel{(44)} { {1\over 4\pi Gc^6}
\Delta (U_{aa} \partial^2_t \chi)} - \stackrel{(45)}{ {1\over 4\pi
Gc^6} \Delta (\partial_t^2 \chi_{aa} U)} + \stackrel{(46)} { {1\over 2\pi
Gc^6} \Delta (K^{aa} U)} -\stackrel{(47)} { {1\over 4\pi Gc^6} \Delta
(U_{aa} U_{bb}) }\nonumber \\ &+& \stackrel{(48)}{ {1\over 2\pi Gc^6}
\Delta (U_{jk} U_{jk})} - \stackrel{(49)}{ {4\over \pi Gc^6} \Delta (U
{\hat X})} - \stackrel{(50)} { {2\over \pi Gc^6} \Delta (U {\hat
Z}_{aa})} \Biggr\}
\label{18a}
\;,\\ {\rm SII} &=& {1\over 2c^2(2l +3)} \sous{\!\!\!\!B= 0}{\rm FP} \int
d^3{\bf x}~\!|\tilde{\bf x}|^B \partial^2_t \Biggl\{ |{\bf x}|^2
\hat{x}_L \biggl[\stackrel{(1)} {\sigma} + \stackrel{(2)} {{4U\over
c^4} \sigma^{aa}} - \stackrel{(3)} {{2\over \pi Gc^4} U_i \partial_t
\partial_i U} \nonumber\\ &-& \stackrel{(4)} { {1\over \pi Gc^4}
(\partial_{ij}^2 U) U_{ij}} - \stackrel{(5)} {{1\over 2\pi Gc^4}
(\partial_t U)^2 }+\stackrel{(6)} { {2\over \pi Gc^4} \partial_i U_j
\partial_j U_i} \biggr] \nonumber\\ &-& \stackrel{(7)}{ {2l+3 \over
\pi Gc^2} \hat{x}_L U^2} - \stackrel{(8)} {{1\over 2\pi Gc^2}
\partial_i [\partial_i (U^2) |{\bf x}|^2 \hat{x}_L - U^2 \partial_i
(|{\bf x}|^2 \hat{x}_L)]} \nonumber \\ &-& \stackrel{(9)}{ {2l+3 \over
\pi Gc^4} \hat{x}_L U \partial_t^2 \chi}- \stackrel{(10)} {{1\over
2\pi Gc^4} \partial_i [\partial_i (U\partial^2_t \chi) |{\bf x}|^2
\hat{x}_L - U\partial^2_t \chi \partial_i (|{\bf x}|^2 \hat{x}_L)]}
\nonumber \\ &-& \stackrel{(11)} { {2l+3 \over \pi Gc^4} \hat{x}_L U
U_{aa}} - \stackrel{(12)} {{1\over 2\pi Gc^4} \partial_i [\partial_i
(U U_{aa}) |{\bf x}|^2 \hat{x}_L - U U_{aa} \partial_i (|{\bf x}|^2
\hat{x}_L)]} \nonumber\\ &-& \stackrel{(13)} {{4(2l+3)\over 3\pi Gc^4}
\hat{x}_L U^3} -\stackrel{(14)}{ {2\over 3\pi Gc^4} \partial_i
[\partial_i (U^3) |{\bf x}|^2 \hat{x}_L - U^3 \partial_i ( |{\bf x}|^2
\hat{x}_L)]} \Biggr\}
\label{18b}\;,\\
{\rm SIII} &=& {1\over 8c^4 (2l +3)(2l +5)} \sous{\!\!\!\!B= 0}{\rm
FP} \int d^3{\bf x}~\!|\tilde{\bf x}|^B \partial_t^4 \Biggl\{
\stackrel{(1)} {|{\bf x}|^4 \hat{x}_L \sigma}\nonumber\\ &-&
\stackrel{(2)} {{2(2l+5) \over \pi Gc^2}|{\bf x}|^2 \hat{x}_L U^2} -
\stackrel{(3)} {{1\over 2\pi Gc^2}\partial_i [\partial_i (U^2) |{\bf
x}|^4 \hat{x}_L - U^2 \partial_i (|{\bf x}|^4 \hat{x}_L)}] \Biggr\}
\label{18c}\;,\\
{\rm SIV} &=& {1\over 48c^6 (2l +3)(2l +5)(2l +7)} \sous{\!\!\!\!B=
0}{\rm FP} \int d^3{\bf x}~\!|\tilde{\bf x}|^B \stackrel{(1)} {|{\bf
x}|^6 \hat{x}_L \partial^6_t \sigma}
\label{18d}\;,\\ 
{\rm VI} &=& {-4 (2l +1)\over c^2 (l +1)(2l +3)} \sous{\!\!\!\!B=
0}{\rm FP} \int d^3{\bf x}~\!|\tilde{\bf x}|^B \hat{x}_{aL} \nonumber
\\ &\times & \partial_t \Biggl\{ \stackrel{(1)} {\sigma_a}
+\stackrel{(2)} { {2\over c^2} \sigma_a U} - \stackrel{(3)} {{2\over
c^2} \sigma U_a} + \stackrel{(4)} {{1\over \pi Gc^2} \partial_k U
\partial_a U_k } + \stackrel{(5)} {{3\over 4\pi Gc^2} \partial_t U
\partial_a U} \nonumber \\ &-& \stackrel{(6)} {{1\over 2\pi Gc^2}
\Delta (U U_a)} +\stackrel{(7)} { {\sigma_a\over c^4} \partial_t^2
\chi } +\stackrel{(8)} { {2\over c^4} \sigma_a U^2} \nonumber \\ &-&
\stackrel{(9)} {{1\over c^4} \sigma \partial_t^2 \chi_a}
-\stackrel{(10)} { {4\over c^4} \sigma {\hat R}_a} +\stackrel{(11)} {
{2\over c^4} U_a \sigma_{ss}} +\stackrel{(12)} { {2\over c^4} U_{ak}
\sigma_k} \nonumber \\ &+&\stackrel{(13)} { {2\over c^4} U_k
\sigma_{ak}} + \stackrel{(14)} {{1\over 2\pi Gc^4} \partial_k U
\partial_a \partial_t^2 \chi_k } \nonumber \\ &+& \stackrel{(15)}
{{1\over 2\pi Gc^4} (\partial_k \partial_t^2 \chi) (\partial_a U_k)
}+\stackrel{(16)} { {1\over 2\pi Gc^4} U_a \partial_t^2 U} \nonumber
\\ &+& \stackrel{(17)} {{3\over 8\pi Gc^4} \partial_t U (\partial_a
\partial_t^2 \chi)} +\stackrel{(18)} { {3\over 8\pi Gc^4}
(\partial_t^3 \chi) \partial_a U} -\stackrel{(19)} { {1\over 2\pi
Gc^4} U \partial_t^2 U_a }+\stackrel{(20)} { {1\over \pi Gc^4}
\partial_t U \partial_t U_a } \nonumber \\ &-& \stackrel{(21)}
{{2\over \pi Gc^4} U_k (\partial_k \partial_t U_a)} + \stackrel{(22)}
{{3\over 2\pi Gc^4} U \partial_t U \partial_a U }-\stackrel{(23)} {
{1\over \pi Gc^4} U_a \partial_k U \partial_k U} \nonumber \\ &+&
\stackrel{(24)} {{3\over 2\pi Gc^4} U_k \partial_a U \partial_k U
}+\stackrel{(25)} { {2\over \pi Gc^4} \partial_k U \partial_a {\hat
R}_k }-\stackrel{(26)} { {1\over \pi Gc^4} U_{kl}~\partial_{kl}^2 U_a}
+\stackrel{(27)} { {1\over \pi Gc^4} \partial_t U_{ak} \partial_k U}
\nonumber \\ &-& \stackrel{(28)} {{1\over \pi Gc^4} \partial_k U_l
\partial_a U_{kl}} + \stackrel{(29)} {{1\over \pi Gc^4} \partial_k
U_{al} \partial_l U_k} -\stackrel{(30)} { {1\over 2\pi Gc^4} \Delta
(U^2 U_a)} -\stackrel{(31)} { {1\over 4\pi Gc^4} \Delta (U
\partial^2_t \chi_a)} \nonumber\\ &-& \stackrel{(32)} {{1\over 4\pi
Gc^4} \Delta (\partial_t^2 \chi U_a)} -\stackrel{(33)} { {1\over \pi
Gc^4} \Delta (U {\hat R}_a)} -\stackrel{(34)} { {1\over 2\pi Gc^4}
\Delta (U_{kk} U_a) }+\stackrel{(35)} { {1\over 2\pi Gc^4} \Delta
(U_{ak} U_k)}\Biggr\}
\label{18e}\;,\\
{\rm VII} &=& {-2(2l +1)\over c^4(l +1)(2l +3)(2l +5)}
\sous{\!\!\!\!B= 0}{\rm FP} \int d^3{\bf x}~\!|\tilde{\bf x}|^B
\partial_t^3 \Biggl\{ |{\bf x}|^2 \hat{x}_{aL} \biggl[ \stackrel{(1)}
{\sigma_a} \nonumber\\ &+& \stackrel{(2)} {{2\over c^2} \sigma_a U
}-\stackrel{(3)} { {2\over c^2} \sigma U_a }+\stackrel{(4)} { {1\over
\pi Gc^2} \partial_k U \partial_a U_k} + \stackrel{(5)} {{3\over 4\pi
Gc^2} \partial_t U \partial_a U } \biggr] \nonumber \\ &-&
\stackrel{(6)} {{2l+5 \over \pi Gc^2} \hat{x}_{aL} U U_a} -\stackrel
{(7)} { {1\over 2\pi Gc^2} \partial_i [\partial_i (U U_a) |{\bf x}|^2
\hat{x}_{aL} - U U_a \partial_i (|{\bf x}|^2 \hat{x}_{aL})]}\Biggr\}
\label{18f}\;,\\
{\rm VIII} &=& {-(2l +1)\over 2c^6(l +1)(2l +3)(2l +5)
(2l +7)} \sous{\!\!\!\!B= 0}{\rm FP} \int d^3{\bf x}~\!|\tilde{\bf x}|^B 
\stackrel{(1)}  {\hat{x}_{aL} 
|{\bf x}|^4 \partial_t^5 \sigma^a}
\label{18g}\;,\\
{\rm TI} &=& {2(2l +1)\over c^4(l +1)(l +2)(2l +5)} \sous{\!\!\!\!B=
0}{\rm FP} \int d^3{\bf x}~\!|\tilde{\bf x}|^B \hat{x}_{abL}
\nonumber\\ &\times& \partial_t^2 \Biggl\{ \stackrel{(1)}{
\sigma_{ab}} + \stackrel{(2)} {{1\over 4\pi G} \partial_a U \partial_b
U }+\stackrel{(3)} { {4\over c^2} \sigma_{ab} U} -\stackrel{(4)} {
{4\over c^2} \sigma_{(a} U_{b)} } \nonumber\\ &+& \stackrel{(5)}
{{1\over 4\pi Gc^2} \partial_{(a} U \partial_{b)} \partial^2_t \chi
}+\stackrel{(6)} { {2\over \pi Gc^2} \partial_{(a} U \partial_t U_{b)}
}-\stackrel{(7)} { {1\over \pi Gc^2} \partial_a U_k \partial_b U_k
}\nonumber \\ &+& \stackrel{(8)} {{2\over \pi Gc^2} \partial_{(a} U_k
\partial_k U_{b)}} -\stackrel{(9)} { {1\over 2\pi Gc^2} \Delta (U_a
U_b)} \Biggr\}
\label{18h}\;,\\
{\rm TII} &=& {2l +1\over c^6 (l +1)(l +2)(2l +5)(2l +7)}
\sous{\!\!\!\!B= 0}{\rm FP} \int d^3{\bf x}~\!|\tilde{\bf x}|^B
\hat{x}_{abL} |{\bf x}|^2 \nonumber\\ & \times & \partial_t^4 \Biggl\{
\stackrel{(1)}{ \sigma_{ab}} +\stackrel{(2)}{ {1\over 4\pi G}
\partial_a U \partial_b U} \Biggr\}
\label{18i}
\;.\end{eqnarray}\end{subequations}$\!\!$
In the case of the 2PN current moment we write similarly

\begin{equation}\label{19}
J_L = \hbox{VI+VII+VIII+TI+TII}+{\cal O}(5) \;.\end{equation} The
expressions of these separate pieces have the same structure as the
corresponding V- and T-terms in the 3PN mass moment $I_L$. The
differences lie only in the over-all coefficient, in the number of
time-derivatives, and in the presence of a Levi-Civita symbol. We have
 
\begin{subequations}\label{20}\begin{eqnarray}
{\rm VI} &=& -\sous{\!\!\!\!B= 0}{\rm FP} \int d^3{\bf x}~\! |\tilde{\bf x}|^B
\varepsilon_{ab<i_l }~\hat{x}_{L-1>b} \Bigl\{ \hbox{same as in the
curly brackets of (\ref{18e})}\Bigr\} \;,\\ {\rm VII} &=& -{1\over
2c^2(2l +3)} \sous{\!\!\!\!B= 0}{\rm FP} \int d^3{\bf x}~\! |\tilde{\bf x}|^B
\varepsilon_{ab<i_l}~\hat{x}_{L-1>b} \partial_t^2 \Bigl\{\hbox{same as in
(\ref{18f})} \Bigr\} \;,\\ {\rm VIII} &=& -{1\over 8c^4 (2l+3)(2l+5)}
\sous{\!\!\!\!B= 0}{\rm FP} \int d^3{\bf x}~\! |\tilde{\bf x}|^B |{\bf x}|^4
\varepsilon_{ab<i_l}~\hat{x}_{L-1>b}\partial_t^4 \Bigl\{\hbox{same as in
(\ref{18g})} \Bigr\} \;,\\ {\rm TI} &=& {2l +1\over c^2(l +2)(2l +3)}
\sous{\!\!\!\!B= 0}{\rm FP} \int d^3{\bf x}~\! |\tilde{\bf x}|^B
\varepsilon_{ac<i_l}~\hat{x}_{L-1>bc} \partial_t \Bigl\{ \hbox{same as in
(\ref{18h})} \Bigr\}\;,\\ {\rm TII} &=& {2l +1 \over 2c^4(l +2)(2l
+3)(2l +5)}\sous{\!\!\!\!B= 0}{\rm FP} \int d^3{\bf x}~\! |\tilde{\bf x}|^B |{\bf
x}|^2 \varepsilon_{ac<i_l }~ \hat{x}_{L-1>bc} \partial_t^3 \Bigl\{
\hbox{same as in (\ref{18i})} \Bigr\}
\;.\end{eqnarray}\end{subequations}$\!\!$

We explained that we denote the terms in the previous formulas by SI(1),
SI(2), ..., SI(50), SII(1), ..., TII(2). Our convention is that this
notation means that the terms involve their complete coefficient in
front; for instance,

\begin{subequations}\label{21}\begin{eqnarray}
{\rm SI(5)} &=& -{1\over \pi Gc^4} \sous{\!\!\!\!B= 0}{\rm FP} \int
d^3{\bf x}~\!|\tilde{\bf x}|^B \hat{x}_L U_{ij} \partial^2_{ij} U
\label{22a}\;,\\ {\rm SII(14)} &=& -{1\over 3\pi Gc^6(2l +3)}
\sous{\!\!\!\!B= 0}{\rm FP} \int d^3{\bf x}~\!|\tilde{\bf x}|^B
\partial^2_t \partial_i \bigl[\partial_i (U^3) |{\bf x}|^2 \hat{x}_L -
U^3 \partial_i ( |{\bf x}|^2 \hat{x}_L)\bigr]\label{22b}\;,\\ {\rm
TI(7)} &=& -{2(2l +1)\over \pi Gc^6(l +1)(l +2)(2l +5)}
\sous{\!\!\!\!B= 0}{\rm FP} \int d^3{\bf x}~\!|\tilde{\bf x}|^B
\hat{x}_{abL}\partial_t^2 \bigl\{ \partial_a U_k \partial_b U_k
\bigr\}\label{22c} \;.\end{eqnarray}\end{subequations}$\!\!$ The
notation means also that the terms include all the post-Newtonian
corrections relevant to obtain the 3PN order in the energy
flux. Consistently with that order we shall have to compute the mass
quadrupole moment $I_{ij}$ to the 3PN order, the mass octupole
$I_{ijk}$ and current quadrupole $J_{ij}$ to the 2PN order only. Look
for instance at the term SI(5) given by Eq. (\ref{22a}): this is a 2PN
term since it carries a factor $1/c^4$. Thus, in the mass quadrupole
$I_{ij}$ we need to compute SI(5) with 1PN relative precision, while
in the mass octupole $I_{ijk}$ the Newtonian precision is sufficient
[the term SI(5) does not exist in the current moments]. Note also that
a term such as SII(14) given by (\ref{22b}) includes in fact two terms
(which come from an operation by parts). Furthermore, since the
different pieces (of types V and T) composing the current moments have
exactly the same structure as in the mass moments, we employ the same
notation for these terms in both $I_L$ and $J_L$. For instance TI(7)
denotes both the term in the mass moment as given by Eq. (\ref{22c})
and the corresponding term in the current moment (with a little
experience there can be no confusion). Finally, in some cases we split
the term into subterms according to the nature of a potential therein,
either ``compact'' or ``non-compact'' potential. The compact
(respectively non-compact) part of a potential is that part which is
generated by a source with compact (non-compact) support. For instance
the term SI(5), which contains the potential $U_{ij}$ given by
Eq. (\ref{14c}), is naturally split into the two contributions

\begin{equation}\label{22}
{\rm SI(5)}={\rm SI(5C)}+{\rm SI(5NC)} 
\;,\end{equation}
where $U_{ij}$ is replaced by its compact (C) or
non-compact (NC) parts given by

\begin{subequations}\label{23}\begin{eqnarray}
U_{ij} &=& U_{ij}^{({\rm C})}+U_{ij}^{({\rm NC})} \;,\\ U_{ij}^{({\rm
C})} &=& \Delta^{-1}\{ -4 \pi G (\sigma_{ij} - \delta_{ij}
\sigma_{kk})\} \;,\\ U_{ij}^{({\rm NC})} &=& \Delta^{-1}\{ - \partial_i U
\partial_j U\}
\;.\end{eqnarray}\end{subequations}$\!\!$
We shall split similarly all the terms containing the potentials
$U_{ij}$, $\chi_{ij}$, ${\hat R}_i$, ${\hat Z}_{ij}$ and ${\hat
X}$. This splitting into C and NC parts is fairly obvious from the
expressions of the potentials: for instance,

\begin{equation}\label{24}
{\hat R}_i^{({\rm NC})} = \Delta^{-1}\Bigl\{ - 2 \partial_k U
\partial_i U_k - {3\over 2} \partial_t U \partial_i U\Bigr\}
\;.\end{equation} When computing the terms in the moments
(\ref{17})-(\ref{20}) we shall separate them into various categories,
according to the way their computation is performed. This entails
introducing some new terminology for the various classes. For instance
we shall consider the compact-support terms like SI(1), or so-called
Y-terms made of the quadratic product of two $U$-type potentials
[examples are VI(4) and also SI(5C)], or so-called non-compact terms
like SI(5NC) or SII(4NC). These categories of terms are defined when
we tackle their computation. The resulting nomenclature is complicated
but turned out to be useful during the explicit computation and the
many associated checks, since it delineates clearly the different
problems posed by the different categories of terms.

\section{Application to point-particles}

Our aim is to compute the multipole moments for a system of two
point-like particles. One is not allowed {\it a priori} to use the
expressions (\ref{5}) as they have been obtained in
Ref. \cite{B98mult} under the assumption of a continuous (smooth)
source. Applying them to a system of point-particles, we find that the
integrals are divergent at the location of the particles, i.e. when
${\bf x} \to {\bf y}_1(t)$ or ${\bf y}_2(t)$, where ${\bf y}_1(t)$ and
${\bf y}_2(t)$ denote the two trajectories. Therefore we must
supplement the computation by a prescription for how to remove the
infinite part of these integrals. In this paper, we systematically
employ the Hadamard regularization \cite{Hadamard,Schwartz} (see
Ref. \cite{Sellier} for an entry to the mathematical literature). The
usefulness of this regularization for problems involving
point-particles in general relativity has been shown by numerous works
(see e.g. \cite{BFP98}). Recently the properties of the Hadamard
regularization have been re-visited and a new set of generalized
functions (distributional forms) associated with this regularization
were introduced \cite{BFreg,BFregM}.

The functions $F({\bf x})$ we need to deal with are smooth on
${\mathbb R}^3$ excised of the two points ${\bf y}_1$ and ${\bf
y}_2$, and admit when $r_1=|{\bf x}-{\bf y}_1|\to 0$ (and similarly
when $r_2=|{\bf x}-{\bf y}_2|\to 0$) a singular expansion of the type

\begin{equation}\label{25}
\forall n\in {\mathbb N}\;,\quad F({\bf x})=\sum_{a_0\leq a\leq n}
r_1^a \!\!\sous{1}{f}_{a}({\bf n}_1)+o(r_1^n) \;,\end{equation} where
the coefficients ${}_1f_a$ of the various powers of $r_1$ in the
expansion depend on the unit direction ${\bf n}_1=({\bf x}-{\bf
y}_1)/r_1$. The powers $a$ of $r_1$ are real, range in discrete steps
(i.e. $a$ belongs to some countable set $(a_i)_{i\in {\mathbb N}}$)
and are bounded from below ($a_0\leq a$). The functions like $F$ are
said to belong to the class of functions ${\cal F}$ (see
Ref. \cite{BFreg} for precise definitions). If $F$ and $G$ belong to
${\cal F}$ so do the ordinary (pointwise) product $FG$ and the
ordinary gradient $\partial_iF$. The Hadamard ``partie finie'' of $F$
at the location of particle 1 is defined as

\begin{equation}\label{26}
(F)_1= \int {d\Omega_1\over 4\pi} \!\!\sous{1}{f}_0({\bf n}_1) 
\;,\end{equation}
where $d\Omega_1= d\Omega ({\bf n}_1)$ is the solid angle
element centered on ${\bf y}_1$ and of direction ${\bf n}_1$. On the
other hand, the Hadamard partie finie (${\rm Pf}$) of the integral
$\int d^3{\bf x}~F$, divergent because of the two
singular points ${\bf y}_1$ and ${\bf y}_2$, is defined by

\begin{eqnarray}\label{27}
{\rm Pf}_{u_1,u_2}\int d^3{\bf x}~ F &=&~\lim_{u\to
0}~\biggl\{\int_{r_1>u\atop r_2>u} d^3{\bf x}~ F\nonumber\\ &&\qquad
+~4\pi\sum_{a+3< 0}{u^{a+3}\over a+3} \left({F\over r_1^a}\right)_1+ 4
\pi \ln\left({u\over u_1}\right) \left(r_1^3 F\right)_1
+1\leftrightarrow 2\biggr\}\;.
\end{eqnarray}
The first term represents the integral on ${\mathbb R}^3$ excluding  
 two spherical volumes of radius $u$ surrounding the
singularities. The other terms are such that they cancel out the
divergent part of the latter integral when $u\to 0$ (the symbol
$1\leftrightarrow 2$ means the terms obtained by exchanging the labels
1 and 2). Notice the presence of a logarithmic term, which depends on
an arbitrary constant $u_1$, and similarly $u_2$ for the other
singularity. In this paper we shall keep the constants $u_1$ and $u_2$
all the way through our calculation. We assume nothing about these
constants, for instance they are different {\it a priori} from  similar
constants $s_1$ and $s_2$ introduced in the equations of motion
(Section II in \cite{BFeom}).  We shall see that the multipole moments
do depend on $u_1$ and $u_2$ (as well as on $r_0$) at the 3PN order.

The strategy we adopt in this paper is to insert into the source
multipole moments (\ref{5}) the following expression of the matter
stress-energy tensor $T^{\mu\nu}$ for two point-masses,

\begin{subequations}\label{29}\begin{eqnarray}
T^{\mu\nu}_{\rm point-particle}&=& m_1 v_1^\mu v_1^\nu \left({dt\over
d\tau}\right)_1 \left({1\over \sqrt{-g}}\right)_1\delta ({\bf x}-{\bf
y}_1) + 1\leftrightarrow 2 \;,\\ \left({dt\over
d\tau}\right)_1&=&{1\over \sqrt{-(g_{\rho\sigma})_1 v_1^\rho
v_1^\sigma/c^2}} \;,\end{eqnarray}\end{subequations}$\!\!$ where $m_1$
is the (Schwarzschild) mass, ${\bf y}_1(t)$ the trajectory, and ${\bf
v}_1(t)=d{\bf y}_1/dt$ the velocity of body 1 [with $v_1^\mu=(c,{\bf
v}_1)$]. This stress-energy tensor constitutes a ``naive'' model to
describe the particles, since the factors of the Dirac distribution
have been evaluated at the point 1 by means of the regularization
defined by Eq. (\ref{26}). However, because of the so-called
non-distributivity of the Hadamard partie finie, other tensors are
possible as well. In particular, we discuss in Section X the effect of
choosing another stress-energy tensor, which is particularly natural
within the context of the Hadamard regularization, and that we
proposed in Ref. \cite{BFregM}. After $T^{\mu\nu}_{\rm
point-particle}$ is substituted inside them, the moments comprise of
many divergent integrals and we define each of these integrals by
means of the Hadamard partie finie (\ref{27}). Therefore our ansatz
for applying the general ``fluid'' formalism to the ill-defined case
of point-particles is

\begin{subequations}\label{30}\begin{eqnarray}
(I_L)_{\rm point-particle} &=& {\rm Pf} \Bigl\{ I_L [T^{\mu\nu}_ {\rm
point-particle}] \Bigr\} \;,\\ (J_L)_{\rm point-particle} &=& {\rm Pf}
\Bigl\{ J_L [T^{\mu\nu}_ {\rm point-particle}] \Bigr\}
\;,\end{eqnarray}\end{subequations}$\!\!$ where the functionals $I_L$
and $J_L$ are exactly the ones given by (\ref{5}) or
(\ref{17})-(\ref{20}) (including in particular the finite part ${\rm
FP}_{B=0}$ at infinity). In what follows we shall carefully apply this
prescription, but in order to reduce clutter we generally omit writing
the partie-finie symbol ${\rm Pf}$.

The relative position and velocity of two bodies in harmonic
coordinates are denoted by

\begin{equation}\label{31}
x^i =y_1^i - y_2^i \ ; \quad\hbox{and}\quad v^i = {dx^i \over dt}
= v_1^i - v_2^i 
\;.\end{equation}
To the 2PN order (only needed in this paper) the relation between the
absolute trajectories in a center-of-mass frame and the relative ones
reads, in the case of a circular orbit (see e.g. Ref. \cite{BDI95}), as

\begin{subequations}\label{32}\begin{eqnarray}
y_1^i &=& {m_2+3\nu~\! \gamma^2 \delta m \over m} x^i + {\cal O}(5)\;,\\
y_1^i &=& {-m_1+3\nu~\! \gamma^2 \delta m \over m} x^i + {\cal O}(5)
\;.\end{eqnarray}\end{subequations}$\!\!$
Here $m_1$ and $m_2$ are the two
masses, with $m=m_1+m_2$, $\nu=m_1 m_2/m^2$ (such that $0<\nu\leq
1/4$) and $\delta m=m_1-m_2$. Furthermore,

\begin{equation}\label{33}
\gamma = {Gm\over rc^2}
\;,\end{equation}
represents a small post-Newtonian parameter of order ${\cal O}(2)$, with
$r=|{\bf x}|$, often also denoted $r_{12}$, the distance between the
two masses in harmonic coordinates.

When computing the multipole moments we get many terms involving
accelerations and derivatives of accelerations. These are reduced to
the consistent post-Newtonian order by means of the binary's equations
of motion. To control the moments at the 3PN order we need the
equations of motion at the 2PN order. For circular orbits these
equations are (see e.g. \cite{BDI95})

\begin{subequations}\label{34}\begin{eqnarray}
{d{\bf v}\over dt} &=& -\omega^2{\bf x}+{\cal O}(5) \label{34a}\;,\\
\omega^2 &=& {Gm\over r^3} \left\{1+\left[-3+\nu\right]\gamma +
\left[6+{41\over 4}\nu + \nu^2 \right] \gamma^2 +{\cal
O}\left(\gamma^3\right)\right\} \label{34b}
\;.\end{eqnarray}\end{subequations}$\!\!$ 
The content of these equations lies in the relation (\ref{34b})
between the orbital frequency $\omega$ and the coordinate separation
$r$ in harmonic coordinates. However, note that the precision given by
the equations (\ref{34}) is insufficient to obtain the (second and
higher) time-derivatives of the moments at the 3PN order. Evidently
for this we need the more accurate 3PN equations of motion. These will
be given in Section XII when we compute the total energy flux [see
Eq. (\ref{121}) below].  In addition, we shall also need for some
intermediate computations the equations of motion for general (not
necessarily circular) orbits but at the 1PN order. These are given by

\begin{eqnarray}\label{34'}
{d{\bf v}_1\over dt} &=& -{Gm_2\over r^2} {\bf n} \nonumber\\
&+&{Gm_2\over c^2r^2} \left\{{\bf n} \left[ -v^2_1 - 2v^2_2 +
4(v_1v_2) +{3\over 2} (nv_2)^2 + 5{Gm_1\over r} + 4{Gm_2\over r}
\right]\right.\nonumber\\ &+&\left. {\bf v} \left[4(nv_1)
-3(nv_2)\right] \right\} +{\cal O}(4)
\end{eqnarray}
(and {\it idem} for $1\leftrightarrow 2$). The notation $(nv_1)$ for
instance means the usual scalar product between the vectors ${\bf
n}={\bf x}/r$ (sometimes denoted also ${\bf n}_{12}$) and ${\bf
v}_1$. With these preliminary inputs in place, we are in a position to
tackle the computation of each of the terms composing the multipole
moments (\ref{17})-(\ref{20}).

\section{Compact terms}

In this category we consider all the terms in (\ref{17})-(\ref{20})
whose integrand involves explicitly the matter densities $\sigma$,
$\sigma_i$ or $\sigma_{ij}$ as a factor, and thus which extend only
over the spatially compact support of the source. For these terms the
finite part operation ${\rm FP}_{B=0}$ (which deals with the bound at
infinity of the integral) can be dropped out. With the present
notation the compact terms are

\medskip\noindent
(i) compact term at Newtonian order: SI(1); 

\medskip\noindent
(ii) compact terms at 1PN order: SII(1), VI(1); 

\medskip\noindent
(iii) compacts at 2PN: SI(3), SIII(1), VI(2), VI(3), VII(1), TI(1);

\medskip\noindent
(iv) compacts at 3PN: SI(13), SI(14), SI(15), SI(16C),
SII(2), SIV(1), VI(7), VI(8), VI(9), VI(10C), VI(11), VI(12C), VI(13),
VII(2), VII(3), VIII(1), TI(3), TI(4), TII(1).

\medskip\noindent As explained earlier, it is convenient, when the
potential is composed of both compact and non-compact parts, to
separate out these pieces. Thus we shall also have the compact terms
involving the non-compact part of a potential, namely

\medskip\noindent
SI(16NC), VI(10NC), VI(12NC).

\medskip\noindent Evidently we have to compute the ``Newtonian'' term
SI(1) with the maximal 3PN precision, while for instance a term which
appears at 3PN needs only the Newtonian precision. We devote this
section to the computation of the Newtonian term SI(1), and to one
example of a compact term with non-compact potential: SI(16NC); the
computation of the other compact terms is similar, or does not present
any difficulty, so we  only list  the final  results in Appendix A.

From the stress-energy tensor (\ref{29}) we find that the matter
source densities (\ref{9}) are given by

\begin{subequations}\label{35}\begin{eqnarray}
\sigma ({\bf x},t) &=& \tilde{\mu}_1 \delta [{\bf x}-{\bf y}_1(t)] +
1\leftrightarrow 2 \;,\\ \sigma_i ({\bf x},t) &=& \mu_1 v_1^i \delta
[{\bf x}-{\bf y}_1 (t)]+ 1\leftrightarrow 2 \;,\\ \sigma_{ij}({\bf x},t)
&=& \mu_1 v_1^i v_1^j \delta [{\bf x} -{\bf y}_1 (t)]+
1\leftrightarrow 2
\;,\end{eqnarray}\end{subequations}$\!\!$
where we have  introduced some ``effective'' masses $\mu_1$ and ${\tilde
\mu}_1$  defined  by

\begin{subequations}\label{36}\begin{eqnarray}
\mu_1(t)&=& m_1 \left({dt\over d\tau}\right)_1\left({1\over
\sqrt{-g}}\right)_1\;,\\ {\tilde\mu}_1(t)&=&\mu_1(t)\left[1+{{\bf
v}_1^2 \over c^2}\right] \;.\end{eqnarray}\end{subequations}$\!\!$
These effective masses are some mere functions of time $t$ through the
dependence over the particle trajectories and velocities (the
accelerations are order-reduced). Notice that, had we used the
stress-energy tensor proposed in Section V of \cite{BFregM} (see also
the discussion in Section X below), we would have found that $\mu_1$
and ${\tilde \mu}_1$ depend both on time and space, as they contain
the factor $1/\sqrt{-g}$ that is given at any field point ${\bf
x}$. Using the metric (\ref{8}), expressed in terms of the retarded
potentials (\ref{10})-(\ref{12}), we find the expressions of the two
required factors entering the effective masses (\ref{36}) up to the 3PN
order; namely

\begin{subequations}\label{37}\begin{eqnarray}
\left({1\over \sqrt{-g}}\right)_1 &=& \left( 1 - {2\over c^2}V +
{1\over c^4} \left [-2\hat{W} + 2V^2 \right]\right.  \nonumber \\ &+&
\left. {1\over c^6} \left [-8 \hat{Z} - 8\hat{X} + 4 V \hat{W} - 8 V_i
V_i - {4\over 3} V^3 \right] \right)_1 + {\cal O}(8)\;,\\
\left({dt\over d\tau}\right)_1 &=& \left( 1 + {1\over c^2} \left[V +
{1\over 2} v_1^2\right] + {1\over c^4} \left[{1\over 2} V^2 + {5\over
2} V v_1^2 - 4 V_i v_1^i + {3\over 8} v_1^4\right] \right. \nonumber
\\ &&\quad + {1\over c^6} \left[4{\hat X} + 4 V_i V_i - 8 {\hat R}_i
v_1^i + 2 {\hat W}_{ij} v_1^i v_1^j - 12V V_i v_1^i \right. \nonumber
\\ &&\quad - \left. \left. 6 V_i v_1^i v_1^2 + {1\over 6} V^3 +
{25\over 4}V^2 v_1^2 + {27\over 8}V v_1^4 + {5\over 16} v_1^6 \right]
\right)_1 + {\cal O}(8) \;,\end{eqnarray}\end{subequations}$\!\!$ where
the subscript 1 means that all the potentials are to be evaluated
following the regularization (\ref{26}). In these expressions there
are no problems associated with the non-distributivity of the Hadamard
partie-finie; that is, we can assume $(FG)_1 = (F)_1(G)_1$ for this
computation (see, however, Section X). Most of the regularized values
of the needed potentials at 1 (for general orbits) have been computed
in Ref. \cite{BFP98} (see the Appendix B there). Here we simply report
the appropriate formulas (where $r_{12}=|{\bf y}_1-{\bf y}_2|$, ${\bf
n}_{12}=({\bf y}_1-{\bf y}_2)/r_{12}$).

\begin{subequations}
\label{39} 
\begin{eqnarray}
(V)_1 &=& {Gm_2\over r_{12}} \left\{1+{1\over c^2}\left[-{3\over
2}{Gm_1\over r_{12}}+2v_2^2-{1\over 2}(n_{12}v_2)^2\right] +{4\over
3}{Gm_1\over r_{12}c^3}(n_{12}v_{12}) \right.\nonumber\\ & &\qquad
+{Gm_1\over r_{12}c^4}\left[ {11\over 2}{Gm_1\over r_{12}}+{5\over
4}{Gm_2\over r_{12}}+{15\over 8}v_1^2-{7\over 4}(v_1v_2)-{25\over
8}v_2^2 \right. \nonumber\\ & &\qquad \qquad \quad \left.+ {1\over
8}(n_{12}v_1)^2-{25\over 4}(n_{12}v_1)(n_{12}v_2)+{33\over
8}(n_{12}v_2)^2\right] \nonumber\\ & &\left. \qquad+{1\over
c^4}\left[2v_2^4-{3\over 2}(n_{12}v_2)^2v_2^2+{3\over
8}(n_{12}v_2)^4\right]\right\}+{\cal O}(5) \;,\\ (V_i)_1 &=& {Gm_2\over
r_{12}}\biggl\{v_2^i +{v_2^i\over c^2}\left[-2{Gm_1\over
r_{12}}+v_2^2-{1\over 2}(n_{12}v_2)^2\right] +{1\over 2}{Gm_1\over
r_{12}c^2}v_1^i \nonumber\\ & &\qquad+{Gm_1\over
r_{12}c^2}n_{12}^i\left[-{3\over 2}(n_{12}v_1)+{1\over
2}(n_{12}v_2)\right]+{\cal O}(3) \;,\\ ({\hat W}_{ij})_1 &=& {Gm_2\over
r_{12}}\biggl\{v_2^{ij}-\delta^{ij}v_2^2+ {Gm_1\over
r_{12}}\left[-2n_{12}^{ij}+\delta^{ij}\right]\nonumber\\
&&\qquad+{Gm_2\over
4r_{12}}\left[n_{12}^{ij}-\delta^{ij}\right]\biggr\}+{\cal O}(1)
\label{39''}\;,\\
({\hat R}_i)_1 &=& {G^2m_1m_2\over r_{12}^2}\left[-{3\over
4}v_1^i+{5\over 4}v_2^i-{1\over 2}(n_{12}v_1)n_{12}^i-{1\over
2}(n_{12}v_2)n_{12}^i\right] \nonumber\\ &+&{G^2m_2^2\over
r_{12}^2}\left[-{1\over 8}v_2^i+{1\over 8}
(n_{12}v_2)n_{12}^i\right]+{\cal O}(1) \;,\\ ({\hat X})_1 &=&
{G^2m_1m_2\over r_{12}^2}\left[-{3\over 2}{Gm_1\over r_{12}}+{1\over
4}v_1^2- 2(v_1v_2)+{9\over 4}v_2^2 \right.\nonumber\\
&&\quad-\left.{11\over 4}(n_{12}v_1)^2+{9\over
2}(n_{12}v_1)(n_{12}v_2) -{11\over 4}(n_{12}v_2)^2\right] \nonumber\\
&+&{G^2m_2^2\over r_{12}^2}\left[{1\over 12}{Gm_2\over r_{12}}-{1\over
8}v_2^2+{1\over 8}(n_{12}v_2)^2\right]+{\cal O}(1)
\;.\end{eqnarray}\end{subequations}$\!\!$
Notice that during the computation of the potential $V$ at the 2PN
order we used the 1PN equations of motion for general orbits: these
are given by Eq. (\ref{34'}). In addition to the above, we need the
trace ${\hat W}={\hat W}_{ii}$ at 1PN order. [To the order considered
in (\ref{39''}) we have $U_{ij}={\hat W}_{ij}$.] By a computation
similar to those of Ref. \cite{BFP98} we get

\begin{eqnarray}\label{40} 
(\hat{W})_1&=& {Gm_2\over r_{12}} \left [{Gm_1\over r_{12}} - {1\over
2} ~{Gm_2\over r_{12}} - 2v_2^2 \right] \nonumber\\
&-&{2G^2m_1m_2\over r_{12}^2c}(n_{12}v_{12}) + {G^2m_1m_2\over
r_{12}^2c^2} \left [-3{Gm_1\over r_{12}} + {1\over 2}~{Gm_2\over
r_{12}} \right. \nonumber \\ &+& \left. {3\over 2}v_1^2 + {13\over
2}v_2^2 + (n_{12}v_1)^2 + 2(n_{12}v_1) (n_{12}v_2) - 2(n_{12}v_2)^2
\right] \nonumber\\ &+& {G^2m_2^2\over r_{12}^2c^2} \left [-{9\over
4}v_2^2 + {3\over 4} (n_{12} v_2)^2 \right] + {Gm_2\over r_{12}c^2}
\left [-2v_2^4 + (n_{12}v_2)^2 v_2^2 \right] + {\cal O}(3)
\;.\end{eqnarray} Inserting these expressions into (\ref{37}) we
obtain the 3PN ${\tilde\mu}_1$ and then straightforwardly compute
SI(1). In the quadrupole case $l=2$ it is given by
  
\begin{eqnarray}\label{41} 
{\rm SI(1)}&=&\int d^3{\bf x}~{\tilde\mu}_1 {\hat
x}^{ij}\delta_1+1\leftrightarrow 2\nonumber\\ &=&{\tilde\mu}_1
y_1^{<i}y_1^{j>}+1\leftrightarrow 2
\;.\end{eqnarray}
The final result for circular orbits [using the relations
(\ref{32})-(\ref{33})] reads then

\begin{eqnarray}\label{42} 
{\rm SI(1)}&=&m\nu \left[1+ {\gamma \over 2}(1-5\nu) - {\gamma^2
\over 8} (13-61\nu +25\nu^2) \right. \nonumber \\ &+&
\left. {\gamma^3\over 16}(149-573\nu +354\nu^2-29\nu^3)
\right]{\hat x}_{ij}
\;.\end{eqnarray}
The sensitivity of this result to the choice of stress-energy tensor
for point-particles (in accordance with the ``non-distributivity'' of
the partie finie) is discussed in Section X.

Other interesting terms in this category are

\begin{equation}\label{42'}
{\rm SI(16NC)} = {4\over c^6} \sous{\!\!\!\!B= 0}{\rm FP} \int d^3{\bf
x}~\!|\tilde{\bf x}|^B \hat{x}_L \sigma _{ab}U_{ab}^{\rm (NC)}
\;,\end{equation}
and the similar VI(10NC) and VI(12NC).
Applying our computation rules we get

\begin{equation}\label{42''}
{\rm SI(16NC)} = {4m_1\over
c^6}v_1^{ab}\left((y_1+r_1n_1)^{<i}(y_1+r_1n_1)^{j>}U_{ab}^{\rm
(NC)}\right)_1+1\leftrightarrow 2 \;,\end{equation} where we have
written $x^i=y_1^i+r_1n_1^i$ valid in the vicinity of the point 1. The
result follows from applying the regularization (\ref{26}), with the
help of the Newtonian approximation of the NC potential. The
interesting point is that the regularized factor in (\ref{42''}) is
different from $y_1^{<i}y_1^{j>}\left(U_{ab}^{\rm (NC)}\right)_1$ as a
consequence of the non-distributivity. See Section X.

\section{Quadratic terms}

In this category we consider all the terms whose support is spatially
non-compact (hence the finite part operation ${\rm FP}_{B=0}$ plays a
crucial role), and which are made of the integral of a product of two
derivatives of compact-support potentials. Furthermore we sub-divide
the quadratic terms into sub-categories $Y$-, $S$- and $T$-terms named
after the functions $Y_L$, $S_L$ and $T_L$ defined below, and we
classify all these terms according to their dominant post-Newtonian
order. The exhaustive list follows.

\medskip\noindent
(i) $Y$-terms at 2PN: 
SI(4), SI(6), SI(7), SII(7), VI(4), VI(5), TI(2);

\medskip\noindent (ii) $Y$-terms at 3PN: SI(31), SI(35C), SI(37C),
SI(38C), VI(16), VI(20), VII(6), VI(19), VI(21), VI(25C), VI(26C),
VI(27C), VI(29C), TI(6), TI(7), TI(8);

\medskip\noindent
(iii) $S$-terms at 3PN: 
SII(3), SII(4C), SII(5), SII(6), SIII(2), VII(4), VII(5), TII(2); 

\medskip\noindent
(iv) $T$-terms at 3PN: 
SI(17), SI(19C), SI(21C), SI(24), SI(25), SII(9), 
VI(14), VI(15), VI(17), VI(18), TI(5).

\medskip\noindent The $Y$- and $S$-terms involve the product of two
compact-support potentials $U$, $U_i$ or $U_{ij}^{({\rm C})}$, while
the $T$-terms involve a product of one of the latter potentials (of
type $U$) and a potential of the type $\chi$, $\chi_i$ or
$\chi_{ij}^{({\rm C})}$ [see (\ref{14})]. Compared to $Y$-terms, the
$S$-terms contain in addition a factor $|{\bf x}|^2$ inside their
integrand. In the two-body case these compact-support $U$-type
potentials read

\begin{subequations}\label{44}\begin{eqnarray}
U &=& {G{\tilde \mu}_1\over r_1}+1\leftrightarrow 2\;,\\ U_i &=&
{G\mu_1\over r_1} v_1^i+1\leftrightarrow 2\;,\\ U_{ij}^{({\rm C})} &=&
{G\mu_1\over r_1}(v_1^{ij}-\delta^{ij}v_1^2) +1\leftrightarrow 2
\;.\end{eqnarray}\end{subequations}$\!\!$
The potentials of type $\chi$ are obtained by replacing $1/r_1$ by
$r_1$ in these expressions.  Then from the structure $\sim
1/r_1+1/r_2$ or $\sim r_1+r_2$ it is not difficult to express all the
$Y$-, $S$- and $T$-terms with the help of three and only three types
of elementary integrals $Y_L$, $S_L$ and $T_L$ respectively (where
$L=i_1i_2\cdots i_l$ denotes the multipolar index). Two examples in
the quadrupole case $ij$ are

\begin{subequations}\label{45}\begin{eqnarray}
{\rm SI(4)} &=& -{4G\over c^4} \mu_1[\tilde{\mu}_2 v_1^a v_2^b
\!\!\!\sous{2}{\partial}_{~\!ab} Y_{ij} + {\dot {\tilde \mu}}_2 v_1^a
\!\!\!\sous{2}{\partial}_{~\!a} Y_{ij}] + 1\leftrightarrow 2 \;,\\
{\rm SII(4C)} &=& {G\over 7c^6} m_1 m_2 \frac{d^2}{dt^2}[
(v_2^{ab}-\delta^{ab} v_2^2) \!\!\!\sous{1}{\partial}_{~\!ab} S_{ij}]
+ 1\leftrightarrow 2 \;.\end{eqnarray}\end{subequations}$\!\!$ Since
SI(4) is a 2PN term it needs the relative 1PN precision (for
simplicity we do not write the post-Newtonian remainders). The
elementary integrals are defined by

\begin{subequations}\label{46}\begin{eqnarray}
Y_L ({\bf y}_1,{\bf y}_2) &=& -{1\over 2\pi}\sous{\!\!\!\!B= 0}{\rm
FP} \int d^3{\bf x}~\! |\tilde{\bf x}|^B {\hat{x}_L \over r_1 r_2}
\label{46a}\;,\\ S_L ({\bf y}_1,{\bf y}_2) &=& -{1\over
2\pi}\sous{\!\!\!\!B= 0}{\rm FP} \int d^3{\bf x}~\! |\tilde{\bf x}|^B
|{\bf x}|^2{\hat{x}_L\over r_1 r_2} \;,\\ T_L ({\bf y}_1,{\bf y}_2) &=&
-{1\over 2\pi}\sous{\!\!\!\!B= 0}{\rm FP} \int d^3{\bf x}~\!
|\tilde{\bf x}|^B \hat{x}_L {r_1 \over r_2}
\;.\end{eqnarray}\end{subequations}$\!\!$
In these definitions, the finite part at infinity is absolutely
crucial (it comes directly from the formalism
\cite{B95,B98mult}). However, it is easily seen that the integrals are
convergent near the two bodies so the Hadamard partie finie is not
needed. The integral $Y_L$ agrees with the definition used in
\cite{B95,BDI95} and is equivalent with the alternative form proposed
in Ref. \cite{DI91a}.

We present several derivations of the closed-form expressions of these
integrals for arbitrary $l$. This permits us to introduce some
techniques which are necessary when we compute some more complicated
integrals in Sections VIII and IX. The first method consists of
writing the multipolarity factor $\hat{x}_L$ in the form

\begin{equation}\label{47}
\hat{x}_L = \sum_{p=0}^l {l \choose p} r_1^{<P} y_1^{L-P>}
\;,\end{equation} where ${l \choose p}$ denotes the binomial
coefficient (and $<>$ refers to the STF projection). Inserting this
into the integral $Y_L$, it is easy to obtain the equivalent
expression

\begin{equation}\label{48}
Y_L = -{1\over 2\pi} \sum_{p=0}^l {l \choose p} {(-)^p\over (2p-1)!!}
y_1^{<L-P} \!\!\!\sous{1}{\partial}_{~\!P>} \left\{\sous{\!\!\!\!B=
0}{\rm FP} \int d^3{\bf x}~\! |\tilde{\bf x}|^B {r_1^{2p-1}\over r_2}
\right\}
\;.\end{equation} 
Next we compute the integral inside the brackets of (\ref{48}). Let us
show that the polar part of this integral when $B\to 0$ is zero. We
replace the integrand by its expansion when $|{\bf x}| \to \infty$
(any pole at $B=0$ necessarily comes from the behaviour of the
integral at infinity), we integrate over the angles and look for
radial integrals of the type $\int^{+\infty} d|{\bf x}|~\!|{\bf
x}|^{B-1}$ which are the only ones to produce a pole. However these
radial integrals do not exist since after the angular integration the
powers of $|{\bf x}|$ are only of the type $B+2k$ where $k$ is an
integer. So the integral in (\ref{48}) can be computed by analytic
continuation down to the value $B=0$. We obtain ($\forall p\in
{\mathbb N}$)

\begin{equation}\label{49}
\sous{\!\!\!\!B= 0}{\rm FP} \int d^3{\bf x}~|{\tilde {\bf x}}|^B
{r_1^{2p-1}\over r_2} = -{2\pi ~\!r_{12}^{2p+1} \over (p+1)(2p+1)}
\;,\end{equation}
which is a particular case of the Riesz formula \cite{Riesz}, valid
for any $a, b\in {\mathbb C}$ except at some isolated poles:
 
\begin{equation}\label{50}
\int d^3{\bf x}~ r_1^a r_2^b = \pi^{3/2}{\Gamma\left({a+3 \over
2}\right)\Gamma\left({b+3 \over 2}\right)\Gamma\left(-{a+b+3 \over
2}\right)\over \Gamma\left(-{a\over 2}\right) \Gamma\left(-{b\over
2}\right) \Gamma\left({a+b+6 \over 2}\right)}r_{12}^{a+b+3}
\end{equation} 
($\Gamma$ denotes the Eulerian function). A closely related reasoning
to prove (\ref{49}) is to replace the regularization factor
$|\tilde{\bf x}|^B$ by its expansion when $B\to 0$, i.e.

\begin{equation}\label{51}
|\tilde{\bf x}|^B = {\tilde r}_1^B \left\{1+{B\over 2} \ln
\left[1+2{(n_1y_1)\over r_1}+{y_1^2\over r_1^2}\right]+{\cal
O}(B^2)\right\} \;.\end{equation} Since the integral does not develop
any pole when $B\to 0$, the term of order $B$ cannot contribute, nor
any of the higher-order terms ${\cal O}(B^2)$. This means that we can
replace the regularization factor $|{\tilde {\bf x}}|^B$ by ${\tilde
r}_1^B$ (where ${\tilde r}_1=r_1/r_0$). From the Riesz formula, with
$a=B+2p-1$ and $b=-1$, and computation of the limit $B\to 0$ we get
the same result.

Thus, plugging (\ref{49}) into (\ref{48}) we find the explicit
expression of $Y_L$ as

\begin{equation}\label{52}
Y_L ({\bf y}_1,{\bf y}_2)= r_{12}\sum^l_{p=0} {l \choose p}
{(-)^p\over p+1} y_1^{<L-P} y_{12}^{P>}
\;,\end{equation}
where $y_{12}^i=y_1^i-y_2^i$ and $r_{12}=|{\bf y}_{12}|$. In terms of
$y_1^i$ and $y_2^i$ the expression is simpler:
 
\begin{equation}\label{53}
Y_L={r_{12}\over l+1} \sum^l_{q=0} y_1^{<L-Q} y_2^{Q>}
\;.\end{equation}
Using exactly the same method we find for the $S_L$-integral, 

\begin{eqnarray}\label{54}
S_L&=& r_{12}\sum^l_{p=0} {l \choose p} (-)^p y_1^{<L-P} y_{12}^{P>}
\left[ {{\bf y}_{12}^2 \over (p+2)(p+3)}\left(p+1-{2l\over 3}\right) -
{2{\bf y}_1.{\bf y}_{12}\over p+2}+{{\bf y}_1^2\over
p+1}\right]\nonumber\\ &=& {r_{12}\over (l+1)(l+2)} \sum^l_{q=0}
y_1^{<L-Q} y_2^{Q>} \left[ (l+1-q) {\bf y}_1^2 - {2\over 3}(q+1) (l
+1-q) {\bf y}_{12}^2 + (q+1) {\bf y}_2^2 \right]\;,\nonumber\\
\end{eqnarray}
and, for the $T_L$-integral,

\begin{eqnarray}\label{55}
T_L&=& {r_{12}^3\over 3} \sum^l_{p=0} {l \choose p} {(-)^p\over p+2}~\!
y_1^{<L-P} y_{12}^{P>}\nonumber\\ &=&{r_{12}^3\over 3(l+1)(l+2)}
\sum^l_{q=0} (q+1) y_1^{<L-Q} y_2^{Q>}
\;.\end{eqnarray}
Notice that $S_L$ can be deduced from $T_L$ and $Y_L$ using the formula

\begin{equation}\label{56}
S_L = (1-2y_1^i \!\!\!\sous{1}{\partial}_{~\!i}) T_L + {\bf y}_1^2~\! Y_L 
\;.\end{equation}
The integrals $Y_L$, $S_L$ and $T_L$ vanish in the limit ${\bf y}_1\to
{\bf y}_2$. As is clear from the defining expressions (\ref{46}) there
is no problem with the latter limit, in the sense that it does not
introduce any singularity at the point 1. This justifies {\it a
posteriori} our neglect of all the ``self'' contributions
(proportional to $m_1^2$ and $m_2^2$) in the quadratic terms; see the
examples given by Eqs. (\ref{45}). However, when we compute the cubic
and non-compact terms in Sections VIII and IX we shall find some
important non-zero self contributions.

Another method for the computation of the integrals (\ref{46}) is
based on the set of functions defined by

\begin{subequations}\label{57}\begin{eqnarray}
g &=& \ln (r_1+r_2+r_{12}) \label{57a}\;,\\ f &=& {1\over
6}(r_1^2+r_2^2-r_{12}^2)\left(g-{1\over 3}\right)+{1\over
6}(r_{12}r_1+r_{12}r_2-r_1 r_2) \;,\\ \stackrel{12}{f} &=& {1\over
6}(r_1^2+r_{12}^2-r_2^2)\left(g-{1\over 3}\right)+{1\over 6}(r_1
r_2+r_{12}r_2-r_1 r_{12}) \;,\\ \stackrel{21}{f} &=& {1\over
6}(r_2^2+r_{12}^2-r_1^2)\left(g-{1\over 3}\right)+{1\over 6}(r_1
r_2+r_{12}r_1-r_2 r_{12})
\;,\end{eqnarray}\end{subequations}$\!\!$
which satisfy, in the sense of distribution theory,

\begin{subequations}\label{58}\begin{eqnarray}
\Delta g &=& {1\over r_1 r_2} ; \quad\Delta_1 g = {1\over r_1 r_{12}}
; \quad\Delta_2 g = {1\over r_2 r_{12}} \;,\\ \Delta f &=& 2g ;
\quad\Delta_1 f = {r_1\over r_{12}} ; \quad\Delta_2 f = {r_2\over
r_{12}} \;,\\ \Delta \stackrel{12}{f} &=& {r_1\over r_2} ;
\quad\Delta_1 \stackrel{12}{f} = 2g ; \quad\Delta_2 \stackrel{12}{f} =
{r_{12}\over r_2} \;,\\ \Delta \stackrel{21}{f} &=& {r_2\over r_1} ;
\quad\Delta_1 \stackrel{21}{f} = {r_{12}\over r_1} ; \quad\Delta_2
\stackrel{21}{f} = 2g \;,\end{eqnarray}\end{subequations}$\!\!$ where
the Laplacians $\Delta=\partial_i\partial_i$,
$\Delta_1={}_{1}{\partial}_{i}{}_{1}{\partial}_{i}$,
$\Delta_2={}_{2}{\partial}_{i}{}_{2}{\partial}_{i}$. Let us take the
example of the integral $Y_L$. With the help of (\ref{57a}) it can be
re-written as

\begin{equation}\label{59}
Y_L = -{1\over 2\pi} \sous{\!\!\!\!B= 0}{\rm FP} \int d^3{\bf x}~
|\tilde{\bf x}|^B \hat{x}_L \Delta g
\;.\end{equation}
We operate the Laplacian by parts, discard the $B$-dependent surface
term which is zero by analytic continuation, and use the formula
$\Delta (|{\bf x}|^B \hat{x}_L)=B(B+2l+1)|{\bf
x}|^{B-2}\hat{x}_L$. Hence,

\begin{equation}\label{60}
Y_L = -{1\over 2\pi} \sous{\!\!\!\!B= 0}{\rm FP} \left\{ B(B+2l+1)
\int d^3{\bf x}~ |\tilde{\bf x}|^B |{\bf x}|^{-2} \hat{x}_L g \right\}
\;.\end{equation} Because there is an explicit factor $B$ in front of
the integral we need to look only at the polar part when $B\to 0$,
which depends only on the behaviour of the integrand at the upper
bound $r\equiv |{\bf x}|\to +\infty$ (this $r$ should not be confused
with $r=r_{12}$ as we sometimes denote the orbital separation). Thus
we are allowed to replace the function $g$ in (\ref{60}) by its
expansion at infinity. It can be checked that the (simple) pole of the
integral in (\ref{60}) is produced exclusively by the term in the
expansion of $g$ of order $r^{-l-1}$. Let us consider the quadrupole
case $l=2$. We have

\begin{eqnarray}\label{61}
g&=&\ln (2r)+{1\over r}\left\{\cdots\right\}+{1\over
r^2}\left\{\cdots\right\}\nonumber\\ &+&{1\over
r^3}\left\{{r_{12}\over
4}\left[(ny_1)^2+(ny_1)(ny_2)+(ny_2)^2\right]+\cdots\right\} +{\cal
O}\left({1\over r^4}\right)
\;,\end{eqnarray}
where the dots indicate some terms which yield no contribution to the
present computation, either because they do not belong to the relevant
order $r^{-3}$ or they will be zero after angular integration. Thus
the formula (\ref{60}) becomes in this case

\begin{equation}\label{62}
Y_{ij} = \sous{\!\!\!\!B= 0}{\rm FP} \left\{ -2B(B+5) \int^{+\infty}
dr~ {\tilde r}^B r^{-1}\int {d\Omega\over 4\pi} ~\!{\hat
n}_{ij}{r_{12}\over 4}\left[(ny_1)^2+(ny_1)(ny_2)+(ny_2)^2\right]
\right\}
\;.\end{equation}
The notation for the radial integral means that only the bound at
infinity contributes to its value. The latter expression is easily
transformed into

\begin{equation}\label{63}
Y_{ij} = {r_{12}\over 3} \left[
y_1^{<ij>}+y_1^{<i}y_2^{j>}+y_2^{<ij>}\right] \;,\end{equation} in
agreement with the more general result (\ref{53}). The same method
works for $S_L$ as well, but one performs two successive integrations
by parts using the functions $g$ and $f$. Concerning $T_L$, one
integration by parts is sufficient but using the function $f^{12}$
(labels such as $12$ are placed at the top when the quantity appears
in an equation and as right-side superscripts when it is within the
text).

With the latter expressions of the elementary integrals $Y_L$, $S_L$
and $T_L$ we obtain all the quadratic terms. The results in the case
of circular orbits are displayed in Appendix A.

\section{Cubic terms}

By cubic terms we refer to all the terms which are made of a product
between three (derivatives of) compact-support potentials $U$ and
$U_i$ [there are no such terms involving the tensor potential
$U_{ij}^{({\rm C})}$]. From (\ref{18}) we can check that the only
cubic terms appear at the 3PN order. These are

\medskip\noindent
SI(26), SI(27), SI(28), SI(29), SI(30), SI(34), SI(36),
SII(13),VI(22), VI(23), VI(24).

\medskip\noindent Let us proceed in a way similar to the computation
of the quadratic terms, i.e. by expressing the terms as functionals of
some elementary integrals that are computed separately. Since the
cubic terms are 3PN, their computation can be done using the Newtonian
potentials

\begin{subequations}\label{64}\begin{eqnarray}
U &=& {G m_1\over r_1}+{\cal O}(2)+1\leftrightarrow 2\;,\\ U_i &=& {G
m_1\over r_1} v_1^i+{\cal O}(2)+1\leftrightarrow 2
\;.\end{eqnarray}\end{subequations}$\!\!$
For simplicity we gather in one computation the sum of all the cubic
terms in SI [and similarly in VI; there is only one cubic term in SII,
which is SII(13)]. In the case of mass-type moments we get

\begin{subequations}\label{65}\begin{eqnarray}
&&{\rm SI(26+27+28+29+30+34+36)} = {G^2m^3_1\over c^6} \left \{
-{32\over 15} v_1^i v_1^j \!\!\!\sous{1}{\partial}_{~\!ij}
\stackrel{(-3,0)}{Y_L} \right.\nonumber \\ && \qquad \left. + {88\over
5} v_1^2 \stackrel{(-5,0)}{Y_L} + {512\over 225}
v_1^{ab}\!\!\!\sous{1}{\partial}_{~\!ab}{\hat y}_1^L \right \}
\nonumber \\ && \qquad+{G^2m_1^2m_2\over c^6} \left \{ \left(-{9\over
2}v_1^i v_2^j - {5\over 2} v_1^i v_1^j \right)
\!\!\!\sous{1}{\partial}_{~\!ij} \stackrel{(-2,-1)}{Y_L} - 8 v_1^i
v_2^j \!\!\!\sous{2}{\partial}_{~\!ij} \stackrel{(-2,-1)}{Y_L}
\right.\nonumber\\ &&\qquad +[-2v_1^i v_2^j - 6 v_1^i v_1^j + 8
\delta^{ij} (v_1v_2) + 8\delta^{ij} v_1^2]
\!\!\!\sous{1}{\partial}_{~\!i} \!\!\!\sous{2}{\partial}_{~\!j}
\stackrel{(-2,-1)}{Y_L} \nonumber \\ && \qquad +\left. [15(v_1v_2) +
3v_1^2] \stackrel{(-4,-1)}{Y_L} \right \} + 1\leftrightarrow 2 \;,\\
&&{\rm SII (13)} = {4G^2\over 3c^6}~{d^2\over dt^2} \left\{m_1^3
\stackrel{(-3,0)}{Y_L}+ 3 m_1^2 m_2 \stackrel{(-2,-1)}{Y_L} \right\} +
1\leftrightarrow 2 \;,\\ &&{\rm VI(22+23+24)} = {8G^2(2l +1)\over
c^6(l+1)(2l+3)} {d\over dt} \left\{ -m^3_1 v_1^a
\stackrel{(-5,0)}{Y_{aL}} \right. \nonumber \\ && \qquad +m_1^2m_2
\left[{3\over 4} (v_1^k-v_2^k)
\!\!\!\sous{1}{\partial}_{~\!a}\!\!\!\sous{2}{\partial}_{~\!k}
\stackrel{(-2,-1)}{Y_{aL}} - v_1^a \!\!\!\sous{1}{\partial}_{~\!k}
\!\!\!\sous{2}{\partial}_{~\!k} \stackrel{(-2,-1)}{Y_{aL}}
\right. \nonumber \\ && \qquad - \left.\left.{3\over
16}(v_1^k-v_2^k)\!\!\!\sous{1}{\partial}_{~\!ak}
\stackrel{(-2,-1)}{Y_{aL}} -\left({3\over 8} v_1^a +{5\over 8}
v_2^a\right) \stackrel{(-4,-1)}{Y_{aL}} \right]\right\} +
1\leftrightarrow 2\label{65c} \;.\end{eqnarray}\end{subequations}$\!\!$
In the case of the current-type moments there are only the VI-terms,
which admit a formula analogous to (\ref{65c}). As we see, we could
express all the cubic terms by means of a single type of elementary
integral,

\begin{equation}\label{67}
\stackrel{(n,p)}{Y_L}({\bf y}_1,{\bf y}_2) = -{1\over
2\pi}\sous{\!\!\!\!B= 0}{\rm FP} \int d^3{\bf x}~\! |\tilde{\bf x}|^B
\hat{x}_L ~r^n_1 r^p_2 \;,\end{equation} of which some particular
cases used in the previous section read $Y_L=Y_L^{(-1,-1)}$ and
$T_L=Y_L^{(1,-1)}$ [we use right-side superscripts $(n,p)$ when the
quantity appears within the text]. The integral (\ref{67}) is
well-defined in the vicinity of the points ${\bf y}_1$ and ${\bf y}_2$
only when $n>-3$ and $p>-3$. When this is not the case -- for instance
the integral $Y^{(-3,0)}_L$ appearing in (\ref{65}) -- one should add
the Hadamard partie-finie operation Pf defined by (\ref{27}) and
depending {\it a priori} on two constants $u_1$ and $u_2$. According
to our convention we generally do not write such parties finies, but
they are always implicitly understood.

The integral $Y^{(-2,-1)}_L$ is perfectly well-behaved near the two
bodies (like $Y_L$, $S_L$ and $T_L$ considered in Section VII), so it
does not need the partie finie. We substitute in it a formula obtained
from (\ref{47}) by exchanging the labels 1 and 2, obtaining

\begin{equation}\label{68}
\stackrel{(-2,-1)}{Y_L} = -{1\over 2\pi} \sum_{p=0}^l {l \choose p}
{(-)^p\over (2p-1)!!} y_2^{<L-P} \!\!\!\sous{2}{\partial}_{~\!P>}
\left\{\sous{\!\!\!\!B = 0}{\rm FP} \int d^3{\bf x}~\! |\tilde{\bf
x}|^B {r_2^{2p-1}\over r_1^2} \right\}
\;.\end{equation} 
Next we replace the regularization factor $|\tilde{\bf x}|^B$ by its
expansion around $B=0$ already written in (\ref{51}). Since the
integral can develop simple poles at most, we can limit ourselves to
the first order in $B$. Then the integral in the brackets of
(\ref{68}) reads

\begin{eqnarray}\label{69}
\sous{\!\!\!\!B= 0}{\rm FP} \int d^3{\bf x}~\! |\tilde{\bf x}|^B
{r_2^{2p-1}\over r_1^2} &=& \sous{\!\!\!\!B= 0}{\rm FP} \int d^3{\bf
x}~\! \tilde{r}_1^B {r_2^{2p-1}\over r_1^2} \nonumber\\ &+&{\rm
FP}_{B=0}\left\{ {B\over 2} \int d^3{\bf x}~\! \tilde{r}_1^B
{r_2^{2p-1}\over r_1^2} \ln \left[1+2{(n_1y_1)\over r_1}+{y_1^2\over
r_1^2}\right]\right\} \;.\end{eqnarray} The first term follows from
the Riesz formula (\ref{50}), and the second term depends only on the
poles developed by the integral at infinity (because of the explicit
factor $B$ in front). Now, contrarily to the case of the integral
$Y_L\equiv Y_L^{(-1,-1)}$ investigated in Section VII, we find that
this second term gives a net contribution to the integral,
straightforwardly obtained from expanding the integrand when $r=|{\bf
x}|\to +\infty$. The final values that we obtain in the quadrupole and
octupole cases ($l=2$ and $l=3$) of interest are

\begin{subequations}\begin{eqnarray}\label{70}
\stackrel{(-2,-1)}{Y_{ij}}&=& y_1^{<ij>} \left [{16\over
15}\ln\tilde{r}_{12} - {188\over 225} \right ] + y_1^{<i}y_2^{j>}
\left [{8\over 15} \ln\tilde{r}_{12}- {4\over 225} \right ]
\nonumber\\ &+& y_2^{<ij>} \left [{2\over 5} \ln\tilde{r}_{12} -
{2\over 25} \right ]\label{70a}\;,\\ \stackrel{(-2,-1)}{Y_{ijk}}&=&
y_1^{<ijk>} \left [{32\over 35} \ln\tilde{r}_{12} - {2552\over 3675}
\right ] + y_1^{<ij}y_2^{k>}\left [{16\over 35} \ln\tilde{r}_{12} +
{124\over 3675} \right] \nonumber\\ &+& y_1^{<i}y_2^{jk>} \left
[{12\over 35} \ln\tilde{r}_{12}+ {66\over 1225} \right] + y_2^{<ijk>}
\left [{2\over 7} \ln\tilde{r}_{12} - {2\over 49} \right ]
\;.\end{eqnarray}\end{subequations}$\!\!$
Note the occurence of some logarithms of ${\tilde
r}_{12}=r_{12}/r_0$. Applying on these values the point-1 Laplacian
$\Delta_1={}_1\partial_{ii}$, and using $\Delta_1 r_1^{-2}=2 r_1^{-4}$
(a statement valid in the sense of distributions), we obtain

\begin{subequations}\label{71}\begin{eqnarray}
\stackrel{(-4,-1)}{Y_{ij}}&=& {1\over r_{12}^2} \left({8\over 3}
y_1^{<ij>} - {4\over 3} y_1^{<i}y_2^{j>} - {1\over 3}y_2^{<ij>}
\right) \;,\\ \stackrel{(-4,-1)}{Y_{ijk}} &=& {1\over r_{12}^2} \left
({16\over 5} y_1^{<ijk>} - {8\over 5}y_1^{<ij}y_2^{k>} - {2\over 5}
y_1^{<i}y_2^{jk>} - {1\over 5} y_2^{<ijk>}\right)
\;.\end{eqnarray}\end{subequations}$\!\!$
Alternatively, the results (\ref{71}) can also be obtained by the same
technique as used previously for $Y^{(-2,-1)}_L$ (i.e. from the Riesz
formula and search for the pole at infinity).

The computation of the integral $Y^{(-3,0)}_L$, defined by

\begin{equation}\label{71'}
\stackrel{(-3,0)}{Y_L}({\bf y}_1) = -{1\over
2\pi}\sous{\!\!\!\!B= 0}{\rm FP} \int d^3{\bf x}~\! |\tilde{\bf x}|^B
{\hat{x}_L\over r_1^3}
\;,\end{equation}
is {\it a priori} more tricky because this integral necessitates the
Hadamard partie finie for curing the divergence at the point ${\bf
y}_1$. Actually, the same method as before, based on the Riesz
formula, could be used because we know that the Hadamard partie finie
can also be obtained as an analytic continuation (see
e.g. \cite{BFreg}). We prefer here to vary the techniques and to
present some other derivations. We split the integration domain
${\mathbb R}^3$ into a ball centered on ${\bf y}_1$ with some fixed
radius ${\cal R}_1$, and the complementary domain, i.e. $r_1>{\cal
R}_1$. The partie finie applies only on the ``inner'' domain,
surrounding the singularity 1, and the finite part ${\rm FP}_{B=0}$
applies only on the integral extending to infinity. Hence,
 
\begin{equation}\label{72}
\stackrel{(-3,0)}{Y_L} = -{1\over 2\pi} {\rm Pf}_{u_1} \int_{r_1<{\cal
R}_1} d^3{\bf x}~\!  {\hat{x}_L\over r^3_1}-{1\over 2\pi}
\sous{\!\!\!\!B= 0}{\rm FP} \int_{r_1>{\cal R}_1} d^3{\bf x}~\!
|\tilde{\bf x}|^B {\hat{x}_L\over r^3_1}
\;.\end{equation}
In the first term we recall that the partie finie depends on a
constant $u_1$ [see the definition (\ref{27})]. For this term we
readily find

\begin{equation}\label{73}
-{1\over 2\pi} {\rm Pf}_{u_1} \int_{r_1<{\cal R}_1} d^3{\bf x}~\!
{\hat{x}_L\over r^3_1}=-2~\!{\hat y}_1^L \ln\left({\cal R}_1\over
u_1\right)
\;.\end{equation}
On the other hand, one must replace into the second term the factor
$|\tilde{\bf x}|^B$ by its $B$-expansion as given by (\ref{51}). This
yields two contributions: one is immediately computed using the
properties of the analytic continuation, the other contains an
explicit factor $B$ and therefore relies on the existence of poles at
infinity:

\begin{eqnarray}\label{74}
&&-{1\over 2\pi} \sous{\!\!\!\!B= 0}{\rm FP} \int_{r_1>{\cal R}_1}
d^3{\bf x}~\! |\tilde{\bf x}|^B {\hat{x}_L\over r^3_1}\nonumber\\
&&\qquad=2~\!{\hat y}_1^L \ln\left({{\cal R}_1\over
r_0}\right)-{1\over 2\pi} {\rm FP}_{B=0}\left\{ {B\over 2}\int_{{\cal
R}_1}^{+\infty} d^3{\bf x} ~{\tilde r}_1^B {\hat{x}_L\over r^3_1} \ln
\left[1+2{(n_1y_1)\over r_1}+{y_1^2\over r_1^2}\right]\right\}
\;.\end{eqnarray}
As expected, the sum of the two contributions (\ref{73}) and
(\ref{74}) is independent of the intermediate length scale ${\cal
R}_1$. Indeed, the integral in the second term of (\ref{74}) does not
in fact depend on ${\cal R}_1$ as it depends only on the infinite
bound. We obtain

\begin{equation}\label{75}
\stackrel{(-3,0)}{Y_L} = 2~\!\hat{y}^L_1 \ln\left({u_1\over
r_0}\right)-{1\over 4\pi} {\rm FP}_{B=0}\left\{ B\int^{+\infty}
d^3{\bf x} ~{\tilde r}_1^B {\hat{x}_L\over r^3_1} \ln
\left[1+2{(n_1y_1)\over r_1}+{y_1^2\over r_1^2}\right]\right\}
\;.\end{equation}
The computation of the second term proceeds along the same line as for
the reduction of $Y_L$ in (\ref{60}). We expand the log-term up for
instance to the order $1/r_1^2$ necessary to get the quadrupole case
$l=2$,

\begin{equation}\label{76}
\ln \left[1+2{(n_1y_1)\over r_1}+{y_1^2\over
r_1^2}\right]=2{(n_1y_1)\over r_1} +{1\over
r_1^2}\left[-2(n_1y_1)^2+y_1^2\right]+{\cal O}\left({1\over
r_1^3}\right)
\;.\end{equation}
Therefore:

\begin{eqnarray}\label{77}
\stackrel{(-3,0)}{Y_{ij}} &=& 2\hat{y}^{ij}_1 \ln\left({u_1\over
r_0}\right)\nonumber\\ &-& {\rm FP}_{B=0}\left\{ B
\int^{+\infty} d r_1 ~{\tilde r}_1^B r_1^{-1}\int {d\Omega_1\over 4\pi}
~\!{\hat
n}_1^{ij}\left[4n_1^{<i}y_1^{j>}(n_1y_1)-2n_1^{<ij>}(n_1y_1)^2\right]\right\}
\;.\end{eqnarray}
The integral follows immediately. This method yields the results
(cases $l=2,3$)

\begin{subequations}\begin{eqnarray}\label{78}
\stackrel{(-3,0)}{Y_{ij}} &=& \left[2\ln\left(u_1\over
r_0\right)+{16\over 15} \right] y_1^{<ij>} \label{78a}\;,\\
\stackrel{(-3,0)}{Y_{ijk}} &=& \left[2\ln\left(u_1\over
r_0\right)+{142\over 105} \right] y_1^{<ijk>}
\;.\end{eqnarray}\end{subequations}$\!\!$
The results depend on the Hadamard-regularization constant
$u_1$.

We present another derivation of the integral $Y^{(-3,0)}_L$, based on
the interesting formula of distribution theory (see
e.g. \cite{Sellier})

\begin{equation}\label{79}
\Delta\left[{1\over r_1}\ln\left({r_1\over u_1}\right)\right]=-{\rm
Pf}_{u_1} \left( {1\over r_1^3}\right) + 4\pi \delta ({\bf x}-{\bf
y}_1) \;.\end{equation} [Notice the sign of the distributional term,
$+4\pi\delta_1$, opposite to the sign in the more famous formula
$\Delta (1/r_1)=-4\pi\delta_1$.] With Eq. (\ref{79}) one can
re-express $Y^{(-3,0)}_L$ in the form

\begin{equation}\label{80}
\stackrel{(-3,0)}{Y_L} = -2~\!\hat{y}^L_1 +{1\over 2\pi} {\rm
FP}_{B=0}\int d^3{\bf x} ~|\tilde{\bf x}|^B \hat{x}_L
\Delta\left[{1\over r_1}\ln\left({r_1\over u_1}\right)\right]
\;.\end{equation}
Here the first term comes from the delta-function in
(\ref{79}). Integrating the second term by parts, we get

\begin{equation}\label{81}
\stackrel{(-3,0)}{Y_L} = -2\hat{y}^L_1 +{1\over 2\pi} {\rm
FP}_{B=0}\left\{ B(B+2l+1)\int^{+\infty} d^3{\bf x}~\! |\tilde{\bf
x}|^B ~\!|{\bf x}|^{-2}{\hat{x}_L\over r_1} \ln\left({r_1\over
u_1}\right) \right\}\;.\end{equation} Following the same principle as
before, we compute the remaining integral by looking at the pole at
infinity. The result is in agreement with the earlier derivation (as
we checked in the case $l=2$). Let us also mention that still another
method to compute $Y^{(-3,0)}_L$ consists of taking the limit ${\bf
y}_2 \to {\bf y}_1$ of the integral $Y^{(-2,-1)}_L$. The limit is
singular since $Y^{(-2,-1)}_L$ diverges when the two particles merge
together. In fact the limit must be taken in the sense of the Hadamard
partie finie (\ref{26}). Indeed, applying Eq. (5.5) in
Ref. \cite{BFreg}, we obtain the following limit relation between
$Y^{(-3,0)}_L$ and $Y^{(-2,-1)}_L$:

\begin{equation}\label{83}
\stackrel{(-3,0)}{Y_L}({\bf y}_1)= \left(\stackrel{(-2,-1)}{Y_L}({\bf
y}_1, {\bf x})-2\left[\ln\left({r_1\over
u_1}\right)-1\right]y_1^{<L>}\right)_1
\;.\end{equation}
Inserting for instance the result for $Y^{(-2,-1)}_{ij}$ obtained in
(\ref{70a}) we recover exactly the function $Y^{(-3,0)}_{ij}$ given by
(\ref{78a}).

Finally it is easy to see that the function $Y^{(-5,0)}_L$, also
needed into the cubic terms (\ref{65}), is identically
zero. We apply the point-1 Laplacian $\Delta_1$ onto the expression of
$Y^{(-3,0)}_L$ using the known formula of distribution theory

\begin{equation}\label{84}
\Delta_1 \left({1\over r_1^3}\right) = {6 \over r_1^5}-{10\pi\over 3}
\Delta_1 \delta_1 \;,\end{equation} and readily obtain, for any $l$,

\begin{equation}\label{85}
\stackrel{(-5,0)}{Y_L} = 0
\;.\end{equation}
The results for the cubic terms in the case of circular orbits are
reported in Appendix A.

\section{Non-compact terms}

The most difficult part of the present analysis is the computation of
the so-called ``non-compact'' terms, which are cubically non-linear
terms (like the cubic terms) made of the product of a compact-support
potential like $U$ and a quadratic ``non-compact'' potential like
$U_{ij}^{({\rm NC})}$.  The complete list of non-compact terms is

\medskip\noindent SI(5NC), SI(19NC), SI(20), SI(21NC), SI(33NC),
SI(35NC), SI(37NC), SI(38NC), SII(4NC), VI(25NC), VI(26NC), VI(27NC),
VI(28NC), VI(29NC).

\subsection{Expressions of the NC terms}

As before, here again our strategy is 
to express the non-compact terms as functionals
of certain elementary integrals, that are computed separately. We
substitute inside the sources of non-compact terms the appropriate
post-Newtonian potentials computed for two particles on a general
orbit. The compact potentials $U$, $U_i$ and $U_{ij}^{({\rm C})}$ (and
similar expressions for the $\chi$'s) were already given by
(\ref{44}). Here we list all the non-compact potentials needed for
this computation [see (\ref{14})-(\ref{15}) for definitions]. The
potential $U_{ij}^{({\rm NC})}$ is the only one which is needed at 1PN
order; the other potentials are Newtonian.

\begin{subequations}\label{86}\begin{eqnarray}
U_{ij}^{({\rm NC})}&=&-{1\over 8}{\tilde
\mu}_1^2\left(\partial_{ij}\ln r_1+{\delta_{ij}\over
r_1^2}\right)-{\tilde \mu}_1{\tilde \mu}_2 ~\!{}_ig_j+1\leftrightarrow 2
\;,\\ \chi_{ij}^{({\rm NC})}&=&-{1\over
4}m_1^2\left(\partial_{ij}\left[{r_1^2\over 6} \left(\ln r_1-{5\over
6}\right)\right]+\delta_{ij}\ln r_1\right)-m_1m_2 {}_if_j+1\leftrightarrow 2
\;,\\ {\hat R}_i^{({\rm NC})}&=&-{1\over
16}m_1^2v_1^k\left(\partial_{ik}\ln r_1+{\delta_{ik}\over
r_1^2}\right)-2m_1m_2 (v_1^k-{3\over 4}v_2^k){}_ig_k +1\leftrightarrow
2 \;,\\ {\hat Z}_{ij}^{({\rm NC})}&=&m_1^2\left\{
a_1^{(i}\!\!\!\sous{1}{\partial}_{~\!j)}\ln r_1+{1\over
8}v_1^2\!\!\!\sous{1}{\partial}_{~\!ij}\ln r_1 +{1\over
32}\delta^{ij}v_1^{km}\!\!\!\sous{1}{\partial}_{~\!km}\ln r_1
\right.\nonumber\\ &+&\left.{1\over 2}{v_1^{ij}\over r_1^2}-{11\over
32}\delta^{ij}{v_1^2\over r_1^2}\right\} \nonumber\\ &+& m_1m_2
\left\{ 2a_1^{(i}g_{j)}+2v_1^{k(i}{}_kg_{j)}-2v_1^{(i}v_2^k{}_kg_{j)}
+(v_1v_2){}_{(i}g_{j)}+v_1^{(i}v_2^{j)}{}_kg_k \right.\nonumber\\
&-&\left.{3\over 4}\delta_{ij}v_1^kv_2^m{}_kg_m
+\delta_{ij}v_1^mv_2^k{}_kg_m -\delta_{ij}(v_1v_2){}_kg_k
\right\}+1\leftrightarrow 2 \label{86d}\;,\\ K_{ij}&=&m_1^2\left\{ \left[
{1\over 48}a_1^k\!\!\!\sous{1}{\partial}_{~\!ijk}+{1\over
96}v_1^{km}\!\!\!\sous{1}{\partial}_{~\!ijkm}\right](r_1^2\ln r_1)
\right.\nonumber\\ &+& \left[ -{1\over
8}\delta^{ij}a_1^k\!\!\!\sous{1}{\partial}_{~\!k} +{1\over
4}a_1^{(i}\!\!\!\sous{1}{\partial}_{~\!j)}-{1\over
16}\delta^{ij}v_1^{km}\!\!\!\sous{1}{\partial}_{~\!km} +{1\over
16}v_1^2\!\!\!\sous{1}{\partial}_{~\!ij}\right](\ln r_1) \nonumber\\
&+& \left.\left[ {1\over 16}\delta^{ij}v_1^2 +{1\over
8}v_1^{ij}\right](r_1^{-2}) \right\} \nonumber\\ &+& m_1m_2 \left[
a_1^k\!\!\!\sous{1}{\partial}_{~\!k}
+v_1^{km}\!\!\!\sous{1}{\partial}_{~\!km}\right]{}_{(i}\stackrel{12}{f}_{j)}
+1\leftrightarrow 2
\;.\end{eqnarray}\end{subequations}$\!\!$
Here, $g$, $f$, $f^{12}$ and $f^{21}$ are defined by (\ref{57}), and
we denote e.g. ${}_ig_j={}_{1}{\partial}_{i}{}_{2}{\partial}_{j}g$
(see Ref. \cite{BFP98} for the expression of ${}_ig_j$); the
acceleration is $a_1^i=dv_1^i/dt$; the parenthesis around indices
denotes the symmetrization (and $G=1$).

Notice that we have chosen to express the non-compact potentials by
means of $g$, $f$, $f^{12}$ and $f^{21}$. But these functions
constitute merely some {\it particular} solutions of the Laplace
equations (\ref{58}) we have to solve, and the question arises of
which solution is the correct one. The most general solution will be
obtained by adding to the particular one an {\it homogeneous} term,
solving a source-free Laplace-type equation. We have checked that the
only possible homogeneous solutions, that are regular at the origin,
are constants or linear functions of the position, and that these are
always either cancelled by some spatial or time derivatives, or
disappear at the end of our computations. This justifies our use of
the particular solutions (\ref{57}). (Similarly, we found that the
same happens in the computation of the 3PN equations of motion, where
these particular solutions are sufficient \cite{BFeom}.)

The potentials (\ref{86}) contain a ``self'' part, proportional to
$m_1^2$ or $m_2^2$ (before replacement of the accelerations), and an
``interaction'' part, proportional to $m_1m_2$. Similarly the sources
of the non-compact terms will involve a self part, proportional to
$m_1^3$ or $m_2^3$, and an interaction part, proportional to
$m_1^2m_2$ or $m_1m_2^2$. At the 2PN level, all the self parts
cancelled out in the multipole moments \cite{BDI95}. At the 3PN level,
we shall find that the self parts bring a contribution to the
moments. [Actually, we shall argue in Section X that the self parts
are unknown.] For treating the NC terms we used the standard
distributional derivative \cite{Schwartz,Sellier}. Thus, we have, for
instance,

\begin{subequations}\label{87}\begin{eqnarray}
\Delta {1\over r_1} &=&-4\pi \delta_1 \;,\\ \partial_{ij}\left({1\over
r_1}\right) &=& {3n_1^{ij} -\delta^{ij} \over r_1^3}-{4\pi\over
3}\delta^{ij}\delta_1 \label{87b}\;,\\ \Delta \left({1\over
r_1^3}\right) &=& {6 \over r_1^5} -{10\pi\over 3} \Delta \delta_1
\;,\\ \partial_{ij}\left({1\over r_1^3}\right) &=&
{15n_1^{ij}-3\delta^{ij} \over r_1^5}-{2\pi\over 5} \delta^{ij}\Delta
\delta_1-{32\pi\over 15} \partial_{ij}\delta_1
\;.\end{eqnarray}\end{subequations}$\!\!$ However, the use of the
standard Schwartz derivative can be justified only when the terms
involved are multiplied by some smooth functions. In the case of the
self parts of NC terms, this will not be true in general, so the
Schwartz derivative gives some ill-defined contributions, composed of
the product of a delta-function and a singular function. In Section X
we consider a well-defined way to do the computation of the self
terms, which is based on the distributional derivatives proposed in
Ref. \cite{BFreg}. From the discussion in Section X we conclude that
one must add to the present computation some undetermined terms taking
into account the ambiguities in the choice of the regularization and
distributional derivatives.  All the expressions in Eq. (\ref{88})
below are modulo these ill-defined contributions and we can safely
proceed with the knowledge that our procedure is unambiguous and
complete. We are securely protected from such ill-defined
contributions at this stage since we shall add such terms with an
arbitrary coefficient in Section X.  We obtain the following
expressions of the non-compact terms, as functionals of several new
types of elementary integrals (we pose $D=
{}_{1}{\partial}_{i}~\!{}_{2}{\partial}_{i}$ and $G=1$).  In the case
of the mass-type moments:

\begin{subequations}\label{88}\begin{eqnarray}
{\rm SI(5NC)} &=& {\tilde{\mu}_1^3\over c^4} \stackrel{(-5,0)}{Y_L}
\nonumber\\ &+& {\tilde{\mu}^2_1 \tilde{\mu}_2\over c^4} \left\{
-{1\over 2} \hat{y}^L_2 r^{-2}_{12} - {1\over 4} D^2
\stackrel{(0,-1)}{N_{L}} - 4 \!\!\!\sous{2}{\partial}_{~\!s} G^s_{L}
\right\} +1\leftrightarrow 2 \;,\\ {\rm SI(19NC)} &=& {m_1^3\over c^6}
\left[-{1\over 30} v_1^{ab} \!\!\!\sous{1}{\partial}_{~\!ab}
\stackrel{(-3,0)}{Y_L} + {2\over 5} v_1^2 \stackrel{(-5,0)}{Y_L}
+{8\over 225}v_1^{ab}\!\!\!\sous{1}{\partial}_{~\!ab}{\hat y}_1^L
\right] \nonumber \\ &+& {m_1^2 m_2\over c^6} \left\{ \left(-{1\over
2}{(n_{12}a_1) \over r_{12}} + {(n_{12} v_1)^2\over r_{12}^2} -
{1\over 2} {v_1^2\over r_{12}^2}\right) \hat{y}^L_2 \right. \nonumber
\\ &-& {1\over 24} (a_1^a \!\!\!\sous{1}{\partial}_{~\!a} + v_1^{ab}
\!\!\!\sous{1}{\partial}_{~\!ab}) D^2\left[\stackrel{(2,-1)}{N_L} -
{5\over 6} \stackrel{(2,-1)}{Y_L} \right] \nonumber\\ &-& 2
\left({d\over dt}\right)^2 \!\!\!\sous{2}{\partial}_{~\!i} F^i_{L} - 4
v_1^j {d\over dt} (\!\!\!\sous{2}{\partial}_{~\!i} F^{ij}_{L}) - 2
a_1^j \!\!\!\sous{2}{\partial}_{~\!i} F_{L}^{ij} - \left. 2v_1^{jk}
\!\!\!\sous{2}{\partial}_{~\!i} F_{L}^{ijk} \right\} +
1\leftrightarrow 2 \;,\\ {\rm SI(20)} &=& {m_1^3\over c^6} \left\{
\left[{1\over 3} a_1^a \!\!\!\sous{1}{\partial}_{~\!a} - {1\over 15}
v_1^{ab} \!\!\!\sous{1}{\partial}_{~\!ab} \right]
\stackrel{(-3,0)}{Y_L} + {4\over 5} v_1^2 \stackrel{(-5,0)}{Y_L}
-{2\over 9}a_1^s\!\!\!\sous{1}{\partial}_{~\!s}{\hat y}_1^L +{16\over
225}v_1^{ab}\!\!\!\sous{1}{\partial}_{~\!ab}{\hat y}_1^L \right\}
\nonumber\\ &+& {m_1^2 m_2\over c^6} \left\{ \left[{1\over 2}
(n_{12}a_1) r_{12}^{-1} - {1\over 2} (n_{12}v_1)^2 r_{12}^{-2} \right]
\hat{y}^L_2 \right.  \nonumber \\ &+& \left[-{1\over 24} a_1^a
\!\!\!\sous{1}{\partial}_{~\!a} - {1\over 48} v_1^{ab}
\!\!\!\sous{1}{\partial}_{~\!ab} \right] D^2\stackrel{(2,-1)}{N_L} +
\left[-{1\over 2} a^a_1 \!\!\!\sous{2}{\partial}_{~\!a} - {1\over 8}
v_1^2 D \right] D \stackrel{(0,-1)}{N_L} - {1\over 4} v_1^{ab}
\!\!\!\sous{2}{\partial}_{~\!ab} \stackrel{(-2,-1)}{Y_L} \nonumber\\
&-& 2 [a_2^a \!\!\!\sous{2}{\partial}_{~\!a} + v_2^{ab}
\!\!\!\sous{2}{\partial}_{~\!ab}] \!\!\!\sous{2}{\partial}_{~\!s}
\stackrel{21}{F^s_{L}} - 2[a^a_1 \!\!\!\sous{1}{\partial}_{~\!a} +
v_1^{ab} \!\!\!\sous{1}{\partial}_{~\!ab}]
\!\!\!\sous{2}{\partial}_{~\!s} \stackrel{12}{F^s_{L}} \nonumber\\ &-&
2 [a^a_1 + 2v_1^{ab} \!\!\!\sous{1}{\partial}_{~\!b}]
\!\!\!\sous{2}{\partial}_{~\!s} \stackrel{12}{F^{sa}_{L}}-2 v_1^{ab}
\!\!\!\sous{2}{\partial}_{~\!s} \stackrel{12}{F^{sab}_{L}} \} +
1\leftrightarrow 2 \;,\\ {\rm SI(21NC)} &=& {m_1^3\over c^6} \left \{
\left [-{1\over 6} a_1^s \!\!\!\sous{1}{\partial}_{~\!s} - {2\over 15}
v_1^{su} \!\!\!\sous{1}{\partial}_{~\!su}\right]
\stackrel{(-3,0)}{Y_L} + {3\over 5} v_1^2 \stackrel{(-5,0)}{Y_L}
+{1\over 9}a_1^s\!\!\!\sous{1}{\partial}_{~\!s}{\hat y}_1^L +{32\over
225}v_1^{ab}\!\!\!\sous{1}{\partial}_{~\!ab}{\hat y}_1^L \right \}
\nonumber \\ &+& {m_1^2 m_2\over c^6} \left \{ -{1\over 8} (a_2^a
\!\!\!\sous{2}{\partial}_{~\!a} + v_2^{ab}
\!\!\!\sous{2}{\partial}_{~\!ab}) D^2 \stackrel{(0,1)}{N_{L}}
\right. \nonumber \\ &-& {1\over 4} (a^a_2
\!\!\!\sous{2}{\partial}_{~\!a} + v_2^{ab}
\!\!\!\sous{2}{\partial}_{~\!ab}) \stackrel{(-2,-1)}{Y_L} + 2a_1^a
\!\!\!\sous{2}{\partial}_{~\!s} H_{L}^{as} \nonumber\\ &-& \left. 2
v_1^{ab} \!\!\!\sous{2}{\partial}_{~\!s} H_{L}^{abs} \right\} +
1\leftrightarrow 2 \;,\\ {\rm SI(33NC)} &=& {m_1^3\over c^6}
\left[{2\over 15} v_1^{ab} \!\!\!\sous{1}{\partial}_{~\!ab}
\stackrel{(-3,0)}{Y_L} - {8\over 5} v_1^2 \stackrel{(-5,0)}{Y_L}
-{32\over 225}v_1^{ab}\!\!\!\sous{1}{\partial}_{~\!ab}{\hat y}_1^L
\right] \nonumber\\ &+& {m_1^2m_2\over c^6} \left\{ -v_1^a v_2^b
[\!\!\!\sous{1}{\partial}_{~\!ab} D \stackrel{(0,-1)}{N_{L}} +
\!\!\!\sous{1}{\partial}_{~\!a} \!\!\!\sous{2}{\partial}_{~\!b}
\stackrel{(-2,-1)}{Y_L}]\right. \nonumber\\ &+& 8 v_1^b \partial_t
[\!\!\!\sous{2}{\partial}_{~\!b} G_{L} +
\!\!\!\sous{2}{\partial}_{~\!a}({}_bG_{L}^a)] \nonumber \\ &+&
\left. 8 v_1^{bc} [\!\!\!\sous{2}{\partial}_{~\!b}(G_{L}^c) +
\!\!\!\sous{2} {\partial}_{~\!a}({}_bG_{L}^{ac})] \right\} +
1\leftrightarrow 2 \;,\\ {\rm SI(35NC)} &=& {m_1^3\over c^6} \left
[-{16\over 3} a_1^i \!\!\!\sous{1}{\partial}_{~\!i}
\stackrel{(-3,0)}{Y_L} + {4\over 5} v_1^{ij} \partial_{1ij}
\stackrel{(-3,0)}{Y_L} - {28\over 5} v_1^2 \stackrel{(-5,0)}{Y_L}
+{32\over 9}a_1^s\!\!\!\sous{1}{\partial}_{~\!s}{\hat y}_1^L -{64\over
75}v_1^{ab}\!\!\!\sous{1}{\partial}_{~\!ab}{\hat y}_1^L \right]
\nonumber \\ &+& {m_1^2m_2\over c^6} [(-5v_1^2 - (n_{12}v_1)^2)
\hat{y}_2^L r^{-2}_{12} + (12(v_1v_2) + 8(n_{12}v_1)(n_{12}v_2))
\hat{y}^L_1 r^{-2}_{12} \nonumber\\ &+& 8 a_1^i
\!\!\!\sous{2}{\partial}_{~\!i} D \stackrel{(0,-1)}{N_{L}} + v_1^2 D^2
\stackrel{(0,-1)}{N_{L}} + 4 v_1^{ij} \!\!\!\sous{2}{\partial}_{~\!ij}
\stackrel{(-2,-1)}{Y_L} + 16 a_1^i \!\!\!\sous{2}{\partial}_{~\!j}
K^{ij}_{L} \nonumber\\ &+& 16(a_2^i - v_2^i(v_1^k-v_2^k)
\!\!\!\sous{2}{\partial}_{~\!k}+(v_1v_2)
\!\!\!\sous{2}{\partial}_{~\!i}) G^i_{L} + 16 v_1^i (v_1^k-v_2^k)
\!\!\!\sous{2}{\partial}_{~\!j}~ _kG^{ij}_{L} \nonumber\\ &+&
\left. 16 v_1^i v_2^j U^{ij}_{L} \right] + 1\leftrightarrow 2 \;,\\
{\rm SI(37NC)} &=& {m_1^3\over c^6} \left [{1\over 15} v_1^{ab}
\!\!\!\sous{1}{\partial}_{~\!ab} \stackrel{(-3,0)}{Y_L} - {4\over 5}
v_1^2 \stackrel{(-5,0)}{Y_L} -{16\over
225}v_1^{ab}\!\!\!\sous{1}{\partial}_{~\!ab}{\hat y}_1^L \right]
\nonumber\\ &+& {m_1^2m_2\over c^6} \left \{{1\over 2} v_1^a v_2^b
(\!\!\!\sous{1}{\partial}_{~\!a} \!\!\!\sous{2}{\partial}_{~\!b} D
\stackrel{(0,-1)}{N_{L}} + \!\!\!\sous{2}{\partial}_{~\!ab}
\stackrel{(-2,-1)}{Y_L}) \right. \nonumber\\ &+& \left. 16
v_1^a(v_1^b-{3\over 4}v_2^b) \!\!\!\sous{2}{\partial}_{~\!b}~G_{L}^a +
16 v_1^a (v_2^b-{3\over 4}v_1^b)
\!\!\!\sous{2}{\partial}_{~\!c}({}_bG_{L}^{ac}) \right\} +
1\leftrightarrow 2 \;,\\ {\rm SI(38NC)} &=& {m_1^3\over c^6}
\left[-{2\over 15} v_1^{ab} \!\!\!\sous{1}{\partial}_{~\!ab}
\stackrel{(-3,0)}{Y_L} + {8\over 5} v_1^2 \stackrel{(-5,0)}{Y_L}
+{32\over 225}v_1^{ab}\!\!\!\sous{1}{\partial}_{~\!ab}{\hat y}_1^L
\right] \nonumber\\ &+& {m_1^2m_2\over c^6} \left\{v_1^av_2^b
(\!\!\!\sous{1}{\partial}_{~\!ab} D \stackrel{(0,-1)}{N_{L}} +
\!\!\!\sous{2}{\partial}_{~\!a} \!\!\!\sous{1}{\partial}_{~\!b}
\stackrel{(-2,-1)}{Y_L}) - 32 v_1^j (v_1^k-{3\over 4} v_2^k)
\!\!\!\sous{2}{\partial}_{~\!k} (G^j_{L}) \right.  \nonumber\\ &-& 32
v_1^j(v_2^k-{3\over 4}v_1^k)
\!\!\!\sous{2}{\partial}_{~\!i}({}_kG^{ij}_{L}) + 32
v_1^j(v_1^k-{3\over 4}v_2^k) \!\!\!\sous{2}{\partial}_{~\!k}(I_{L(j)})
\nonumber\\ &+& \left. 32 v_1^j(v_2^k-{3\over 4}v_1^k)
\!\!\!\sous{2}{\partial}_{~\!i}({}_kI^i_{L(j)}) \right\} +
1\leftrightarrow 2 \;,\\ {\rm SII(4NC)} &=& {m_1^3\over c^6}
\partial_t^2 \left[{1\over 14} \stackrel{(-5,0)}{S_{ij}} \right]
\nonumber\\ &+& {m_1^2m_2\over c^6} \partial_t^2 \left[-{1\over 28}
\hat{y}_2^{ij} {\bf y}^2_2 r^{-2}_{12} - {1\over 56} D^2
\stackrel{(0,-1)}{M_{ij}}- {2\over 7}
\!\!\!\sous{2}{\partial}_{~\!b}(Q_{ij}^b) \right] + 1\leftrightarrow 2
\;,\\ {\rm VI(25NC)} &=& {m_1^3\over c^6}~{d\over dt} \left[{2\over
63} v_1^l \!\!\!\sous{1}{\partial}_{~\!al} \stackrel{(-3,0)}{Y_{aij}}
- {8\over 21} v_1^a \stackrel{(-5,0)}{Y_{aij}} -{32\over
945}v_1^k\!\!\!\sous{1}{\partial}_{~\!ak}{\hat y}_1^{aij} \right]
\nonumber\\ &+& {m_1^2m_2\over c^6}~{d\over dt} \left\{ -{5\over 21}
v_1^l \!\!\!\sous{1}{\partial}_{~\!al} D \stackrel{(0,-1)}{N_{aij}}-
{5\over 21} v_1^l \!\!\!\sous{1}{\partial}_{~\!a}
\!\!\!\sous{2}{\partial}_{~\!l} \stackrel{(-2,-1)}{Y_{aij}}\right.
\nonumber\\ &+& {160\over 21} (v_2^l-{3\over 4}v_1^l)
\!\!\!\sous{2}{\partial}_{~\!k}(_l G^{ak}_{aij}) + {160\over 21}
(v_1^l -{3\over 4}v_2^l ) \!\!\!\sous{2}{\partial}_{~\!l}(G^a_{aij})
\nonumber\\ &-& \left. {160\over 21} (v_2^l -{3\over 4} v_1^l )
\!\!\!\sous{2}{\partial}_{~\!k} (_l I^k_{aij(a)}) - {160\over 21}
(v_1^l -{3\over 4}v_2^l) \!\!\!\sous{2}{\partial}_{~\!l}I_{aij(a)}
\right\} + 1\leftrightarrow 2 \;,\\ {\rm VI(26NC}&+&{\rm
27NC+28NC+29NC)} \nonumber\\ &=& {m_1^3\over c^6}{d\over dt}
\left[-{2\over 63} v_1^s \!\!\!\sous{1}{\partial}_{~\!sa}
\stackrel{(-3,0)}{Y_{aij}} - {4\over 7} v_1^a
\stackrel{(-5,0)}{Y_{aij}} +{32\over
945}v_1^k\!\!\!\sous{1}{\partial}_{~\!ak}{\hat y}_1^{aij} \right]
\nonumber\\ &+& {m_1^2m_2\over c^6}{d\over dt} \left \{ {10\over 21}
v_2^a \hat{y}_2^{aij} r^{-2}_{12} + {5\over 21} v_2^a D^2
\stackrel{(0,-1)}{N_{aij}} \right. \nonumber \\ &+& {5\over 21} v_1^s
\!\!\!\sous{1}{\partial}_{~\!as}~D \stackrel{(0,-1)}{N_{aij}} +
{5\over 21} v_1^s \!\!\!\sous{1}{\partial}_{~\!s}
\!\!\!\sous{2}{\partial}_{~\!a} \stackrel{(-2,-1)}{Y_{aij}} \nonumber
\\ &+& {5\over 21} v_2^k \!\!\!\sous{1}{\partial}_{~\!a}
\!\!\!\sous{2}{\partial}_{~\!k} \stackrel{(-2,-1)}{Y_{aij}}- {5\over
21} v_2^k \!\!\!\sous{1}{\partial}_{~\!k}
\!\!\!\sous{2}{\partial}_{~\!a} \stackrel{(-2,-1)}{Y_{aij}} \nonumber
\\ &+& {80\over 21} v_1^a \!\!\!\sous{2}{\partial}_{~\!l}(G_{aij}^l) -
{40\over 21} v_1^l \!\!\!\sous{2}{\partial}_{~\!l}(G_{aij}^a) -
{40\over 21} v_1^l \!\!\!\sous{2}{\partial}_{~\!k}({}_lG_{aij}^{ka})
\nonumber \\ &-& {40\over 21}{d\over dt} [
\!\!\!\sous{2}{\partial}_{~\!a} (G_{aij}) +
\!\!\!\sous{2}{\partial}_{~\!k} (_aG_{aij}^k)] + {40\over 21} v_1^l
\!\!\!\sous{2}{\partial}_{~\!l}~I_{aij(a)} \nonumber \\ &+& {40\over
21} v_1^l \!\!\!\sous{2}{\partial}_{~\!k} ({}_lI^k_{aij(a)}) -
{40\over 21} v_1^ l \!\!\!\sous{2}{\partial}_{~\!a}(I_{aij(l)})
\nonumber \\ &-& \left. {40\over 21} v_1^l
\!\!\!\sous{2}{\partial}_{~\!k} ({}_aI_{aij(l)}^k) \right \}+
1\leftrightarrow 2 \;.\end{eqnarray}\end{subequations}$\!\!$ We have
similar expressions (involving VI-type terms) for the current moments.
The elementary integrals parametrizing the NC terms include some
generalizations of the integrals already introduced in Section VIII,

\begin{subequations}\label{90}\begin{eqnarray}
\stackrel{(n,p)}{Y_L} &=& -{1\over 2\pi}\sous{\!\!\!\!B= 0}{\rm FP}
\int d^3{\bf x}~\! |\tilde{\bf x}|^B \hat{x}_L r^n_1 r^p_2 \;,\\
\stackrel{(n,p)}{S_L} &=& -{1\over 2\pi} \sous{\!\!\!\!B= 0}{\rm FP}
\int d^3{\bf x}~\! |\tilde{\bf x}|^B \hat{x}_L |{\bf x}|^2 r_1^n r_2^p
\;,\\ \stackrel{(n,p)}{N_L}&=& -{1\over 2\pi} \sous{\!\!\!\!B= 0}{\rm FP}
\int d^3{\bf x}~\! |\tilde{\bf x}|^B \hat{x}_L r_1^n r^p_2 \ln {\tilde
r}_1 \label{90c} \;,\\ \stackrel{(n,p)}{M_L}&=& -{1\over 2\pi}
\sous{\!\!\!\!B= 0}{\rm FP} \int d^3{\bf x}~\!  |\tilde{\bf x}|^B
\hat{x}_L |{\bf x}|^2 r_1^n r^p_2 \ln {\tilde r}_1 \label{90d}
\;.\end{eqnarray}\end{subequations}$\!\!$
As usual the Hadamard partie finie ${\rm Pf}$ is to be added when the
integral diverges near the particles. The logarithms in
(\ref{90c})-(\ref{90d}) contain the constant $r_0$ through the
notation ${\tilde r}_1=r_1/r_0$. In addition we have the more involved
integrals

\begin{subequations}\label{90'}\begin{eqnarray}
G^P_L &=& -{1\over 2\pi} \sous{\!\!\!\!B= 0}{\rm FP} \int d^3{\bf
x}~\! |\tilde{\bf x}|^B \hat{x}_L \partial_{aP}\left({1\over
r_1}\right){}_ag \label{90g} \;,\\ _bG^{aP}_L &=& -{1\over 2\pi}
\sous{\!\!\!\!B= 0}{\rm FP} \int d^3{\bf x}~\! |\tilde{\bf x}|^B
\hat{x}_L \partial_{aP}\left({1\over r_1}\right){}_bg \;,\\ K^P_{L} &=&
-{1\over 2\pi} \sous{\!\!\!\!B= 0}{\rm FP} \int d^3{\bf x}~\!
|\tilde{\bf x}|^B \hat{x}_L \partial_P \left({1\over r_1}\right) g
\label{90k} \;,\\ U_{L}^P &=& -{1\over 2\pi} \sous{\!\!\!\!B= 0}{\rm FP}
\int d^3{\bf x}~\! |\tilde{\bf x}|^B \hat{x}_L \partial_P
\left({1\over r_1}\right){}_kg_k \label{90u} \;,\\ F^P_{L} &=& -{1\over
2\pi} \sous{\!\!\!\!B= 0}{\rm FP} \int d^3{\bf x}~\! |\tilde{\bf x}|^B
\hat{x}_L \partial_{aP} \left({1\over r_1}\right) {}_af \;,\\
\stackrel{12}{F^P_L} &=& -{1\over 2\pi} \sous{\!\!\!\!B= 0}{\rm FP}
\int d^3{\bf x}~\! |\tilde{\bf x}|^B \hat{x}_L \partial_{aP}
\left({1\over r_1}\right) \stackrel{12}{_af} \;,\\ \stackrel{21}{F^P_L}
&=& -{1\over 2\pi} \sous{\!\!\!\!B= 0}{\rm FP} \int d^3{\bf x}~\!
|\tilde{\bf x}|^B \hat{x}_L \partial_{aP} \left({1\over r_1}\right)
\stackrel{21}{_af} \;,\\ H^P_{L} &=& -{1\over 2\pi} \sous{\!\!\!\!B=
0}{\rm FP} \int d^3{\bf x}~\! |\tilde{\bf x}|^B \hat{x}_L
\partial_{iP}~(r_1)~_ig \;,\\ Q^P_{L} &=& -{1\over 2\pi} \sous{\!\!\!\!B=
0}{\rm FP} \int d^3{\bf x}~\! |\tilde{\bf x}|^B \hat{x}_L |{\bf x}|^2
\partial_{aP} \left({1\over r_1}\right) {}_ag \;,\\ {}_kI^i_{L(j)} &=&
-{1\over 2\pi} \sous{\!\!\!\!B= 0}{\rm FP} \int d^3{\bf x}~\!
|\tilde{\bf x}|^B \hat{x}_L \partial_j \left[\partial_i \left({1\over
r_1}\right) {}_kg \right] \;,\\ I_{L(j)} &=& -{1\over 2\pi}
\sous{\!\!\!\!B= 0}{\rm FP} \int d^3{\bf x}~\! |\tilde{\bf x}|^B
\hat{x}_L \partial_j \left [\partial_i \left({1\over r_1}\right) {}_ig
\right]
\;.\end{eqnarray}\end{subequations}$\!\!$
The notation is e.g. ${}_af={}_1\partial_af$, $g_b={}_2\partial_bg$,
${}_ag_b={}_1\partial_a~\!{}_2\partial_bg$ (notably ${}_kg_k=Dg$). The
last two integrals are related to some previous ones by

\begin{subequations}\label{91}\begin{eqnarray}
{}_kI^i_{L(j)}&=&-(\!\!\!\sous{1}{\partial}_{~\!j}+\!\!\!\sous{2}
{\partial}_{~\!j}){}_kG^i_L \;,\\
I_{L(j)}&=&-(\!\!\!\sous{1}{\partial}_{~\!j}+\!\!\!\sous{2}{\partial}_{~\!j})
G_L
\;.\end{eqnarray}\end{subequations}$\!\!$

\subsection{Computation of the elementary integrals}

The techniques developed in Sections VII and VIII can be used to
compute many of these integrals. Concerning $S_L^{(n,p)}$ we need only
the particular case $l=2$ and $(n,p)=(-5,0)$. It is computed by the
same methods as used for $Y_{ij}^{(-3,0)}$; we find

\begin{equation}\label{92}
\stackrel{(-5,0)}{S_{ij}} = \left[ {14\over 3} \ln\left(u_1\over
r_0\right)+{8\over 5} \right] y_1^{<ij>} 
\;.\end{equation}
Next the group of integrals constituted by the $N^{(n,p)}_{L}$'s and
$M^{(n,p)}_{L}$'s is obtained in a fashion similar to the one employed
for $Y^{(-2,-1)}_L$ in Section VIII, i.e. basically by application of
the Riesz formula. The logarithms in these integrals are included by
differentiating with respect to the complex parameter $B$. The relevant
results are

\begin{subequations}\label{93}\begin{eqnarray}
\stackrel{(0,-1)}{N_{ij}} &=& y_1^{<ij>} \left\{ {8\over 105} r_{12}^2
\left[ \ln {\tilde r}_{12}-{247 \over 210}\right] -{2\over 105} y^2_1
-{4\over 35} (y_1y_2) + {4\over 105} y_2^2 \right\} \nonumber \\ &+&
y_1^{<i} y_2^{j>} \left\{ {4\over 35} r_{12}^2 \left[ \ln {\tilde
r}_{12}-{59 \over 70}\right] +{2\over 105} y_1^2 - {4\over 35}
(y_1y_2)\right\} \nonumber \\ &+& y_2^{<ij>} \left\{ {1\over 7}
r_{12}^2 \left[ \ln {\tilde r}_{12}-{9 \over 14}\right] -{1\over 35}
y_1^2\right\} \;,\\ \stackrel{(0,-1)}{N_{ijk}} &=& y_1^{<ijk>} \left\{
{16\over 315} r_{12}^2 \left[ \ln {\tilde r}_{12}-{811 \over
630}\right] -{4\over 945}y_1^2 - {88\over 945}(y_1y_2) + {4\over 105}
y_2^2 \right\}\nonumber \\ &+& y_1^{<ij} y_2^{k>} \left\{ {8\over 105}
r_{12}^2 \left[ \ln {\tilde r}_{12}-{601 \over 630}\right] + {2\over
63}y_1^2 - {4\over 35} (y_1y_2) + {4\over 105} y_2^2 \right\}
\nonumber\\ &+& y_1^{<i} y_2^{jk>} \left\{ {2\over 21} r_{12}^2 \left[
\ln {\tilde r}_{12}-{95 \over 126}\right] + {1\over 35} y_1^2 -
{2\over 21}(y_1y_2)\right\} \nonumber\\ &+&
y_2^{<ijk>} \left\{ {1\over 9} r_{12}^2 \left[ \ln {\tilde r}_{12}-{11
\over 18}\right] - {1\over 63} y_1^2 \right\} \;,\\
\stackrel{(2,-1)}{N_{ij}} &=& y_1^{<ij>} \left\{ {4\over 315} r_{12}^4
\left[\ln {\tilde r}_{12}- {887\over 1260}\right] -{13\over 945} y_1^4
- {4\over 315} y_1^2y_2^2 \right.\nonumber\\
&-& \left. {2\over 315} y_2^4 
+{4\over 945} y_1^2(y_1y_2) +{4\over 315} (y_1y_2)^2 
\right\} \nonumber\\
&+& y_1^{<i}y_2^{j>} \left\{ {2\over 63} r_{12}^4 \left[\ln {\tilde
r}_{12}- {127\over 252} \right] - {16\over 945} y_1^4 - {2\over 35}
y_1^2y_2^2 + {8\over 315}y_1^2 (y_1y_2)
+{4\over 63} y_2^2(y_1y_2) 
\right\} \nonumber \\ &+&
y_2^{<ij>} \left\{ {1\over 18} r_{12}^4 \left[\ln {\tilde r}_{12}-
{13\over 36}\right] -{59\over 1260} y_1^4 + {1\over 63}y_1^2y_2^2 +
{1\over 7}y_1^2(y_1y_2) - {1\over 9} (y_1y_2)^2 \right\} \;,\\ D^2
\stackrel{(2,-1)}{N_{ij}} &=& {8\over 5} y_1^{<ij>} \left[\ln {\tilde
r}_{12}+{7\over 15}\right] - {16\over 5} y_1^{<i}y_2^{j>} \left[\ln
{\tilde r}_{12}-{6\over 5}\right] \nonumber\\ &+& {68\over 5}
y_2^{<ij>} \left[\ln {\tilde r}_{12}+{71\over 170}\right] \;,\\
D\stackrel{(0,-1)}{N_{ij}} &=& - {8\over 15} y_1^{<ij>} \left[\ln
{\tilde r}_{12}+{7\over 15}\right] - {4\over 15} y_1^{<i}y_2^{j>}
\left[\ln {\tilde r}_{12}+{37\over 15}\right] \nonumber\\ &-& {6\over
5} y_2^{<ij>} \left[\ln {\tilde r}_{12}+{2\over 15}\right] \;,\\ D^2
\stackrel{(0,-1)}{N_{ij}} &=& {1\over r_{12}^2}\left\{ {4\over
3}y_1^{<ij>} - {8\over 3} y_1^{<i} y_2^{j>} + {10\over 3}
y_2^{<ij>}\right\} \;,\\ \stackrel{(0,1)}{N_{ij}} &=& y_1^{<ij>} \left\{
{4\over 315} r_{12}^4 \left[\ln {\tilde r}_{12}-{1937\over
1260}\right]+{11\over 945} y_1^4 + {4\over 945}y_1^2 (y_1y_2)
\right.\nonumber\\ && \quad \left.  +{44\over 945} (y_1y_2)^2 -
{2\over 135} y_1^2y_2^2 - {4\over 105} y_2^2(y_1y_2) + {2\over 315}
y_2^4\right\} \nonumber\\ &+& y_1^{<i}y_2^{j>} \left\{ {4\over 315}
r_{12}^4 \left[\ln {\tilde r}_{12}- {1517\over 1260}\right] - {1\over
135}y_1^4 - {4\over 189} y_1^2(y_1y_2) \right.\nonumber\\ && \quad
\left.  + {4\over 105} (y_1y_2)^2 - {8\over
315}y_2^2(y_1y_2)\right\}\nonumber \\ &+& y_2^{<ij>} \left\{{1\over
126} r_{12}^4 \left[\ln {\tilde r}_{12}-{253\over 252}\right] -
{1\over 3780}y_1^4 - {1\over 105} y_1^2 (y_1y_2) \right. \nonumber\\
&& \quad \left.  + {1\over 63} (y_1y_2)^2 - {2\over 315} y_1^2y_2^2
\right\} \;,\\ D^2\stackrel{(0,-1)}{M_{ij}} &=& y_1^{<ij>} \left\{
-{88\over 15} \ln {\tilde r}_{12}+ r_{12}^{-2} \left[-{676\over 225}
y_1^2 + {2072\over 225}(y_1y_2) - {1096\over 225} y_2^2\right]
\right\} \nonumber \\ &+& y_1^{<i}y_2^{j>} \left\{ -{16\over 5} \ln
{\tilde r}_{12}+ r_{12}^{-2}\left[-{184\over 25} y_1^2 + {1024\over
75} (y_1y_2) - {224\over 25} y_2^2 \right] \right\} \nonumber \\ &+&
y_2^{<ij>} \left\{ - {48\over 5} \ln {\tilde r}_{12}+ r_{12}^{-2}
\left[-{536\over 75} y_1^2 + {304\over 25} (y_1y_2) - {42\over
25}y_2^2 \right] \right\}
\;.\end{eqnarray}\end{subequations}$\!\!$

The remaining integrals, defined by (\ref{90'}), are more difficult,
but we have been able to obtain all of them using several different
methods, adapted to the computation of each of these integrals
separately. We shall not present all the details of these computations
but simply outline some examples. Consider the integral $K_{L}$
defined by (\ref{90k}) with $p=0$, i.e.

\begin{equation}\label{94}
K_{L} = -{1\over 2\pi} \sous{\!\!\!\!B= 0}{\rm FP} \int d^3{\bf x}~\!
|\tilde{\bf x}|^B \hat{x}_L {g\over r_1}
\;.\end{equation}
Using the fact that $g/r_1$ is a Laplacian,

\begin{equation}\label{95}
{g\over r_1}=\Delta\left[{r_1+r_{12}\over 2}g-{r_1\over 4}-{r_2\over
2}\right]
\;,\end{equation}
we can integrate by parts and transform $K_{L}$ into an integral
containing an explicit $B$-factor,

\begin{equation}\label{96}
K_{L} = -{1\over 2\pi} \sous{\!\!\!\!B= 0}{\rm FP} \left\{ B(B+2l+1)
\int d^3{\bf x}~\! |\tilde{\bf x}|^B |{\bf x}|^{-2} \hat{x}_L
\left[{r_1+r_{12}\over 2}g-{r_1\over 4}-{r_2\over 2}\right] \right\}
\;.\end{equation} From a previous argument, the value of the integral
depends only on the possible occurence of a pole $\sim 1/B$ at
infinity. As the pole is easily computed from expanding the integrand
at infinity, we obtain in this way the expression of $K_L$. Next, from
the formula

\begin{equation}\label{97}
\partial_a\left({1\over r_1}\right){}_ag = -{1\over
2}\Delta_1\left({1\over r_1}g\right) +{g\over 2}\Delta_1\left({1\over
r_1}\right)+{1\over 2r_1}\Delta_1g
\;,\end{equation} 
where one should be careful about considering $\Delta_1r_1^{-1}$ in
the sense of distributions [i.e. $\Delta_1r_1^{-1}=-4\pi\delta_1$], we
deduce $G_{L}$ from the Laplacian of $K_{L}$. Indeed, as a consequence
of (\ref{97}),

\begin{equation}\label{98}
G_{L} = -{1\over 2}\Delta_1 K_L+\Big({\hat x}_L g\Big)_1+{1\over
2r_{12}}\stackrel{(-2,0)}{Y_L}
\;,\end{equation}
and we can easily show that $Y^{(-2,0)}_L$ is actually
zero. Alternatively, one can prove also that

\begin{equation}\label{99}
G_{L} = {1\over 2}\stackrel{(-3,0)}{Y_L}- {1\over
2r_{12}}\stackrel{(-3,1)}{Y_L} +{1\over 2r_{12}}\stackrel{(-2,0)}{Y_L}
\;.\end{equation}
This provides a check of the computation.

To compute $G_{L}^s$ (in the quadrupole case $L=ij$, say) we use a
different method. We remark that $G_{ij}^s$ obeys a Laplace equation,
with respect to the point 2, with known source:

\begin{equation}\label{100}
\Delta_2 ~G_{ij}^s = \!\!\!\sous{1}{\partial}_{~\!a}\left({1\over
r_{12}}\right)\!\!\!\sous{1}{\partial}_{~\!as}Y_{ij} \;.\end{equation}
Here, $Y_{ij}$ is known from (\ref{53}). The right-hand-side of
Eq. (\ref{100}) is expanded, and we obtain a particular solution of
this equation by integrating each of the terms. Now $G_{ij}^s$ is
necessarily equal to this particular solution plus some solution,
regular at the origin, of the homogeneous equation. Taking into
account the index structure of $G_{ij}^s$ and the fact that it has the
dimension of a length, we find that the homogeneous solution is
parametrized by solely two numerical constants $a$ and $b$. At this
stage we have

\begin{eqnarray}\label{101}
G^s_{ij} = &-&{1\over 30}y_{12}^{<ij>s} r_{12}^{-2} + {1\over 6}
y_{12}^{s<i} y_1^{j>}r_{12}^{-2} -{1\over 15} y_{12}^{<i} \delta^{j>s}
\ln {\tilde r}_{12} -{4\over 3} y_1^{<i} \delta^{j>s} \ln {\tilde
r}_{12}\nonumber\\ &+& a~\!y_1^{<i} \delta^{j>s}+ b~\!y_2^{<i}
\delta^{j>s} \;.\end{eqnarray} Incidentally, this expression already
gives the complete result for the gradients ${}_1\partial_s G^s_{ij}$
and ${}_2\partial_s G^s_{ij}$, because the gradients of the
homogeneous terms are zero. To compute the constants $a$ and $b$ we
need some extra information, which is provided by the contracted
product between $y_{12}^s$ and $G_{ij}^s$. Indeed this contraction is
a known quantity thanks to the identity

\begin{equation}\label{102}
y_{12}^s ~G_{ij}^s = -\left(1+y_{12}^s\!\!\!\sous{1}{\partial}_{~\!s}
\right)G_{ij}+{1\over 4}\Delta_1 \!\!\stackrel{(-2,1)}{Y_{ij}}
\;,\end{equation}
where $G_{ij}$ has just been obtained previously. Here,
$Y^{(-2,1)}_{ij}$ can be computed from the Riesz formula exactly like
for $Y^{(-2,-1)}_{ij}$ in Section VIII. [When deriving (\ref{102}) we
take account of the fact that $Y^{(-2,0)}_{ij}=0$.] Comparing the
result for $y_{12}^s G_{ij}^s$ with the one obtained directly from
(\ref{101}) we find {\it three} equations for the {\it two} unknown
constants $a$ and $b$. This overdetermined system fixes uniquely the
constants to the values $a=63/100$ and $b=-257/900$.

The preceding method was successfully applied to several integrals of
the type (\ref{90'}): that is, we (i) compute the ``source'' of the
Laplace equation satisfied by the integral with respect to the point 2
(the source is computable because $\Delta_2$ applies only on the part
of the integrand containing the functions $g$, $f$, etc., and we can
make use of Eqs. (\ref{58}); with respect to the point 1 this would
not work), (ii) compute a particular solution of this equation, (iii)
write down the most general form of the homogeneous solution in terms
of a few arbitrary coefficients (this works only when the dimension of
the integral is a small power of a length so that the number of
unknown coefficients is small), (iv) compute the coefficients using
the extra information provided by the contraction with respect to
${\bf y}_{12}$. Alternatively to (iv) one can use an angular average
with respect to ${\bf n}_{12}$ [see (\ref{111}) below].

As a verification let us introduce the new integral

\begin{equation}\label{103}
R_{ij} = -{1\over 2\pi} \sous{\!\!\!\!B= 0}{\rm FP} \int d^3{\bf x}~\!
|\tilde{\bf x}|^B \hat{x}_{ij} \partial_{a}\left({1\over
r_1}\right)g_a
\;.\end{equation}
From the easily checked formula

\begin{equation}\label{104}
(\Delta_1-\Delta)\left({g\over r_1}\right)={1\over
r_1^2r_{12}}-{1\over r_1^2r_2}+2 \partial_{a}\left({1\over
r_1}\right)g_a
\;,\end{equation}
we deduce a relation between $R_{ij}$ and some computable quantities,

\begin{eqnarray}\label{105}
R_{ij} &=& {1\over 2}\Delta_1 K_{ij}+{1\over
2}\stackrel{(-2,-1)}{Y_{ij}}\nonumber\\ &+&{1\over 4\pi}
\sous{\!\!\!\!B= 0}{\rm FP} \left\{ B(B+5) \int d^3{\bf x}~\!
|\tilde{\bf x}|^B |{\bf x}|^{-2} \hat{x}_{ij} {g\over r_1} \right\}
\;.\end{eqnarray}
The value of the last integral comes from the pole at infinity -- same
method as before. Having obtained $R_{ij}$, the verification is that
${}_2\partial_s G^s_{ij}$, which on one hand is computed from
(\ref{101}), on the other hand should be given by the following
alternative expression:

\begin{equation}\label{106}
\sous{2}{\partial}_{~\!s} G^s_{ij} = -{1\over 2}\Delta_1
R_{ij}-\Bigl(\partial_a ({\hat x}_{ij} g_a)\Bigr)_1
\;,\end{equation}
which is obtained by some integrations by parts inside the integrand
of ${}_2\partial_s G^s_{ij}$. Of course, the value of $R_{ij}$
computed by (\ref{105}) is such that (\ref{106}) is also satisfied.

Once $G_{ij}^s$ is known we can deduce another needed integral,
i.e. ${}_{2}{\partial}_{s} K_{ij}^{as}$, from the identity

\begin{equation}\label{107}
\partial_{as}\left({1\over r_1}\right)(g_s+{}_sg)
=-{1\over 2}\Delta \left[ \partial_a \left({1\over r_1}\right)g \right] +{1\over
2}\partial_a\left(\Delta{1\over r_1}\right)g
-{1\over 4}\!\!\sous{1}{\partial}_{~\!a}\left( {1\over r_1}\Delta g\right)
\;,\end{equation}
which implies

\begin{eqnarray}\label{108}
\sous{2}{\partial}_{~\!s} K_{ij}^{as} &=& -G^a_{ij}-{1\over 4}
\!\!\!\sous{1}{\partial}_{~\!a}
\stackrel{(-2,-1)}{Y_{ij}}-\Bigl(\partial_a ({\hat x}_{ij} g)\Bigr)_1
\nonumber\\ &+&{1\over 4\pi} \sous{\!\!\!\!B= 0}{\rm FP} \left\{
B(B+5) \int d^3{\bf x}~\! |\tilde{\bf x}|^B |{\bf x}|^{-2}
\hat{x}_{ij} \partial_a\left({1\over r_1}\right)g \right\}
\;.\end{eqnarray}
Again the last integral makes no problem. Next, from both $R_{ij}$ and
${}_{2}{\partial}_{s}K_{ij}^{bs}$, we can further deduce
${}_{2}{\partial}_{a}({}_bG_{ij}^a)$. Indeed the other identity

\begin{equation}\label{109}
\partial_a\left({1\over r_1}\right){}_bg_a =
\!\!\!\sous{1}{\partial}_{~\!b} \left[\partial_a\left({1\over
r_1}\right)g_a\right] +\partial_{ab}\left({1\over r_1}\right)g_a
\;,\end{equation}
implies

\begin{equation}\label{110}
\sous{2}{\partial}_{~\!a} ({}_bG_{ij}^a) =
\!\!\sous{1}{\partial}_{~\!b} R_{ij} +\!\!\sous{2}{\partial}_{~\!s}
K_{ij}^{bs}
\;.\end{equation}

Some other integrals are connected directly to the simpler
$Y$-type integrals. For instance, the integral (\ref{90u}) is given by

\begin{equation}\label{110'}
U^{ab}_{ij}={3\over 16}\!\!\!\sous{1}{\partial}_{~\!ab}
\!\!\stackrel{(-2,-1)}{Y_{ij}}-{1\over
8}\delta_{ab}\!\!\stackrel{(-4,-1)}{Y_{ij}}-{1\over
2r_{12}}\!\!\!\sous{1}{\partial}_{~\!ab}
\!\!\stackrel{(-1,-1)}{Y_{ij}}
\end{equation}
(using the facts that $Y^{(-2,0)}_{ij}=0=Y^{(-4,0)}_{ij}$). Once the
value of this integral is obtained, we can check that its trace
$U^{aa}_{ij}=\delta_{ab}U^{ab}_{ij}$ is especially simple:
$U^{aa}_{ij}=-y_1^{<ij>}/r_{12}^2$. This is in perfect agreement with

\begin{equation}\label{110''}
U^{aa}_{ij}= -{1\over 2\pi} \sous{\!\!\!\!B= 0}{\rm FP} \int d^3{\bf
x}~\!  |\tilde{\bf x}|^B \hat{x}_{ij} \Delta \left({1\over
r_1}\right){}_kg_k =2\Bigl({\hat x}_{ij}~\! {}_kg_k\Bigr)_1
\;,\end{equation}
the final reduction being obtained thanks to the known
formula (see e.g. \cite{BFP98})

\begin{equation}
{}_kg_k={1\over 2}\left({1\over r_1r_2}-{1\over r_1r_{12}}-{1\over
r_2r_{12}}\right)
\;.\end{equation}

Still another method is useful in our computation. All the integrals
are certain functions of the two points ${\bf y}_1$ and ${\bf y}_2$,
and it is advantageous to consider their angular average with respect
to the relative direction ${\bf n}_{12}$ between the points, with the
vector ${\bf y}_1$ being fixed. As it turns out, the average is much
easier to compute (using some methods similar as before) than the
integral itself. On the other hand, once we have obtained a result, we
can compute its average, so the comparison leads to an interesting
check of the calculation. Let us see on the example of $G_L$ how one
performs this angular average. From (\ref{90g}) we write

\begin{equation}\label{111}
\int {d\Omega_{12}\over 4\pi} G_{L} = -{1\over 2\pi} \sous{\!\!\!\!B=
0}{\rm FP} \int d^3{\bf x}~\! |\tilde{\bf x}|^B \hat{x}_L
\partial_a\left({1\over r_1}\right)\int {d\Omega_{12}\over 4\pi} {}_ag
\;,\end{equation}
in which we commuted the angular average (where $d\Omega_{12}$ denotes
the solid angle element in the direction ${\bf n}_{12}$) with the
integral sign and the terms depending only on ${\bf y}_1$. This is
correct because ${\bf y}_1$ is kept fixed in the process; for
instance, the average of ${\bf y}_2$ is ${\bf y}_1$, which is obtained
by writing ${\bf y}_2={\bf y}_1-r_{12}{\bf n}_{12}$ and averaging over
${\bf n}_{12}$ with fixed $r_{12}$ and ${\bf y}_1$. In practice,
computing the average (\ref{111}) is not too complicated because the
average of ${}_ag$ is rather simple,

\begin{equation}\label{112}
\int {d\Omega_{12}\over 4\pi} {}_ag=\left\{ \vcenter{\vskip
0.1pc\hbox{ $\left( {r_1\over 6r_{12}^2}-{1\over 2r_{12}}\right) n_1^a
\quad$ when $r_1 \leq r_{12}~,$}\vskip 0.5pc\hbox{ $\left( -{1\over
2r_1}+{r_{12}\over 6r_1^2}\right) n_1^a \quad$ when $r_1 >
r_{12}~.$}\vskip 0.1pc} \right.
\end{equation}
[A more complicated example, that was useful for us, is

\begin{equation}\label{113}
\int {d\Omega_{12}\over 4\pi} {}_ag_b=\left\{ \vcenter{\vskip
0.1pc\hbox{ $-{r_1^2\over 20r_{12}^4}n_1^{ab}+ \left( {r_1^2\over
60r_{12}^4}-{1\over 6r_{12}^2}\right) \delta^{ab}
\qquad\quad\qquad\qquad\quad$ when $r_1 \leq r_{12}~,$}\vskip
0.5pc\hbox{ $ \left(-{1\over 4r_1^2}+{r_{12}\over 5r_1^3}\right)
n_1^{ab}+ \left( -{1\over 3r_1r_{12}}+{1\over 4r_1^2}-{r_{12}\over
15r_1^3} \right) \delta^{ab} \quad$ when $r_1 > r_{12}~.$]}\vskip
0.1pc} \right.
\end{equation}
According to (\ref{112}), we must split the integration over $d^3{\bf
x}$ into two ``near-zone'' and ``far-zone'' contributions,

\begin{eqnarray}\label{114}
\int {d\Omega_{12}\over 4\pi} G_{L} = &-&{1\over
2\pi}\int_{r_1<r_{12}} d^3{\bf x}~{\hat x}_L \left( -{1\over
6r_1r_{12}^2}+{1\over 2r_1^2r_{12}}\right)\nonumber\\ &-&{1\over 2\pi}
\sous{\!\!\!\!B= 0}{\rm FP} \int_{r_1>r_{12}} d^3{\bf x}~\!
|\tilde{\bf x}|^B \hat{x}_L \left({1\over 2r_1^3}-{r_{12}\over
6r_1^4}\right)
\;.\end{eqnarray}
The finite part at $B=0$ is necessary only for the far-zone
integral. Both integrals in (\ref{114}) are now evaluated using
standard methods. In the case $l=2$ we find

\begin{equation}\label{115}
\int {d\Omega_{12}\over 4\pi} G_{ij} = y_1^{<ij>}\left( \ln {\tilde
r_{12}}+{1\over 30}\right) \;.\end{equation} This is in agreement with
the average of $G_{ij}$ computed directly with the result calculated
from (\ref{98}) or (\ref{99}). This method of averaging has been
applied for checking many other integrals. Even, in several cases, the
method has been employed in order to determine some unknown
coefficients. However, for this purpose the method is less powerful
that the method of contraction with the vector ${\bf y}_{12}$, since
the latter method yields in general a redundant determination of the
coefficients.

The complete list of the results for the elementary integrals is as
follows.

\begin{subequations}\label{116}\begin{eqnarray}
G_{ij} &=& y_1^{<ij>}\left[\ln {\tilde r}_{12}- {23\over 60}\right] +
{1\over 3} y_1^{<i}y_2^{j>} + {1\over 12} y_2^{<ij>} \;,\\
\!\!\!\sous{2}{\partial}_{~\!b}(G_{ij}) &=& -y_1^{<ij>} y_{12}^b
r^{-2}_{12} + {1\over 2} \delta^{b<i} y_1^{j>} - {1\over 6}
\delta^{b<i} y^{j>}_{12} \;,\\ \!\!\!\sous{2}{\partial}_{~\!b}(G^c_{ij})
&=& -{1\over 15} y_{12}^{<ij>bc} r^{-4}_{12} + {1\over 30}
y_{12}^{<ij>} \delta^{bc} r^{-2}_{12} + {1\over 15} (y_{12}^{b<i}
\delta^{j>c} + y_{12}^{c<i} \delta^{j>b}) r^{-2}_{12} \nonumber \\ &+&
{1\over 3} y_{12}^{bc<i} y_1^{j>} r^{-4}_{12} - {1\over 6} y^{<i}_{12}
y_1^{j>} \delta^{bc} r_{12}^{-2} + {4\over 3} y_{12}^b y_1^{<i}
\delta^{j>c} r^{-2}_{12} - {1\over 6} y^c_{12}y_1^{<i} \delta^{j>b}
r^{-2}_{12} \nonumber \\ &+& {1\over 15} \delta^{b<i} \delta^{j>c}
\left[\ln {\tilde r}_{12}- {257\over 60}\right] \;,\\
\!\!\!\sous{2}{\partial}_{~\!a}({}_bG^{ac}_{ij}) &=& {1\over 60}
y_{12}^{<ij>bc} r^{-4}_{12} - {1\over 20} y_{12}^{<ij>}\delta^{bc}
r^{-2}_{12} + {7\over 30} y_{12}^{b<i} \delta^{j>c} r^{-2}_{12} +
{1\over 15} y_{12}^{c<i} \delta^{j>b} r^{-2}_{12} \nonumber \\ &+&
{1\over 2} y^{<i}_{12} y_1^{j>} \delta^{bc} r^{-2}_{12} - {1\over 2}
y_{12}^c y^{<i}_1 \delta^{j>b} r^{-2}_{12} + {3\over 4} y_{12}^{bc}
y_1^{<ij>} r^{-4}_{12} - {1\over 4} \delta^{bc} y_1^{<ij>} r^{-2}_{12}
\nonumber \\ &+& {1\over 15} \left[\ln {\tilde r}_{12}-{317\over
60}\right] \delta^{b<i} \delta^{j>c} \;,\\
\!\!\!\sous{2}{\partial}_{~\!a}(_bG^a_{ij}) &=& {2\over 15}
y_{12}^{<ij>b} r^{-2}_{12} - {1\over 2} y_{12}^{b<i}y_1^{j>}
r^{-2}_{12} + {2\over 3} \delta^{b<i} y_1^{j>} \nonumber\\ &+&
\delta^{b<i} y^{j>}_{12} \left[-{1\over 15} \ln {\tilde r}_{12}-
{103\over 900}\right] \;,\\ \!\!\!\sous{2}{\partial}_{~\!s}(G_{ij}^s)
&=& \left({7\over 6}y_1^{<ij>} - {4\over 3}y_1^{<i}y_2^{j>} + {1\over
6} y_2^{<ij>}\right) r^{-2}_{12} \;,\\ G^s_{ij} &=& -{1\over
30}y_{12}^{<ij>s} r^{-2}_{12} + {1\over 6} y_{12}^{s<i} y_1^{j>}
r^{-2}_{12} \nonumber\\ &+& y_{12}^{<i} \delta^{j>s} \left[-{1\over
15} \ln {\tilde r}_{12} + {257\over 900}\right] + y_1^{<i}
\delta^{j>s} \left[-{4\over 3} \ln {\tilde r}_{12}+ {31\over
90}\right] \;,\\ G_{ijk} &=& y_1^{<ijk>} \left[\ln {\tilde r}_{12}-
{307\over 840}\right] + {3\over 8} y_1^{<ij}y_2^{k>} \nonumber \\ &+&
{1\over 8} y_1^{<i} y_2^{jk>} + {1\over 24} y_2^{<ijk>} \;,\\ G^s_{ijk}
&=& \left({1\over 56} y_{12}^{s<ijk>} - {1\over 10} y_{12}^{s<ij}
y_1^{k>} + {1\over 4} y_{12}^{s<i} y_1^{jk>}\right) r^{-2}_{12}
\nonumber \\ &+& \delta^{s<i} y_{12}^{jk>} \left[{1\over 35} \ln
{\tilde r}_{12} - {2843\over 29400}\right] + \delta^{s<i} y_{12}^j
y_1^{k>} \left[-{1\over 5} \ln {\tilde r}_{12}+ {1739\over
2100}\right] \nonumber \\ &+& \delta^{s<i} y_1^{jk>} \left[-2 \ln
{\tilde r}_{12}+ {97\over 420}\right] \;,\\
\!\!\!\sous{2}{\partial}_{~\!b}(_sG^b_{ijk}) &=& \left(-{123\over 280}
y_1^{<ijk>} + {61\over 280} y_1^{<ij} y_2^{k>} + {37\over 280}
y_1^{<i} y_2^{jk>} + {5\over 56} y_2^{<ijk>}\right) y_{12}^s
r^{-2}_{12} \nonumber \\ &+& y_1^{<ij} \delta^{k>s} \left[-{1\over 7}
\ln {\tilde r}_{12}+ {699\over 980}\right] + y_1^{<i} y_2^j
\delta^{k>s} \left[{3\over 35} \ln {\tilde r}_{12} + {2963\over
14700}\right] \nonumber \\ &+& y_2^{<ij} \delta^{k>s} \left[{2\over
35} \ln {\tilde r}_{12}+ {313\over 3675}\right] \;,\\
\!\!\!\sous{2}{\partial}_{~\!b}(_aG^b_{aij}) &=& r^{-2}_{12}
y_1^{<ij>} \left\{ -{1\over 5} r_{12}^2 \ln {\tilde r}_{12}+ {147\over
200} y_1^2 - {463\over 300}(y_1y_2) + {109\over 120}y_2^2\right\}
\nonumber \\ &+& r^{-2}_{12} y_1^{<i}y_2^{j>} \left\{ {3\over 25}
r_{12}^2 \ln {\tilde r}_{12} + {167\over 750}y_1^2 - {449\over
750}(y_1y_2) + {22\over 125}y_2^2 \right\} \nonumber \\ &+&
r^{-2}_{12} y_{2}^{<ij>} \left\{ {2\over 25} r_{12}^2 \ln {\tilde
r}_{12} + {577\over 3000} y_1^2 - {79\over 500} (y_1y_2) + {197\over
3000} y_2^2 \right\} \;,\\
\!\!\!\sous{2}{\partial}_{~\!b}(_sG^{bu}_{ijk}) &=& -{1\over 56}
y_{12}^{<ijk>su} r_{12}^{-4} + {3\over 140} \delta^{us}
y_{12}^{<ijk>}r_{12}^{-2} - {1\over 28} \delta^{s<i} y_{12}^{jk>u}
r_{12}^{-2} - {31\over 280} \delta^{u<i} y_{12}^{jk>s} r_{12}^{-2}
\nonumber \\ &+& {1\over 20} y_1^{<i} y_{12}^{jk>su}r_{12}^{-4} -
{3\over 20} \delta^{su}y_1^{<i} y_{12}^{jk>}r_{12}^{-2} + {1\over 5}
\delta^{s<i} y_1^j y_{12}^{k>u}r_{12}^{-2} + {7\over 10} \delta^{u<i}
y_1^j y_{12}^{k>s}r_{12}^{-2} \nonumber \\ &+& {3\over 4} \delta^{su}
y_1^{<ij} y_{12}^{k>} r_{12}^{-2} - {3\over 4} \delta^{s<i} y_1^{jk>}
y_{12}^u r_{12}^{-2} + {3\over 4} y_1^{<ijk>} y_{12}^{su} r_{12}^{-4}
- {1\over 4} y_1^{<ijk>} \delta^{su}r_{12}^{-2} \nonumber\\ &+&
\delta^{s<i} y_{12}^j \delta^{k>u} \left[{1\over 35} \ln {\tilde
r}_{12} + {1361\over 14700}\right] + \delta^{s<i} y_1^j \delta^{k>u}
\left[{1\over 5} \ln {\tilde r}_{12} - {2159\over 2100}\right] \;,\\
\!\!\!\sous{2}{\partial}_{~\!b}(_sG_{aij}^{ab}) &=& -{11\over 75}
y_{12}^{<ij>s} r_{12}^{-2} + {1\over 300} (y_1y_{12}) y_{12}^{<ij>s} r_{12}^{-4}
- {23\over 300} y_1^s y_{12}^{<ij>} r_{12}^{-2} + {71\over 75} y_1^{<i}
y^{j>s}_{12} r^{-2}_{12} \nonumber\\ &+& {7\over 10} y_1^{s<i}
y_{12}^{j>} r^{-2}_{12} - {1\over 4} y_1^{<ij>s} r^{-2}_{12} + {3\over
4} (y_1y_{12}) y_1^{<ij>}y_{12}^s r^{-4}_{12} - {3\over 10} y_1^2
y_1^{<i} y_{12}^{j>s} r^{-4}_{12} \nonumber \\ &+& {2\over
25}(y_1y_{12}) \delta^{s<i}y_{12}^{j>} r^{-2}_{12} - {7\over
10}(y_1y_{12}) \delta^{s<i}y_1^{j>} r^{-2}_{12} + {1\over 10}
y_1^2\delta^{s<i} y_1^{j>} r^{-2}_{12} \nonumber\\ &+& \delta^{s<i}
y_{12}^{j>} \left[{1\over 25}\ln {\tilde r}_{12}+{51\over 500}\right]
+ \delta^{s<i} y_1^{j>} \left[{7\over 25} \ln {\tilde
r}_{12}-{2029\over 1500}\right] \;,\\ U^{ab}_{ij} &=& y_{12}^{ab}
\left(-{7\over 30}y_1^{<ij>}-{1\over 30}y_1^{<i}y_2^{j>} + {1\over 60}
y_2^{<ij>}\right) r^{-4}_{12} \nonumber \\ &+& \delta^{ab}
\left(-{3\over 10}y_1^{<ij>} + {1\over 10}y_1^{<i}y_2^{j>} - {1\over
20} y_2^{<ij>}\right) r^{-2}_{12} \nonumber \\ &+& {1\over 15}
\delta^{a<i} y_{12}^{j>b} r^{-2}_{12} + {1\over 15} \delta^{b<i}
y_{12}^{j>a} r^{-2}_{12} +\delta^{a<i} \delta^{j>b} \left[{2\over 5}
\ln {\tilde r}_{12} -{97\over 150}\right] \;,\\
\!\!\!\sous{2}{\partial}_{~\!s}(K^{as}_{ij}) &=& \left(-{9\over
10}y_1^{<ij>} - {1\over 30} y_1^{<i}y_2^{j>} - {1\over
15}y_2^{<ij>}\right )y^a_{12} r^{-2}_{12} \nonumber \\ &+&
\delta^{a<i}y_1^{j>} \left[-{17\over 15}\ln {\tilde r}_{12}- {851\over
900}\right] + \delta^{a<i} y_2^{j>} \left[-{1\over 5} \ln {\tilde
r}_{12}- {13\over 300}\right] \;,\\
\!\!\!\sous{2}{\partial}_{~\!b}(Q^b_{ij}) &=& y_1^{<ij>} \left\{
-{8\over 5} \ln {\tilde r}_{12} - {11731\over 4200} + r^{-2}_{12}
\left[{243\over 70}y_1^2 - {22\over 7} (y_1y_2) + {88\over
105}y_2^2\right]\right\} \nonumber\\ &+& y_1^{<i}y_2^{i>}
\left\{-{8\over 15} \ln {\tilde r}_{12}- {5603\over 6300} +
r^{-2}_{12} \left[-{134\over 105}y_1^2-{8\over 21}(y_1y_2) + {34\over
105} y_2^2\right]\right\} \nonumber \\ &+& y_2^{<ij>} \left\{ -{1\over
5} \ln {\tilde r}_{12}- {1777\over 4200} + r^{-2}_{12} \left[{29\over
210}y_1^2 - {1\over 7} (y_1y_2) + {6\over 35} y_2^2\right]\right\} \;,\\
\!\!\!\sous{2}{\partial}_{~\!s}(H^s_{ij}) &=& y_1^{<ij>} \left[{8\over
15} \ln {\tilde r}_{12}- {227\over 1800}\right] \nonumber \\ &+&
y_1^{<i}y_2^{j>} \left[{4\over 15} \ln {\tilde r}_{12}- {33\over
900}\right] + y_2^{<ij>} \left[{1\over 5} \ln {\tilde r}_{12}-
{97\over 1800}\right] \;,\\ \!\!\!\sous{2}{\partial}_{~\!b}(H^{bs}_{ij})
&=& {1\over 30} y_{12}^{s<ij>} r^{-2}_{12} - {1\over 6}
y_{12}^{s<i}y_1^{j>}r^{-2}_{12} - {1\over 2}y^s_{12} y_1^{<ij>}
r^{-2}_{12} \nonumber \\ &+& y_{12}^{<i} \delta^{j>s} \left[{1\over
15}\ln {\tilde r}_{12}- {107\over 900}\right] + y_1^{<i} \delta^{j>s}
\left[-{2\over 3} \ln {\tilde r}_{12}+ {19\over 45}\right] \;,\\
\!\!\!\sous{2}{\partial}_{~\!s}(H_{ij}^{abs}) &=& -{3\over 40}
y_{12}^{ab<ij>} r^{-4}_{12} + {7\over 120} \delta^{ab} y_{12}^{<ij>}
r^{-2}_{12} + {7\over 60} (\delta^{a<i} y_{12}^{j>b}+\delta^{b<i}
y_{12}^{j>a}) r^{-2}_{12} \nonumber\\ &+& {1\over 3} y_{12}^{ab<i}
y_1^{j>} r^{-4}_{12} + {1\over 3} \delta^{ab} y_{12}^{<i}y_1^{j>}
r^{-2}_{12} + {1\over 3} \left(\delta^{a<i} y_1^{j>}y_{12}^b
+\delta^{b<i} y_1^{j>}y_{12}^a\right) r^{-2}_{12} \nonumber\\ &+&
\delta^{a<i} \delta^{j>b} \left[{8\over 15} \ln {\tilde r}_{12}-
{241\over 900}\right] \;,\\ \!\!\!\sous{2}{\partial}_{~\!s}(F^s_{ij}) &=&
y_1^{<ij>} \left[\ln {\tilde r}_{12}+{2\over 3}\ln 2+ {29\over
360}\right]- {1\over 12} y_1^{<i}y_2^{j>} + {1\over 24} y_2^{<ij>} \;,\\
\!\!\!\sous{2}{\partial}_{~\!b}(F^{bs}_{ij}) &=& -{1\over 30}
y_{12}^{s<ij>} r^{-2}_{12} - {1\over 2} y_{12}^{s<i} y_1^{j>}
r^{-2}_{12} - {1\over 2} y_{12}^sy_1^{<ij>} r^{-2}_{12} \nonumber \\
&+& y^{<i}_{12} \delta^{j>s} \left[-{1\over 15} \ln {\tilde r}_{12}-
{43\over 900}\right] + y_1^{<i} \delta^{j>s} \left[-2\ln {\tilde
r}_{12}-{4\over 3}\ln 2+ {28\over 90}\right] \;,\\
\!\!\!\sous{2}{\partial}_{~\!s}(F^{abs}_{ij}) &=& - {1\over 120}
y_{12}^{ab<ij>} r_{12}^{-4} + {23\over 120} \delta^{ab} y^{<ij>}_{12}
r^{-2}_{12} + {23\over 60} \left(\delta^{a<i} y_{12}^{j>b} +
\delta^{b<i} y_{12}^{j>a}\right) r^{-1}_{12} \nonumber \\ &+& {2\over
3} y_{12}^{ab<i} y_1^{j>} r^{-4}_{12} + {2\over 3} \delta^{ab}
y_{12}^{<i} y_1^{j>} r^{-2}_{12} + {2\over 3}
\left(y^b_{12}\delta^{a<i} + y^a_{12} \delta^{b<i} \right) y_1^{j>}
r^{-2}_{12} \nonumber \\ &+& \delta^{a<i} \delta^{j>b} \left[{32\over
15} \ln {\tilde r}_{12}+{4\over 3}\ln 2 - {409\over 900}\right] \;,\\
\!\!\!\sous{2}{\partial}_{~\!s}(\stackrel{12}{F^s_{ij}}) &=&
y_1^{<ij>} \left[-{7\over 15} \ln {\tilde r}_{12}-{2\over 3}\ln 2-
{79\over 100}\right] + y_1^{<i}y_2^{j>} \left[{4\over 15} \ln {\tilde
r}_{12}+ {148\over 225}\right] \nonumber \\ &+& y_2^{<ij>}
\left[{1\over 5} \ln {\tilde r}_{12}+ {13\over 300}\right] \;,\\
\!\!\!\sous{2}{\partial}_{~\!b}(\stackrel{12}{F^{bs}_{ij}}) &=&
-{2\over 15} y_{12}^{s<ij>} r^{-2}_{12} + y_{12}^{s<i} y_1^{j>}
r^{-2}_{12} - y^s_{12}y_1^{<ij>} r^{-2}_{12} +y_1^{<i}
\delta^{j>s}\left[{4\over 3} \ln 2+ {13\over 45}\right] \nonumber \\
&+& y^{<i}_{12} \delta^{j>s} \left[{2\over 5} \ln {\tilde r}_{12}+
{44\over 75}\right] \;,\\
\!\!\!\sous{2}{\partial}_{~\!s}(\stackrel{12}{F^{abs}_{ij}}) &=&
-{7\over 20} y_{12}^{ab<ij>} r^{-4}_{12} + {1\over 20}
\delta^{ab}y_{12}^{<ij>} r^{-2}_{12} - {17\over 30} (\delta^{a<i}
y_{12}^{j>b} + \delta^{b<i} y_{12}^{j>a}) \nonumber \\ &+& {5\over
3}(y_{12}^b \delta^{a<i} + y_{12}^a \delta^{b<i} ) y_1^{j>}
r^{-2}_{12} + {5\over 3}y_{12}^{ab<i}y_1^{j>} r^{-4}_{12} \nonumber \\
&-& {1\over 3} \delta^{ab}y_{12}^{<i} y_1^{j>} r^{-2}_{12} - {5\over
2} y_{12}^{ab}y_{1}^{<ij>} r^{-4}_{12} + {3\over 2}
\delta^{ab}y_1^{<ij>} r^{-2}_{12} \nonumber\\ &+& \delta^{a<i}
\delta^{j>b} \left[{4\over 15} \ln {\tilde r}_{12}-{4\over 3}\ln
2-{229\over 450}\right] \;,\\
\!\!\!\sous{2}{\partial}_{~\!s}(\stackrel{21}{F^s_{ij}}) &=&
y_1^{<ij>} \left[{11\over 15} \ln {\tilde r}_{12}-{2\over 3}\ln 2-
{787\over 900}\right] + y_1^{<i}y_2^{j>} \left[-{4\over 5}\ln {\tilde
r}_{12}+ {196\over 225}\right] \nonumber \\ &+& y_2^{<ij>}
\left[{1\over 15} \ln {\tilde r}_{12}- {77\over 900}\right]
\;.\end{eqnarray}\end{subequations}$\!\!$
Inserting these elementary integrals into the expressions of
non-compact terms [see (\ref{88})], and reducing to the case of
circular orbits, we obtain the results reported in Appendix A.

\section{Point-mass regularization ambiguities}

The computation of the multipole moments we performed so far has been
carried out with standard techniques: standard Hadamard regularization
[see Section V], and Schwartz distributions [see
e.g. Eqs. (\ref{87})]. The result we obtained depends on three
arbitrary constants: the two Hadamard regularization constants $u_1$
and $u_2$ introduced in Eq. (\ref{27}), and the constant $r_0$
entering the definition of the source multipole moments through the
analytic-continuation factor $|\tilde{\bf x}|^B = |{\bf x}/r_0|^B$
[see Eqs. (\ref{5})]. The constant $r_0$ is not a problem since we
know that in this formalism the multipole expansion of the field
exterior to any source is actually independent of $r_0$
\cite{B98mult}. Indeed we shall check in Section XII that $r_0$
disappears from the final expression of the energy flux (the constant
$r_0$ in the source moments is cancelled by the same constant present
in the contribution of ``tails of tails'' in the wave zone; see
Eq. (\ref{118'}) below). However, it will turn out that the constants
$u_1$ and $u_2$, which encode some arbitrariness of the Hadamard
regularization, lead {\it a priori} to two undetermined purely
numerical parameters in the expression of the 3PN quadrupole
moment. In addition, we shall argue that because of some delicate
problems linked with the use of the Hadamard regularization at the 3PN
order, we should consider {\it a priori} a third undetermined
parameter in the quadrupole moment. However, the important point is
that these three parameters combine to yield {\it one and only one}
undetermined constant, that we shall call $\theta$, in the third
time-derivative of the moment which is needed to compute the physical
energy flux for circular orbits. Furthermore, we shall find that the
constant $\theta$ enters the energy flux at the same level as the
constant $\lambda$ coming from the equations of motion (see below), so
that the energy flux depends {\it in fine} merely on one combination
of $\theta$ and $\lambda$.

The equations of motion of compact objects at the 3PN order have been
investigated using the ADM-Hamiltonian formulation of general
relativity \cite{JaraS98,JaraS99}, and by integrating the field
equations in harmonic coordinates \cite{BF00,BFeom}. In both
approaches the compact objects are modelled by point-like particles
described by delta-functions, and the self-field of the particles is
removed by a Hadamard regularization. It was shown that the
regularization permits the determination of the full equations of
motion at the 3PN order except for one undetermined coefficient,
$\lambda$ in the harmonic-coordinate approach and $\omega_{\rm
static}$ in the ADM-Hamiltonian. Very likely the unknown coefficient
accounts for a physical incompleteness of the point-mass
regularization. Actually two unknown coefficients were originally
introduced in \cite{JaraS98,JaraS99}, but one of them was shown later
\cite{JaraS00,DJS00} to be fixed to a unique value by requiring, in an
{\it ad hoc} manner, the global Poincar\'e invariance of the
Hamiltonian. On the other hand, in the harmonic-coordinate approach
\cite{BF00,BFeom} a new Hadamard-type regularization was developed in
order to account for the mathematical ambiguities of the standard
Hadamard regularization \cite{BFreg,BFregM}. A characteristic of this
regularization is the systematic use of a theory of generalized
functions. The regularization is defined in a Lorentz-invariant way,
but was ultimately shown to yield incomplete results for the equations
of motion, in the sense that there remained the unknown numerical
coefficient $\lambda$. The complete physical equivalence between the
harmonic-coordinate \cite{BF00,BFeom} and ADM-Hamiltonian
\cite{JaraS98,JaraS99,JaraS00,DJS00} formalisms has been established
\cite{DJS01,ABF}. Indeed a unique ``contact'' transformation of the
particles motion which changes the harmonic-coordinate Lagrangian (as
given in Ref. \cite{ABF}) into the ADM-Hamiltonian obtained in
Ref. \cite{DJS00} exists. The equivalence holds if and only if the
harmonic-coordinate constant $\lambda$ is related to the
ADM-Hamiltonian static ambiguity by

\begin{equation}\label{200}
\lambda = -{3\over 11} \omega_{\rm static}-{1987\over 3080}
\;.\end{equation} Recently, the value $\omega_{\rm static}=0$ has been
obtained by means of a different regularization (dimensional) within
the ADM-Hamiltonian approach \cite{DJS01b}. This result would mean
that $\lambda = -1987/3080$.  Note that a feature of the
harmonic-coordinate equations of motion derived in
\cite{BF00,BFeom,ABF} is the dependence, in addition to $\lambda$, on
two arbitrary constants ${r'}_1$ and ${r'}_2$ parametrizing some
logarithmic terms. However, contrary to $\lambda$ which is a true
physical ambiguity, the constants ${r'}_1$ and ${r'}_2$ can be removed
by a coordinate transformation and therefore represent merely some
unphysical gauge constants. For instance these constants cancel out in
the center-of-mass invariant energy of circular binaries \cite{BF00}.

\subsection{Hadamard-regularization constants}

The first problem in the present calculation lies in the {\it a
priori} unknown relation between the Hadamard regularization constants
$u_1$ and $u_2$ introduced by Eqs. (\ref{27}) and the two gauge
constants ${r'}_1$ and ${r'}_2$ which parametrize the
harmonic-coordinate equations of motion. Let us investigate more
precisely the dependence of the quadrupole moment on the constants
$u_1$ and $u_2$. Inspection of our computation shows that these
constants come only from the cubic and non-compact terms obtained in
Sections VIII and IX. More precisely, we find that the whole
computation depends on $u_1$, $u_2$ only through the elementary
integrals $Y_L^{(-3,0)}$ and $S_L^{(-5,0)}$, which parametrize the
``self'' parts, proportional to $m_1^3$ or $m_2^3$, of the cubic and
non-compact terms (recall also that $Y_L^{(-5,0)}$ is zero). See for
instance the expressions (\ref{88}) of NC terms. The relevant
$Y_L^{(-3,0)}$ and $S_L^{(-5,0)}$ were obtained in (\ref{78}) and
(\ref{92}). The dependence on $u_1$ and $u_2$ therein is

\begin{subequations}\label{201}\begin{eqnarray}
\stackrel{(-3,0)}{Y_{ij}} &=& 2 \ln\left({u_1\over r_0}\right)
y_1^{<ij>}+\cdots\;,\\ \stackrel{(-5,0)}{S_{ij}} &=& {14\over 3}
\ln\left({u_1\over r_0}\right) y_1^{<ij>}+\cdots
\;.\end{eqnarray}\end{subequations}$\!\!$
The dots indicate the terms independent of $u_1$ and $u_2$. We take
all the cubic and NC terms given by (\ref{65}) and (\ref{88}) [only
the mass quadrupole is to be considered], plug into them the results
(\ref{201}) and find after summation the following part of the
quadrupole moments depending on these constants (for general orbits):

\begin{equation}\label{202}
I_{ij}[u_1,u_2] = \left( -{44\over 3}\frac{G^2
m_1^3}{c^6}\ln\left({u_1\over r_0}\right)~a_1^{<i}y_1^{j>}+ 1
\leftrightarrow 2 \right) +\cdots \;.\end{equation} By
$I_{ij}[u_1,u_2]$ we mean the quadrupole obtained from summing all the
terms computed in the previous sections, i.e. depending on the
Hadamard-regularization constants $u_1$, $u_2$ (as well as, of course,
the constant $r_0$). On the other hand, we found that many of the
``interaction'' terms, proportional to $m_1^2m_2$ or $m_1 m_2^2$,
depend on time-dependent logarithms of the ratio
$\tilde{r}_{12}=r_{12}/r_0$, where $r_0$ is the constant dealing with
the behaviour of the moments at infinity. See for instance the
elementary integrals (\ref{93}). The effect of the result (\ref{202})
is to ``replace'' a part of the latter logarithms of $\tilde{r}_{12}$
by some corresponding logarithms of the ratio $r_{12}/u_1$ (and ditto
with $u_2$). The remaining logarithms stay as they are as logarithms
of the ratio $\tilde{r}_{12}$. Thus we can re-write the dependence of
the quadrupole on $u_1$ and $u_2$ through the logarithms of
$r_{12}/u_1$ and $r_{12}/u_2$ in the form

\begin{equation}\label{203}
I_{ij}[u_1,u_2] = \left( {44\over 3}\frac{G^2
m_1^3}{c^6}\ln\left({r_{12}\over u_1}\right)~a_1^{<i}y_1^{j>}+ 1
\leftrightarrow 2 \right) +\cdots \;.\end{equation} All the other
logarithms, present in the dots of Eq. (\ref{203}), are of the type
$\ln\left({r_{12}\over r_0}\right)$. In this paper we assumed nothing
about the values of $u_1$ and $u_2$. In particular we did not assume
any relation between $u_1$, $u_2$ and the gauge constants ${r'}_1$,
${r'}_2$ that parametrize the final equations of motion in harmonic
coordinates \cite{BF00,BFeom}. However, when computing the energy flux
we shall need to obtain the third time-derivative of the quadrupole
moment, and for that purpose we shall replace the accelerations by
their expressions obtained from the 3PN equations of motion, depending
on ${r'}_1$, ${r'}_2$. As a result the third time-derivative of the
moment will depend on $u_1$, $u_2$ as well as on ${r'}_1$,
${r'}_2$. Therefore, we definitely need to control the relation
between $u_1$, $u_2$ and ${r'}_1$, ${r'}_2$; then we shall have the
quadrupole moment expressed solely in terms of ${r'}_1$ and ${r'}_2$
and we shall check that the latter constants can be removed by the
same coordinate transformation as in the equations of motion, and thus
that the final expression of the physical energy flux must be
independent of these constants. From Eq. (\ref{203}) we can write

\begin{equation}\label{204}
I_{ij}[u_1,u_2] = I_{ij}[{r'}_1,{r'}_2] + {44\over 3}\frac{G^2
m_1^3}{c^6}\ln\left({{r'}_1\over u_1}\right)~a_1^{<i}y_1^{j>}+ 1
\leftrightarrow 2 \;.\end{equation} The notation for
$I_{ij}[{r'}_1,{r'}_2]$ is clear: we mean the sum of all the
contributions obtained in the previous sections, but computed with
${r'}_1$, ${r'}_2$ in place of the regularization constants $u_1$,
$u_2$.

We shall now look for the most general $\ln\left({{r'}_1\over
u_1}\right)$ that is allowed by physical requirements. In this
connection recall the spirit of the regularization: the constants
$u_1$ and $u_2$ reflect some incompleteness of the process, that may
or may not be fixed in a given computation, and therefore they should
be kept completely arbitrary unless there are some physical arguments
to restrict their form. In particular, when used in different
computations, these regularization constants have no reason {\it a
priori} to be the same. For instance, in the present computation of
the moments, the constants $u_1$ and $u_2$ are {\it a priori}
different from the constants $s_1$ and $s_2$ which were originally
used in the 3PN equations of motion (see Eq. (2.3) in
\cite{BFeom}). They are {\it a fortiori} different from the constants
${r'}_1$ and ${r'}_2$ chosen to parametrize the final equations of
motion (Eq. (7.16) in \cite{BFeom}). See also the discussion in
Section VII in Ref. \cite{BFeom}, where we determined the general form
of the relation between $s_1$, $s_2$ and ${r'}_1$, ${r'}_2$ by
imposing the polynomial mass dependence of the equations of motion,
the correct perturbative limit, and the existence of a conserved
energy. Here we shall basically do the same in order to restrict the
form of the relation between $u_1$, $u_2$ and ${r'}_1$, ${r'}_2$. Note
that {\it a priori} the logarithms $\ln\left({{r'}_1/u_1}\right)$ and
$\ln\left({{r'}_2/u_2}\right)$ can depend on the masses $m_1$ and
$m_2$. To determine just what combination of masses is allowed we make
(similarly to the equations of motion) two physical requirements: (i)
that the quadrupole moment be a polynomial function of the two masses
$m_1$, $m_2$ when taken separately; (ii) that the perturbative limit
(corresponding to $\nu\to 0$) not be affected by this possible
dependence over the masses. Because of the factor $m_1^3$ in front of
the log-term in (\ref{204}), and because the acceleration $a_1^i$
brings another factor $m_2$, the most general solution for this
logarithm in order to satisfy the requirement (i) is to be composed
of: a pure numerical constant (say $\xi$), plus a pure constant (say
$\kappa$) times the mass ratio $m/m_1$, plus a constant times $m/m_2$,
next five terms involving the mass ratios $m^2/m_1^2$, $m^2/m_1/m_2$,
$m^3/m_1^3$, $m^3/m_1^2/m_2$ and $m^4/m_1^3/m_2$. Each of these terms
must be such that it does not violate the perturbative limit [our
requirement (ii)]. This means that they should involve, in a
center-of-mass frame, a factor $\nu^2$ at least. We readily find that
the only two admissible terms in this respect are the first two in the
previous list (with constants $\xi$ and $\kappa$). So we end up with
the most general admissible solution

\begin{equation}\label{205}
\ln\left({{r'}_1\over u_1}\right)=\xi + \kappa {m_1+m_2\over
m_1}\qquad\hbox{(and {\it idem} with $1\leftrightarrow 2$)}
\;,\end{equation}
where $\xi$ and $\kappa$ denote some arbitrary purely numerical
constants (for instance rational fractions). This result is similar to
the one obtained in the 3PN equations of motion, concerning the
relation between $s_1$, $s_2$ and ${r'}_1$, ${r'}_2$. See Eqs. (7.9)
in Ref. \cite{BFeom}, where the determination of the constant
analogous to $\xi$ was possible from the requirement of existence
of a conserved energy (and Lagrangian) for the equations of motion.

We now check that the logarithms of $r_{12}/{r'}_1$ and
$r_{12}/{r'}_2$ in the quadrupole moment, which are of the form

\begin{equation}\label{206}
I_{ij}[{r'}_1,{r'}_2] = \left( {44\over 3}\frac{G^2
m_1^3}{c^6}\ln\left({r_{12}\over {r'}_1}\right)~a_1^{<i}y_1^{j>}+ 1
\leftrightarrow 2 \right) +\cdots
\;,\end{equation}
can be eliminated by the {\it same} coordinate transformation as found
in Ref. \cite{BFeom} for the logarithms in the harmonic-coordinate
equations of motion. [As concerns the logarithms of $r_{12}/r_0$ in
the moment they cannot be eliminated by a change of coordinates but
will match precisely with corresponding logarithms present in the
``tails of tails'' at infinity.] We look for a coordinate change of
the type considered in Section VI.A of \cite{BFeom}: namely $\delta
x^\mu=\xi^\mu$, where $\xi_\mu=\eta_{\mu\nu}\xi^\nu$ is a 3PN gauge
vector given by

\begin{equation}\label{207}
\xi_\mu=\frac{G^3m^3}{c^6}\partial_\mu
\left(\frac{\epsilon_1}{r_1}+\frac{\epsilon_2}{r_2}\right)
\;.\end{equation}
We have factorized out $m^3$ (where $m=m_1+m_2$) so that $\epsilon_1$
and $\epsilon_2$, which are constants or mere functions of time $t$,
be dimensionless. The corresponding change of the particle's
trajectories is given to this order by the regularized value of the
gauge vector at the location of the particle (see Section VI.A in
\cite{BFeom}). We obtain

\begin{subequations}\label{208}\begin{eqnarray}
\delta_\xi y_1^i &=& -\epsilon_2 \frac{G^3 m^3}{c^6 r_{12}^3}y_{12}^i
\;,\\ \delta_\xi y_2^i &=& \epsilon_1 \frac{G^3 m^3}{c^6
r_{12}^3}y_{12}^i
\;.\end{eqnarray}\end{subequations}$\!\!$
Since the quadrupole moment starts at the Newtonian level with the
usual $m_1 y_1^{<ij>}+1\leftrightarrow 2$, we easily find its
coordinate change as

\begin{eqnarray}\label{209}
\delta_\xi I_{ij} &=& 2~\!m_1 y_1^{<i}\delta_\xi y_1^{j>} + 1
\leftrightarrow 2 \nonumber\\ &=& -2~\!m_1 \epsilon_2 \frac{G^3
m^3}{c^6 r_{12}^3}y_1^{<i}y_{12}^{j>} + 1 \leftrightarrow 2
\;.\end{eqnarray}
By comparing this with Eq. (\ref{206}) (using the Newtonian particles
acceleration), we find that the gauge transformation required to
eliminate the logarithms is

\begin{subequations}\label{210}\begin{eqnarray}
\epsilon_2 &=&
-\frac{22}{3}\frac{m_1^2m_2}{m^3}\ln\left(\frac{r_{12}}{r'_1}\right)\;,\\
\epsilon_1 &=&
-\frac{22}{3}\frac{m_1m_2^2}{m^3}\ln\left(\frac{r_{12}}{r'_2}\right)
\;,\end{eqnarray}\end{subequations}$\!\!$
in complete agreement with Eq (7.2) in Ref. \cite{BFeom}. In summary, not
only these logarithms disappear when considering physical quantities
associated with the equations of motion (such as the invariant
energy), but they will also cancel from physical quantities associated
with the wave field at infinity, {\it viz} the invariant energy flux
we compute in Section XII.

\subsection{Special features of the regularization}

We now discuss some subtleties of the Hadamard regularization which
motivate the introduction in the quadrupole moment, in addition to
$\xi$ and $\kappa$ considered in Eq. (\ref{205}), of still another
constant (however, see below for the definition of a single constant
$\theta$).

\medskip\noindent {\it Non-distributivity of the Hadamard partie
finie}. By ``non-distributivity'' we mean the fact that the
regularization of a product of two functions $F$ and $G$, singular in
the sense of (\ref{25}), does not equal, in general, the product of
the regularized functions: $(FG)_1 \not= (F)_1(G)_1$. For instance,
with $U=G m_1/r_1+G m_2/r_2$ the Newtonian potential, we have
$(U^n)_1=[(U)_1]^n$ for $n=1,2,3$, but
$(U^4)_1=[(U)_1]^4+2[(U)_1]^2[(U)_2]^2$. An immediate consequence is
that the product of a singular function $F$ with a delta-function does
not equal, in general, the product of its regularized value with the
delta-function: $F\delta_1 \not= (F)_1 \delta_1$. Here we are assuming
that the three-dimensional integral of the product of $F$ with
$\delta_1\equiv \delta ({\bf x}-{\bf y}_1)$ gives back the regularized
value $(F)_1$. Notice that only at the 3PN order does the
non-distributivity play a role. Up to the 2PN order, the
distributivity holds for all the functions encountered in the problem
(hence the computation of the moments as was done in \cite{BDI95} is
correct).

The non-distributivity at 3PN has an important  bearing  on the choice
of the stress-energy tensor for describing point-particles. In this
paper, we adopted the most naive choice for the stress-energy
tensor. See Eq. (\ref{29}) above, which is equivalent, at 3PN order, to

\begin{equation}\label{211}
T^{\mu\nu} = {m_1 v_1^\mu v_1^\nu \over \sqrt{(gg_{\rho\sigma})_1
v_1^\rho v_1^\sigma/c^2}}\delta ({\bf x}-{\bf y}_1) + 1\leftrightarrow
2
\;.\end{equation}
Namely, we assumed that the whole factor of the delta-function
consists of a regularized value at point 1. But because $F\delta_1
\not= (F)_1 \delta_1$, we could obtain a different result by choosing
another stress-energy tensor, defined by replacing the factor of the
delta-function in (\ref{211}), or part of it, by a function depending
on any field point ${\bf x}$ and such that its regularized value when
${\bf x}\to {\bf y}_1$ is the same. In fact, a specific form of the
stress-energy tensor of point-particles, compatible with the Hadamard
regularization, was advocated in Ref. \cite{BFregM} and used to compute the
3PN equations of motion \cite{BFeom}. This form, given by Eq. (5.11)
in Ref. \cite{BFregM}, reads

\begin{equation}\label{212}
{\hat T}^{\mu\nu} = {m_1 v_1^\mu v_1^\nu \over
\sqrt{-(g_{\rho\sigma})_1 v_1^\rho v_1^\sigma/c^2}}{\delta ({\bf
x}-{\bf y}_1)\over \sqrt{-g({\bf x},t)}} + 1\leftrightarrow 2
\;.\end{equation}
Choosing one or the other form of stress-energy tensor does make a
difference in our computation. Consider for instance the term ${\rm
SI(1)}=\int d^3{\bf x}~\!{\hat x}^{ij}\sigma$. We find that the result
for this term, when computed using the tensor (\ref{212}) i.e. using
${\hat \sigma}={\hat T}^{00}+{\hat T}^{ii}$, differs from the original
result by the amount

\begin{equation}\label{213}
\Delta{\rm SI(1)} = \frac{G^2 m_1^3}{c^6}\left[\frac{2}{3}
a_1^{<i}y_1^{j>}-\frac{1}{5} v_1^{<i}v_1^{j>}\right]+ 1
\leftrightarrow 2
\;.\end{equation}
There is also a modification $\Delta{\rm SII(1)}$ but which is of the
same structure (with different numerical coefficients).

On the other hand, some terms in our computation would be different if
the regularization would be distributive. For instance, if for
computing the term SI(16NC) we take into account the
non-distributivity (as we did), we find the result (\ref{42''}), namely

\begin{equation}\label{300}
{\rm SI(16NC)} = {4m_1\over c^6}v_1^{ab}\left(x^{<i}x^{j>}U_{ab}^{\rm
(NC)}\right)_1+1\leftrightarrow 2
\;.\end{equation}
If instead we incorrectly assume that the partie finie is
distributive, then we get

\begin{equation}\label{301}
{\rm SI}{\rm (16NC)}_{\rm distr} = {4m_1\over
c^6}y_1^{<i}y_1^{j>}v_1^{ab}\left(U_{ab}^{\rm
(NC)}\right)_1+1\leftrightarrow 2
\;.\end{equation}
The difference between the two results is not zero:

\begin{equation}\label{302}
\Delta{\rm SI(16NC)} = -\frac{2}{15}\frac{G^2
m_1^3}{c^6}v_1^{<i}v_1^{j>}+ 1 \leftrightarrow 2 \;.\end{equation} The
same happens with the other terms VI(10NC) and VI(12NC); each time the
structure of the difference is the same as in (\ref{213}) or
(\ref{302}).

\medskip\noindent {\it Violation of the Leibniz rule by the
distributional derivative}. In Ref. \cite{BFreg} a new kind of
distributional derivative of singular functions of the type $F$ was
introduced. It was found that it is impossible to define a derivative
satisfying the Leibniz rule for the derivation of the product, but
that a mathematical structure exists when we replace the Leibniz rule
by the weaker rule of ``integration by parts''. The latter rule can be
seen as an integrated version of the Leibniz rule (see Section VII.A
in \cite{BFreg}). More precisely, two different distributional
derivatives were proposed in \cite{BFreg}: a ``particular''
derivative, and a ``correct'' one. Both derivatives reduce to the
derivative of the standard distribution theory \cite{Schwartz} when
applied to smooth test functions with compact support. The particular
derivative is simpler to use in practical computations, but the
correct one is more satisfying because successive derivatives to any
order commute.

Previously we performed numerous simplifications, with the help of
the Leibniz rule, to arrive at the form of multipole moments given by
Eqs. (\ref{18}). Thus we made some errors because of the violation of the
Leibniz rule by the distributional derivative. The strategy adopted in
Ref. \cite{BFeom} was to keep track of all these error terms and to compute
them using the particular and correct derivatives of \cite{BFreg}. In
the present paper we shall proceed differently. We simply give an
example. When simplifying the moment to arrive at the simple-looking
term SI(39) in Eq. (\ref{18a}), we ``forgot'' to include the error term

\begin{equation}\label{214}
\Delta {\rm SI(39)}={2\over 3\pi G c^6} \sous{\!\!\!\!B= 0}{\rm FP}
\int d^3{\bf x}~\!|\tilde{\bf x}|^B \hat{x}_{ij}
\left[\Delta(U^4)-4U^3\Delta U-12U^2\partial_aU\partial_aU\right]
\;.\end{equation}
Clearly this term would be zero for any derivative satisfying the
Leibniz rule (in a distributional sense). However, computing it by
means of for instance the ``particular'' derivative (defined by
Eq. (7.7) in Ref. \cite{BFreg}), we find that it is not zero, but

\begin{equation}\label{215}
\Delta {\rm SI(39)} = \frac{64}{3}\frac{G^2 m_1^3}{c^6}
a_1^{<i}y_1^{j>} + 1 \leftrightarrow 2
\;.\end{equation}
Again this result has the same type of structure as found
previously. We have checked that all the terms coming from the
violation of the Leibniz rule have the same structure, either of type
$m_1^3 a_1^{<i}y_1^{j>}$ like in (\ref{215}) or of type $m_1^3
v_1^{<i}v_1^{j>}$.

\medskip\noindent {\it Cubically non-linear self-interaction
terms}. We take the example of the self contribution in the term
SI(35NC). This term is

\begin{equation}\label{216}
{\rm SI(35NC)} = -{4\over \pi Gc^6} \sous{\!\!\!\!B= 0}{\rm FP} \int
d^3{\bf x}~\!|\tilde{\bf x}|^B \hat{x}_L {\hat Z}_{ij}^{{\rm
(NC)}}\partial^2_{ij} U
\;.\end{equation}
The ``self'' part of this term corresponds to that part in ${\hat
Z}_{ij}^{{\rm (NC)}}$ which is proportional to $m_1^2$, in the sense
that

\begin{eqnarray}\label{217}
{\hat Z}_{ij}^{{\rm (NC)}}&=&G^2 m_1^2\left\{
a_1^{(i}\!\!\!\sous{1}{\partial}_{~\!j)}\ln r_1+{1\over
8}v_1^2\!\!\!\sous{1}{\partial}_{~\!ij}\ln r_1 +{1\over
32}\delta^{ij}v_1^{km}\!\!\!\sous{1}{\partial}_{~\!km}\ln r_1
\right.\nonumber\\ &+&\left.{1\over 2}{v_1^{ij}\over r_1^2}-{11\over
32}\delta^{ij}{v_1^2\over r_1^2}\right\}+{\cal O}(m_2)
\end{eqnarray}
[see Eq. (\ref{86d})], and that part of $U$ due to 1 itself, i.e. $U=G
m_1/r_1+{\cal O}(m_2)$. The resulting term, proportional to $m_1^3$,
is ill-defined in distribution theory because the delta-function,
coming from the distributional derivative of $1/r_1$ as given by
Eq. (\ref{87b}), is multiplied by the terms in (\ref{217}) which are
singular at point 1. The partie finie pseudo-functions and their
derivatives proposed in Ref. \cite{BFreg} permit us to give a
mathematical meaning to such ill-defined terms. The ``particular''
derivative of $1/r_1$ reads
$\partial_{ij}^2(1/r_1)=\partial_{ij}^2(1/r_1)_{\rm
ordinary}+D_{ij}[1/r_1]$, where the purely distributional
part is

\begin{equation}\label{218}
D_{ij}\left[{1\over r_1}\right] =
-\frac{4\pi}{3}\left(\delta^{ij}+\frac{15}{2}{\hat
n}_1^{ij}\right)\delta_1
\;.\end{equation}
[Compare this with the result (\ref{87b}) of distribution theory.] We
easily compute the effect of this new derivative on the self
part of the term (\ref{216}). Once  again we find the same type of structure
as before:

\begin{equation}\label{219}
\Delta{\rm SI(35NC)} = \frac{G^2 m_1^3}{c^6}\left[-\frac{64}{3}
a_1^{<i}y_1^{j>}+\frac{38}{15} v_1^{<i}v_1^{j>}\right]+ 1
\leftrightarrow 2
\;.\end{equation}
Similarly we checked that all other self-interaction contributions take the
same form with simply different numerical coefficients.

\subsection{Definition of the $\theta$-ambiguity}

As we have seen the structure of the possible terms associated with
the previous subtleties in the Hadamard regularization is limited to
only two types, either $m_1^3 a_1^{<i}y_1^{j>}$ or $m_1^3
v_1^{<i}v_1^{j>}$. The first type was already considered in
Eqs. (\ref{204})-(\ref{205}), where it yielded the arbitrary constant
$\xi$. Thus, modulo a redefinition of $\xi$, we do not need to
consider this term. The other type, given by $m_1^3 v_1^{<i}v_1^{j>}$,
was not considered earlier. Therefore, motivated by the previous
discussion, we shall from now on add such a term to the multipole
moment, with a new constant in front, say $\zeta$. In summary, we
consider three types of ``ambiguous'' terms (in the sense of
\cite{JaraS98,JaraS99}), parametrized by the two constants $\xi$,
$\kappa$ of Eq. (\ref{204}) and the $\zeta$. The quadrupole moment we
finally consider in this paper is thus

\begin{equation}\label{220}
I_{ij} = I_{ij}[{r'}_1, {r'}_2] + \Delta I_{ij} \;,\end{equation}
where $I_{ij}[{r'}_1, {r'}_2]$ denotes the computation we have done in
Sections VI-IX (i.e. the sum of all the terms, defined for general
orbits, and given for circular orbits in Appendix A), when expressed
by means of the {\it same} regularization constants ${r'}_1$, ${r'}_2$
as the ones appearing in the 3PN equations of motion (we know that
these constants are pure gauge). Now the undetermined part reads as
 
\begin{eqnarray}\label{221}
\Delta I_{ij} = {44\over 3}\frac{G^2
m_1^3}{c^6}\left[\left(\xi+\kappa\frac{m}{m_1}\right)~a_1^{<i}y_1^{j>}
+\zeta~v_1^{<i}v_1^{j>}\right]+ 1 \leftrightarrow 2 \;.\end{eqnarray}
In a center-of-mass frame we get

\begin{equation}\label{222}
\Delta I_{ij} = {44\over 3}\frac{G^2
m^3\nu^2}{c^6}\left[\left(\xi+2\kappa\right)~a^{<i}x^{j>}
+\zeta~v^{<i}v^{j>}\right]
\end{equation}
(where $x^i=y_1^i-y_2^i$, $v^i=dx^i/dt$ and $a^i=dv^i/dt$).  The
constants $\xi$, $\kappa$ and $\zeta$ will be left unspecified in the
present paper. It could be possible that the more sophisticated
regularization procedure of \cite{BFreg,BFregM} determines some of
these constants. However, the point for our purpose is that we are
going to show that the physical energy flux for circular orbits
depends only on one parameter. Indeed, the flux depends on the third
time-derivative of the quadrupole, and by a straightforward
computation (using the Newtonian equations of motion) we find that, in
the case of circular orbits, the third time-derivative of (\ref{222})
is

\begin{equation}\label{223}
\Delta I^{(3)}_{ij} = {352\over 3}~\!{G m^2\nu^2\over r^3} \gamma^3
~\!\theta~x^{<i}v^{j>}
\;,\end{equation}
where $\theta=\xi+2\kappa+\zeta$ is a single unknown
constant. Therefore, the ambiguous part of the physical 3PN flux, as
concerns this effect, depends in fact only on $\theta$. It is given
(for circular orbits) by

\begin{equation}\label{225}
\Delta {\cal L} = {2G\over 5c^5} I^{(3)}_{ij}\Delta I^{(3)}_{ij} =
{32c^5\over 5G}\nu^2 \gamma^5\left\{
-\frac{88}{3}\theta\nu\gamma^3\right\} \;.\end{equation} In addition
to $\theta$, the flux will depend also on the constant $\lambda$
coming from the equations of motion \cite{BF00,BFeom}. However, we
shall find that, in the case of circular orbits, both $\theta$ and
$\lambda$ enter the flux at the same level, so the flux depends only
on one combination of these constants: $\lambda-\frac{2}{3}\theta$,
from the end result (\ref{128}) below. Further work, supplementing the
Hadamard self-field regularization by suitable extensions and
alternative methods, may be required to determine the constants
$\theta$ and $\lambda$.

\section{The binary's multipole moments}

The computation of the moments is now almost complete. The remaining
terms are

\medskip\noindent (i) the ``odd'' terms: SI(11), SI(12). These terms
involve the fifth (odd) power of $1/c$ (2.5PN order). They appear
because of the expansion of retardations in the potentials (\ref{13});
they are pure functions of time, parametrized by $Q_{ij}(t)$ and
$Q(t)$ [see (\ref{16})]. The sum of the two odd terms has been
computed in the equations (4.9) and (4.12) of Ref. \cite{B96}. With
the present notation, in the quadrupole case, we have

\begin{equation}
{\rm SI(11)}+{\rm SI(12)}={G\over c^5}\left\{-{8\over
7}Q^{(3)}_{k<i}Q_{j>k}-{10\over 7}Q^{(3)}_{<ij>}Q-{2\over
21}Q_{<ij>}Q^{(3)}\right\}
\;.\end{equation}
These terms do not contribute to the flux for circular orbits.

\medskip\noindent (ii) the ``divergence'' terms: SI(2), SI(8), SI(9),
SI(10), SI(39), SI(40), SI(41), SI(42), SI(43), SI(44), SI(45),
SI(46), SI(47), SI(48), SI(49), SI(50), SII(8), SII(10), SII(12),
SII(14), SIII(3), VI(6), VI(30), VI(31), VI(32), VI(33), VI(34),
VI(35), VII(7), T1(9). The integrand of these terms is made of the
product of $|\tilde{\bf x}|^B$ and a pure divergence $\partial_iA$ or
$\Delta A$. Their computation makes use of the same techniques as
those employed in Sections VII-IX, but with the notable simplification
that because of the divergence one can perform an integration by
parts, and that as a result the elementary integrals contain
explicitly a factor $B$ (due to the differentiation of $|\tilde{\bf
x}|^B$) so their computation is quite easy. See the results in
Appendix A.

\medskip\noindent (iii) four particular terms that we have left out
because their sum is in fact zero:

\begin{equation}\label{116'}
{\rm SI(22)}+{\rm SI(23)}+{\rm SI(32)}+{\rm SII(11)}=0
\;.\end{equation}

We sum up all the terms given in Appendix A, plus the undetermined
correction given by (\ref{222}), and obtain the expressions of the 3PN
mass-type quadrupole moment, 2PN mass-type octupole moment and 2PN
current-type quadrupole moment of the compact binary moving on a
circular orbit. (Note that most of the investigation of this paper is
valid for general orbits, but we are interested in inspiralling
binaries whose orbit is quickly circularized by radiation reaction.)
The 3PN mass quadrupole reads

\begin{equation}\label{117}
I_{ij} = \mu \left(A~\!{\hat x}_{ij}+B~\!{r^2\over c^2}{\hat v}_{ij} +
{48\over 7}~\!{r\over c} x_{<i}v_{j>}\nu\gamma^2\right)+{\cal O}(7)
\;,\end{equation}
where the third term is the 2.5PN odd term, and where

\begin{subequations}\label{117'}\begin{eqnarray}
A &=& 1 + \gamma \left(-{1\over 42}-{13\over 14}\nu \right) + \gamma^2
\left(-{461\over 1512} -{18395\over 1512}\nu - {241\over 1512}
\nu^2\right) \nonumber \\ &+&\gamma^3 \left({395899\over
13200}-{428\over 105}\ln \left({r\over r_0}\right) +\left[{139675\over
33264}- {44\over 3}(\xi+2\kappa) -{44\over 3}\ln \left({r\over
{r'}_0}\right)\right]\nu \right.\nonumber\\ &+&\left. {162539\over
16632} \nu^2 + {2351\over 33264}\nu^3 \right) \;,\\ B &=& {11\over
21}-{11\over 7}\nu + \gamma \left({1607\over 378}-{1681\over 378} \nu
+{229\over 378}\nu^2\right) \nonumber \\ &+&\gamma^2
\left(-{357761\over 19800}+{428\over 105} \ln \left({r\over r_0}
\right)+\left[-{75091\over 5544}+{44\over 3}\zeta\right]\nu +
{35759\over 924} \nu^2 + {457\over 5544} \nu^3 \right)
\;.\end{eqnarray}\end{subequations}$\!\!$
The mass parameters are: $m=m_1+m_2$, $\delta m=m_1-m_2$,
$\mu=m_1m_2/m$ and $\nu=\mu/m$. The post-Newtonian parameter is
$\gamma=G m/(rc^2)={\cal O}(2)$ [see Eq. (\ref{33})].  The logarithms
depend either on the constant $r_0$ associated with the finite part at
infinity (recall $|\tilde{\bf x}|^B=|{\bf x}/r_0|^B$) or on the
``logarithmic barycenter'' ${r'}_0$ of the regularization constants
${r'}_1$ and ${r'}_2$ (see Section X), defined by
$m\ln{r'}_0=m_1\ln{r'}_1+m_2\ln{r'}_2$. We shall investigate in
Section XII the fate of these constants $r_0$ and ${r'}_0$. In
addition the moment depends on the unknown constants $\xi$, $\kappa$
and $\zeta$ introduced in Eq. (\ref{222}). The 2PN mass-octupole and 2PN
current-quadrupole are free of any of such constants and given by

\begin{subequations}\label{117''}\begin{eqnarray}
I_{ijk} &=& \mu {\delta m\over m} \hat{x}_{ijk} \left[-1 + \gamma \nu
+ \gamma^2 \left({139\over 330}+{11923\over 660}\nu +{29\over
110}\nu^2\right) \right]\nonumber\\ &+& \mu {\delta m\over m} x_{<i}
v_{jk>}{r^2\over c^2}\left[-1 + 2\nu + \gamma \left(-{1066\over
165}+{1433\over 330}\nu -{21\over 55} \nu^2\right) \right]+{\cal O}(5)
\;,\\ J_{ij} &=& \mu {\delta m\over m} \varepsilon_{ab<i} x_{j>a}v_b
\left[-1 +\gamma \left(-{67\over 28}+{2\over 7}\nu \right) + \gamma^2
\left(-{13\over 9} +{4651\over 252}\nu +{1\over 168}\nu^2 \right)
\right] \nonumber\\&+&{\cal O}(5)
\;.\end{eqnarray}\end{subequations}$\!\!$
The higher multipole moments which are needed in the 3PN energy flux
are the 1PN current octupole, 1PN mass $2^4$-pole, Newtonian current
$2^4$-pole and Newtonian mass $2^5$-pole. For these moments we simply
report the expressions already obtained in Ref. \cite{BDI95}.

\begin{subequations}\label{117'''}\begin{eqnarray}
I_{ijkl} &=& \mu ~\! {\hat x}_{ijkl}\left[1 - 3\nu + \gamma
\left({3\over 110}-{25\over 22}\nu +{69\over 22} \nu^2\right)
\right]\nonumber\\ &+& {78\over 55} \mu ~\! x_{<ij} v_{kl>}{r^2\over
c^2} (1-5\nu +5\nu^2)+{\cal O}(3) \;,\\ J_{ijk} &=& \mu
~\!\varepsilon_{ab<i} x_{jk>a}v_b \left[1-3\nu + \gamma
\left({181\over 90}-{109\over 18}\nu + {13\over 18}\nu^2\right)\right]
\nonumber\\ &+& {7\over 45} \mu ~\! (1-5\nu +5\nu^2)
\varepsilon_{ab<i} v_{jk>b}x_a {r^2\over c^2}+{\cal O}(3) \;,\\ I_{ijklm}
&=& \mu {\delta m\over m}(-1+2\nu){\hat x}_{ijklm}+{\cal O}(1) \;,\\
J_{ijkl} &=& \mu {\delta m\over m} (-1+2\nu) \varepsilon_{ab<i}
x_{jkl>a}v_b +{\cal O}(1)
\;.\end{eqnarray}\end{subequations}$\!\!$

As proved in Refs. \cite{B96,B98mult} the multipole moments $I_L$ and $J_L$
are not the only source moments entering the radiation field. However,
the other moments, denoted $W_L$, $X_L$, $Y_L$ and $Z_L$, parametrize
a (linearized) gauge transformation in the exterior field, and as a
result make a contribution to the non-linear radiation field at a
quite high post-Newtonian order: 2.5PN. It is always possible to
re-express the radiation field in terms of solely two sets of moments,
denoted $M_L$ and $S_L$, given by some non-linear functionals of the
moments $I_L$, $J_L$, $W_L$, $X_L$, $Y_L$ and $Z_L$, but differing
from $I_L$ and $J_L$ starting at the 2.5PN order (see Section VI in
\cite{B98mult} for a discussion). From the equations (4.20)-(4.24) in
\cite{B96} the 3PN quadrupole $M_{ij}$ is related to $I_{ij}$ by

\begin{subequations}\label{118}\begin{eqnarray}
M_{ij} &=& I_{ij}-{4G\over c^5}\left[ W^{(2)} I_{ij}-W^{(1)}
I_{ij}^{(1)}\right] + {\cal O}(7) \;,\\ W &=& {1\over 3}\int d^3{\bf
x}~\!  x_i\sigma_i = {1\over 3} m_1 (y_1v_1) +{\cal O}(2) + 1
\leftrightarrow 2
\;.\end{eqnarray}\end{subequations}$\!\!$
For the other moments there is no correction to be made at this order
[for instance $S_{ij}=J_{ij}+{\cal O}(5)$]. Actually we observe that
$W$ is zero for circular orbits, and thus we shall from now on replace
all the moments $I_L$ and $J_L$ by the corresponding $M_L$ and $S_L$.

Finally we need to relate the moments $M_L$, $S_L$ to the
``radiative'' moments, say $U_L$ (mass-type) and $V_L$ (current-type),
which play the role of observables associated with the radiation field
at infinity. Since such a relation has already been worked out at the
3.5PN level in Ref. \cite{B98tail}, we simply report the main result,
which concerns the mass-quadrupole radiative moment $U_{ij}$, that is

\begin{eqnarray}\label{118'}
U_{ij}(t) &=& M^{(2)}_{ij}(t) + {2G M\over c^3} \int^{+\infty}_0 d \tau
M^{(4)}_{ij} (t-\tau) \left[ \ln \left( {c\tau\over 2r_0} \right) +
{11\over 12} \right] \nonumber\\ &+& {2G^2 M^2\over c^6}
\int^{+\infty}_0 d \tau M^{(5)}_{ij}(t-\tau) \left[ \ln^2 \left(
{c\tau\over 2r_0} \right) + {57\over 70} \ln \left( {c\tau\over 2r_0}
\right) + {124627\over 44100} \right] \nonumber\\ &+& {1\over
c^5}\left\{\cdots\right\}+{1\over c^7}\left\{\cdots\right\}+{\cal O}
(8) \;.\end{eqnarray} This formula is valid through 3.5PN order,
modulo the odd-order 2.5PN and 3.5PN terms that we do not show because
they do not contribute to the flux for circular orbits. The only
contributions coming from Eq. (\ref{118'}) are the 1.5PN tail, and 3PN
``tail of tail'' integrals. In the flux we shall derive below the
terms at the orders 2.5PN and 3.5PN are due to the tail integrals in
higher multipole moments (see Ref. \cite{B98tail} for details).

\section{The energy flux of circular compact binaries}

For general sources, the total energy flux (or gravitational
luminosity ${\cal L}$) to the 3PN order is composed of an
``instantaneous'' contribution --- i.e. a functional of the multipole
moments $M_L$ and $S_L$ at the same instant ---, and a ``tail''
contribution. We shall now follow the study in \cite{B98tail} of the
occurence of non-linear effects in ${\cal L}$ up to 3.5PN
order. Following the equation (4.18) in Ref. \cite{B98tail} we split
${\cal L}$ into an instantaneous part, a tail part, a tail square
part, and a tail of tail part:

\begin{equation}\label{119}
{\cal L} = {\cal L}_{\rm inst}+{\cal L}_{\rm tail}+{\cal L}_{\rm
(tail)^2}+{\cal L}_{\rm tail(tail)}
\;.\end{equation}
As all the parts involving tails have already been computed for
circular binaries \cite{B98tail}, we need only to compute the
instantaneous part which is given by

\begin{eqnarray}\label{120}
 {\cal L}_{\rm inst} &=& {G\over c^5} \left\{ {1\over 5} M^{(3)}_{ij}
 M^{(3)}_{ij} + {1\over c^2} \left[ {1\over 189} M^{(4)}_{ijk}
 M^{(4)}_{ijk} + {16\over 45} S^{(3)}_{ij} S^{(3)}_{ij}\right]
 \right. \nonumber \\ &+& {1\over c^4} \left[ {1\over 9072}
 M^{(5)}_{ijkm} M^{(5)}_{ijkm} + {1\over 84} S^{(4)}_{ijk}
 S^{(4)}_{ijk}\right] \nonumber \\ &+& \left. {1\over c^6} \left[
 {1\over 594000} M^{(6)}_{ijkmn} M^{(6)}_{ijkmn} + {4\over 14175}
 S^{(5)}_{ijkm} S^{(5)}_{ijkm}\right] + {\cal O}(8) \right\}
\end{eqnarray}
(see e.g. (4.15) in \cite{B98tail}). To obtain ${\cal L}_{\rm inst}$
we compute the time derivatives of the multipole moments. At this
stage we need a new input, namely the 3PN equations of motion of
circular binaries which are crucial in the differentiation of the 3PN
quadrupole moment. As recently obtained \cite{BF00,BFeom} the 3PN
orbital frequency of the circular motion reads as

\begin{eqnarray}\label{121}
\omega^2 &=& {Gm\over r^3} \Biggl\{1+\left[-3+\nu\right]\gamma +
\left[6+{41\over 4} \nu + \nu^2 \right] \gamma^2 \nonumber \\ &+&
\left[-10+\left(22 \ln \left( r \over {r'}_0\right)+{41\pi^2\over
64}-{67759\over 840}+{44\lambda\over 3}\right)\nu +{19\over 2}\nu^2
+\nu^3\right] \gamma^3 +{\cal O}(\gamma^4)\Biggr\}
\;.\end{eqnarray} 
The inverse of this formula gives the post-Newtonian parameter
$\gamma$ as a function of the frequency related parameter $x =
(Gm\omega/c^3)^{2/3}$,

\begin{eqnarray}\label{123}
\gamma &=& x \Biggl\{1+\left[1-{\nu\over 3}\right]x + \left[1-{65\over
12} \nu \right] x^2 \nonumber \\ &+& \left[1+\left(-{22\over
3}\ln\left(r\over {r'}_0\right)-{41\pi^2\over 192}-{10151\over
2520}-{44\lambda\over 9}\right)\nu +{229\over 36}\nu^2 +{1\over
81}\nu^3\right]x^3 + {\cal O}(x^4)\Biggr\}
\;.\end{eqnarray} 
Note that (\ref{121}) or (\ref{123}) involve the same constant
${r'}_0$ as in the 3PN mass quadrupole moment
(\ref{117})-(\ref{117'}).

Taking all the expressions of the multipole moments found in Section
XI, computing their time-derivatives according to the latter
circular-orbit 3PN equations of motion, and inserting them into
(\ref{120}) we then arrive at the following instantaneous part of the
flux,

\begin{eqnarray}\label{124}
{\cal L}_{\rm inst} &=& {32c^5\over 5G} \gamma^5 \nu^2
\left\{1+\left(-{2927\over 336} -{5\over 4}\nu \right) \gamma +
\left({293383\over 9072}+{380\over 9}\nu \right) \gamma^2 \right.
\nonumber \\ &+& \left.  \left({53712289\over 1108800} -{1712\over
105}\ln \left({r\over r_0}\right) \right.\right.\nonumber\\ &+&
\left.\left.  \left[-{332051\over 720}+{110\over 3}\ln \left({r\over
{r'}_0}\right) + {123\pi^2\over 64}+44\lambda-{88\over
3}\theta\right]\nu -{383\over 9}\nu^2\right) \gamma^3 + {\cal
O}(\gamma^4) \right\} \;,\end{eqnarray} where we recall that
$\theta=\xi+2\kappa+\zeta$. Next we simply add the known other
contributions. The tail one is due to such terms as the 1.5PN integral
appearing in Eq. (\ref{118'}) [and other equations corresponding to
higher multipole moments]. The result is derived to the 3.5PN order in
Eq. (5.5a) in Ref. \cite{B98tail}:

\begin{eqnarray}\label{125}
{\cal L}_{\rm tail}&=&{32c^5\over 5G} \gamma^5 \nu^2 \left\{4\pi
  \gamma^{3/2}+\left(-{25663\over 672}-{125\over 8}\nu\right) \pi
  \gamma^{5/2} \right. \nonumber \\ &+&\left.\left({90205\over
  576}+{505747\over 1512}\nu +{12809\over
  756}\nu^2\right)\pi\gamma^{7/2} +{\cal O}(\gamma^4)\right\}
  \;.\end{eqnarray} 
Second, the tail of tail comes from the 3PN term in Eq. (\ref{118'}),
and the tail square from the square of the 1.5PN term. The sum of
these parts reads, following Eq. (5.9) in \cite{B98tail},
  
\begin{equation}\label{126}
{\cal L}_{(\rm tail)^2+\rm tail(\rm tail)}={32c^5\over 5G} \gamma^5
\nu^2 \left\{ \left( -{1712\over 105} \left[C+\ln \left({4\omega r_0}
\over c\right)\right] +{16\pi^2\over 3} -{116761\over
3675}\right)\gamma^3 +{\cal O}(\gamma^4)\right\}
\;,\end{equation}
where $C$ denotes the Euler constant ($C=0.577\cdots$), and where the
constant $r_0$ is the same as the $r_0$ occuring in the mass
quadrupole moment (\ref{117})-(\ref{117'}). Thus, the energy flux,
complete up to the 3.5PN order, reads

\begin{eqnarray}\label{127}
{\cal L} &=& {32c^5\over 5G} \gamma^5 \nu^2 \left\{1+\left(-{2927\over
336} -{5\over 4}\nu \right) \gamma + 4\pi \gamma^{3/2} +
\left({293383\over 9072}+{380\over 9}\nu \right) \gamma^2
\right.\nonumber \\ &+& \left(-{25663\over 672} -{125\over 8}\nu
\right)\pi \gamma^{5/2} + \left({129386791\over 7761600}+{16\pi^2\over
3}-{1712\over 105}C -{856\over 105}\ln (16\gamma) \right.\nonumber \\
&+& \left.\left[-{332051\over 720}+{110\over 3}\ln \left({r\over
{r'}_0}\right) + {123\pi^2\over 64}+44\lambda-{88\over
3}\theta\right]\nu -{383\over 9}\nu^2\right) \gamma^3 \nonumber \\ &+&
\left. \left({90205\over 576}+{505747\over 1512}\nu +{12809\over
756} \nu^2\right)\pi \gamma^{7/2} + {\cal O}(\gamma^4) \right\}
\;.\end{eqnarray} We observe that the constants $r_0$ have cancelled
out between the instantaneous flux ${\cal L}_{\rm inst}$ and the part
${\cal L}_{(\rm tail)^2+\rm tail(\rm tail)}$. This cancellation is to
be expected for any source: see a proof in \cite{B98tail} (Eqs. (4.14)
there) where it is shown that the tails of tails at the 3PN order
depend on $r_0$ through the effective quadrupole moment $M^{\rm
eff}_{ij}=M_{ij}+{214\over 105}\ln r_0~\!{G^2m^2\over
c^6}~\!M^{(2)}_{ij}$. Using our explicit result (\ref{117'}) for
$I_{ij}=M_{ij}+{\cal O}(5)$ we find that indeed the $r_0$'s cancel
out. The fact that we have recovered the expected dependence on $r_0$
of the source quadrupole moment is a good check of the computation.

On the other hand, the point-mass regularization constant ${r'}_0$
still remains in the flux (\ref{127}). This is because the energy flux
is not yet expressed in a coordinate-independent way, as the
post-Newtonian parameter $\gamma$ depends on the distance between the
masses in harmonic coordinates. To find a truly coordinate-independent
result we must replace $\gamma$ by its expression given by
(\ref{123}) in terms of the frequency-related parameter $x$.  With this
change of variable,  at long last  we   obtain  
our end   result: 

\begin{eqnarray}\label{128}
{\cal L} &=& {32c^5\over 5G} x^5 \nu^2 \left\{1+\left(-{1247\over 336}
-{35\over 12}\nu \right)x + 4\pi x^{3/2} + \left(-{44711\over
9072}+{9271\over 504}\nu +{65\over 18}\nu^2\right)x^2 \right.\nonumber
\\ &+& \left(-{8191\over 672} -{583\over 24}\nu \right)\pi x^{5/2} +
\left({6643739519\over 69854400}+{16\pi^2\over 3}-{1712\over 105}C
-{856\over 105}\ln (16x) \right.\nonumber \\ &+&
\left.\left[-{11497453\over 272160}+ {41\pi^2\over 48}+{176\over
9}\lambda-{88\over 3}\theta\right]\nu -{94403\over 3024}\nu^2
-{775\over 324} \nu^3\right)x^3 \nonumber \\ &+&
\left. \left(-{16285\over 504}+{214745\over 1728}\nu +{193385\over 3024}
\nu^2\right)\pi x^{7/2} + {\cal O}(x^4) \right\}
\;.\end{eqnarray}
In the above expression the constant ${r'}_0$ has cleanly disappeared. 
Of course, this was to be expected because we have seen that ${r'}_0$ is
pure-gauge; nevertheless this cancellation constitutes a
satisfactory test of the algebra. 
However, the result still depends on one physical
undetermined numerical coefficient, which is a linear combination of
the equation-of-motion-related constant $\lambda$ and the
multipole-moment-related constant $\theta$. On the other hand, our
final expression (\ref{128}) is in perfect agreement, in the test-mass
limit $\nu\to 0$, with the result of black-hole perturbation theory
which is already known to a very high post-Newtonian order
\cite{TSasa94,TTS96}.

\acknowledgments Many of the algebraic computations reported in this
paper were checked with the aid of the software Mathematica and using
some programs developed by Guillaume Faye.

\appendix

\section{Results for all the terms}

For the mass quadrupole we factorize out a factor $\mu=m\nu$ in front
of all the terms. We denote $v^i=\sqrt{\frac{G m}{r^3}}w^i$, so for
instance $\hat{w}_{ij}=\frac{r^3}{G m}v^{<i}v^{j>}$ (and
$\hat{x}_{ij}=x^{<i}x^{j>}$). For the current quadrupole all the terms
have to be multiplied by $\delta m/m~\!L_{<i}x_{j>}$, where $\delta
m=m_1-m_2$ and $L_i=\mu~\!\varepsilon_{ijk}x^jv^k$ is the angular
momentum. For the mass octupole we factorize out $\mu\delta
m/m=\nu\delta m$. For simplicity the constants $r_0$ and ${r'}_0$ in
the logarithms are set to one. In the case of the 3PN mass quadrupole,
to the sum of all these terms one must add the undetermined
contribution given by Eq. (\ref{222}) in the text.
 
\subsection{The 3PN mass quadrupole}

\noindent
Miscellaneous:

\begin{subequations}\label{a1}\begin{eqnarray}
{\rm SI(22+23+32)}+{\rm SII(11)}&=&0\;,\\
{\rm VI(16+20)}+{\rm VII(6)}&=&{\rm VI(19)}
\;.\end{eqnarray}\end{subequations}$\!\!$
Compact term at Newtonian order:

\begin{subequations}\label{a2}\begin{eqnarray}
{\rm SI(1)}&=&\left[1+ {\gamma \over 2}(1-5\nu) - {\gamma^2
\over 8} (13-61\nu +25\nu^2) \right. \nonumber \\ &+&
\left. {\gamma^3\over 16}(149-573\nu +354\nu^2-29\nu^3)
\right] \hat{x}_{ij}
\;.\end{eqnarray}\end{subequations}$\!\!$
Compact terms at 1PN:

\begin{subequations}\label{a3}\begin{eqnarray}
{\rm SII(1)}&=&{1\over 56}\gamma \left[(-8+24\nu + \gamma (20-52\nu
-20\nu^2) \right.\nonumber \\ &+& \left.\gamma^2(-23-17\nu +
160\nu^2-55\nu^3))\hat{x}_{ij} \right.\nonumber \\ &+& \left. (8-24\nu +
\gamma(4-28\nu+44\nu^2) \right. \nonumber \\ &+& \left. \gamma^2(-13-9\nu +
238\nu^2+35\nu^3)) \hat{w}_{ij}\right] \;,\\ {\rm VI(1)}&=& {1\over 21}\gamma
\left[(-8+24\nu + \gamma (28-92\nu + 20\nu^2) \right.\nonumber \\ &+&\left.
\gamma^2(-75+259\nu - 176\nu^2+13\nu^3)) \hat{x}_{ij} \right.\nonumber \\ &+&
\left.(8-24\nu + \gamma(-4+12\nu + 4\nu^2) \right.\nonumber \\ &+&\left.
\gamma^2(15-157\nu + 414\nu^2+7\nu^3)) \hat{w}_{ij}\right]
\;.\end{eqnarray}\end{subequations}$\!\!$
Compact terms at 2PN:

\begin{subequations}\label{a4}\begin{eqnarray}
{\rm SI(3)}&=& 2\gamma^2\left[2-8\nu + 4\nu^2+ \gamma(-7+28\nu -
12\nu^2-4\nu^3)\right] \hat{x}_{ij}\;,\\ {\rm SIII(1)}&=& {\gamma^2\over
126}\left[(2-10\nu + 10\nu^2+ \gamma(-11+55\nu -
56\nu^2+3\nu^3))\hat{x}_{ij} \right. \nonumber \\ &+& \left. (-2+10\nu - 10\nu^2+
\gamma(5-23\nu + 16\nu^2+7\nu^3)) \hat{w}_{ij}\right]\;,\\ {\rm
VI(2)}&=&-{8\over 21}\gamma^2\left[(2-8\nu + 4\nu^2+ \gamma(-7+28\nu -
12\nu^2 -4\nu^3)) \hat{x}_{ij} \right.\nonumber \\ &+& \left.(-2+8\nu - 4\nu^2+
\gamma (1-2\nu - 8\nu^2+8\nu^3)) \hat{w}_{ij}\right]\;,\\ {\rm VI(3)}&=&
{8\over 21} \gamma^2\nu\left[(-2+4\nu + \gamma(5-6\nu - 4\nu^2))
\hat{x}_{ij} \right.\nonumber \\ &+& \left.(2-4\nu + \gamma (1-8\nu + 8\nu^2))
\hat{w}_{ij}\right]\;,\\ {\rm VII(1)}&=& {8\over 189} \gamma^2\left[(2-10\nu +
10\nu^2+ \gamma(-13+69\nu - 84\nu^2+17\nu^3)) \hat{x}_{ij} \right.\nonumber
\\ &+& \left.(-2+10\nu - 10\nu^2+ \gamma (7-37\nu + 44\nu^2-7\nu^3))
\hat{w}_{ij}\right]\;,\\ {\rm TI(1)}&=& {1\over 54} \gamma^2\left[(2-10\nu +
10\nu^2+ \gamma(-13+69\nu - 84\nu^2+17\nu^3)) \hat{x}_{ij} \right.\nonumber
\\ &+& \left.(-2+10\nu - 10\nu^2+ \gamma(7-37\nu + 44\nu^2-7\nu^3))
\hat{w}_{ij}\right]
\;.\end{eqnarray}\end{subequations}$\!\!$
Compact terms at 3PN:

\begin{subequations}\label{a5}\begin{eqnarray}
{\rm SI(13)}&=&16\gamma^3\nu^2 \hat{x}_{ij}\;,\\ {\rm SI(14)}&=& 8
\gamma^3(1-5\nu + 5\nu^2) \hat{x}_{ij}\;,\\ {\rm SI(15)}&=&
-2\gamma^3\nu(1-4\nu + 2\nu^2) \hat{x}_{ij}\;,\\ {\rm SI(16C)}&=&0\;,\\ {\rm
SI(16NC)}&=&(-1+9\nu - 17\nu^2)\gamma^3 \hat{x}_{ij}+{2\over
15}\gamma^3\nu\hat{w}_{ij}\;,\\ {\rm SII(2)}&=&{4\over 7}
\gamma^3(-1+2\nu )(1-4\nu + \nu^2)(\hat{x}_{ij} - \hat{w}_{ij})\;,\\ {\rm
SIV(1)}&=&{2\over 2079} \gamma^3(-1+7\nu -
14\nu^2+7\nu^3)(\hat{x}_{ij} - \hat{w}_{ij})\;,\\ {\rm VI(7)}&=&{8\over
21} \gamma^3\nu (1-4\nu + 2\nu^2)(\hat{x}_{ij}-\hat{w}_{ij})\;,\\ {\rm
VI(8)}&=&-{16\over 21} \gamma^3(1-5\nu +
5\nu^2)(\hat{x}_{ij}-\hat{w}_{ij})\;,\\ {\rm VI(9)}&=&-{8\over 21}
\gamma^3\nu (1+\nu )(-1+2\nu )(\hat{x}_{ij}-\hat{w}_{ij})\;,\\ {\rm
VI(10C)}&=&{32\over 21} \gamma^3\nu (-1+2\nu
)(\hat{x}_{ij}-\hat{w}_{ij})\;,\\ {\rm VI(10NC)}&=&-{4\over 21}\nu
\left({8\over 5}-11\nu \right)\gamma^3 (\hat{w}_{ij}-\hat{x}_{ij})\;,\\
{\rm VI(11)}&=&{16\over 21} \gamma^3\nu (1-4\nu +
2\nu^2)(\hat{x}_{ij}-\hat{w}_{ij})\;,\\ {\rm VI(12C)}&=&0\;,\\ {\rm
VI(12NC)}&=&-{4\over 21}\left(1-{38\over 5}\nu +
17\nu^2\right)\gamma^3 (\hat{w}_{ij} -\hat{x}_{ij})\;,\\ {\rm
VI(13)}&=&{16\over 21}\gamma^3\nu (1-4\nu +
2\nu^2)(\hat{x}_{ij}-\hat{w}_{ij})\;,\\ {\rm VII(2)}&=&-{32\over
189}\gamma^3(-1+2\nu )(1-4\nu + \nu^2) (\hat{x}_{ij}-\hat{w}_{ij})\;,\\
{\rm VII(3)}&=&{32\over 189} \gamma^3\nu (1-4\nu +
2\nu^2)(\hat{x}_{ij}-\hat{w}_{ij})\;,\\ {\rm VIII(1)}&=&{16\over 2079}
\gamma^3(-1+7\nu -14\nu^2+7\nu^3) (\hat{x}_{ij}-\hat{w}_{ij})\;,\\ {\rm
TI(3)}&=&-{4\over 27} \gamma^3(-1+2\nu )(1-4\nu + \nu^2)(\hat{x}_{ij}
-\hat{w}_{ij})\;,\\ {\rm TI(4)}&=&{4 \over 27}\gamma^3\nu (1-4\nu +
2\nu^2)(\hat{x}_{ij}-\hat{w}_{ij})\;,\\ {\rm TII(1)}&=&{2 \over
297}\gamma^3(-1+7\nu - 14\nu^2+7\nu^3) (\hat{x}_{ij}-\hat{w}_{ij})
\;.\end{eqnarray}\end{subequations}$\!\!$
$Y$-terms at 2PN:

\begin{subequations}\label{a6}\begin{eqnarray}
{\rm SI(4)}&=&{8\over 3}\gamma^2\nu \left[(2-4\gamma \nu ) \hat{w}_{ij}
+(1-3\nu + \gamma (-3+8\nu + 3\nu^2)) \hat{x}_{ij}\right]\;,\\ {\rm
SI(5C)}&=&{1\over 3}\gamma^2\left[(4-8\nu + \gamma(-2+16\nu^2))
\hat{w}_{ij} +(2-10\nu \right.\nonumber \\ &-& \left. 12\nu^2+ \gamma (-7+35\nu +
34\nu^2+12\nu^3)) \hat{x}_{ij}\right]\;,\\ {\rm SI(6)}&=&{1 \over 3}\gamma^2\nu
\left[(-2+ \gamma(-1+6\nu ))\hat{w}_{ij} +(2-6\nu + \gamma (-5+11\nu +
12\nu^2))\hat{x}_{ij}\right]\;,\\ {\rm SI(7)}&=& {4\over 3}\gamma^2\nu \left[(2+
\gamma(-1-2\nu )) \hat{w}_{ij} +(-2+6\nu+ \gamma (7-21\nu ))
\hat{x}_{ij}\right]\;,\\ {\rm SII(7)}&=&{2 \over 3}\gamma^2(-1+3\nu
)\left[(-2+\gamma(-1+6\nu )) \hat{w}_{ij} +(2+ \gamma(-5-4\nu ))
\hat{x}_{ij}\right]\;,\\ {\rm VI(4)}&=&{4 \over 63} \gamma^2\left[(28-110 \nu +
24\nu^2+ \gamma (-14+15\nu + 160\nu^2-48\nu^3)) \hat{w}_{ij})\right.
\nonumber \\ &+& \left.(-28+110\nu - 24\nu^2+ \gamma(98-373\nu +
22\nu^2+24\nu^3)) \hat{x}_{ij}\right]\;,\\ {\rm VI(5)}&=&{1 \over
42}\gamma^2(-12+62\nu - 24\nu^2)\left[(2+\gamma(1-6\nu )) \hat{w}_{ij}
+(-2+\gamma (5+4\nu )) \hat{x}_{ij}\right]\;,\\ {\rm TI(2)}&=&{2 \over
63}\gamma^2(2-7\nu + \nu^2)\left[(-2+\gamma(-1+6\nu )) \hat{w}_{ij}
+(2+\gamma(-5-4\nu )) \hat{x}_{ij}\right]
\;.\end{eqnarray}\end{subequations}$\!\!$
$Y$-terms at 3PN:

\begin{subequations}\label{a7}\begin{eqnarray}
{\rm SI(31)}&=&{8 \over 3}\gamma^3\nu \left[(1-2\nu-3\nu^2)
\hat{x}_{ij}-\nu \hat{w}_{ij}\right]\;,\\ {\rm SI(33C)}&=&-8\gamma^3\nu^2
\hat{x}_{ij}\;,\\ {\rm SI(35C)}&=&-{8 \over 3}\gamma^3\left[-(1-6\nu + 3\nu^2)
\hat{x}_{ij}+(-2+6\nu )\hat{w}_{ij}\right]\;,\\ {\rm SI(37C)}&=&{16 \over
3}\gamma^3\nu \left[(1-3\nu ) \hat{x}_{ij}+2\hat{w}_{ij}\right]\;,\\ {\rm
SI(38C)}&=&{32 \over 3}\gamma^3\nu\left[(-1+3\nu )
\hat{x}_{ij}+\hat{w}_{ij}\right]\;,\\ {\rm VI(16+20)}+{\rm VII(6)}&=&{2 \over
63}\gamma^3(12-49\nu + 10\nu^2+24\nu^3) (\hat{x}_{ij}-\hat{w}_{ij})\;,\\
{\rm VI(19)}&=&{2 \over 63}\gamma^3(12-49\nu + 10\nu^2+24\nu^3)
(\hat{x}_{ij}-\hat{w}_{ij})\;,\\ {\rm VI(21)}&=&{8 \over 63}\gamma^3\nu
(-1+2\nu )(5+12\nu )(-\hat{x}_{ij} +\hat{w}_{ij})\;,\\ {\rm
VI(25C)}&=&{16 \over 63}\gamma^3(14-55\nu + 12\nu^2)(-\hat{x}_{ij}
+\hat{w}_{ij})\;,\\ {\rm VI(26C)}&=&{4 \over 63}\gamma^3\nu (5-70\nu +
24\nu^2)(-\hat{x}_{ij} +\hat{w}_{ij})\;,\\ {\rm VI(27C)}&=&{4 \over
63}\gamma^3(-5+21\nu ) (\hat{x}_{ij}-\hat{w}_{ij})\;,\\ {\rm
VI(28C)}&=&0\;,\\ {\rm VI(29C)}&=&{4 \over 63}\gamma^3\nu (-1+12\nu
)(\hat{x}_{ij}-\hat{w}_{ij})\;,\\ {\rm TI(6)}&=&{1 \over
189}\gamma^3(-16-21\nu + 174\nu^2+56\nu^3)(\hat{x}_{ij}
-\hat{w}_{ij})\;,\\ {\rm TI(7)}&=&{16 \over 63}\gamma^3\nu (2-7\nu +
\nu^2)(\hat{x}_{ij}-\hat{w}_{ij})\;,\\ {\rm TI(8)}&=&{1 \over
27}\gamma^3\nu (13-46\nu + 8\nu^2)(-\hat{x}_{ij}+\hat{w}_{ij})
\;.\end{eqnarray}\end{subequations}$\!\!$
$S$-terms at 3PN:

\begin{subequations}\label{a8}\begin{eqnarray}
{\rm SII(3)}&=&{2 \over 21}\gamma^3\nu (1+2\nu +
12\nu^2)(\hat{w}_{ij}-\hat{x}_{ij})\;,\\ {\rm SII(4C)}&=&{1 \over
42}\gamma^3(1-2 \nu )(1+2\nu - 12\nu^2)(\hat{w}_{ij} -\hat{x}_{ij})\;,\\
{\rm SII(5)}&=&{1 \over 126}\gamma^3\nu (-11+6\nu +
36\nu^2)(\hat{w}_{ij} -\hat{x}_{ij})\;,\\ {\rm SII(6)}&=&-{2 \over
63}\gamma^3\nu (-11+6\nu + 36\nu^2)(\hat{w}_{ij} -\hat{x}_{ij})\;,\\ {\rm
SIII(2)}&=&-{2 \over 63}\gamma^3(3-22\nu +
36\nu^2)(\hat{w}_{ij}-\hat{x}_{ij})\;,\\ {\rm VII(4)}&=&{8 \over
189}\gamma^3(3-2\nu - 34\nu^2+8\nu^3)(\hat{w}_{ij} -\hat{x}_{ij})\;,\\
{\rm VII(5)}&=&-{2\over 63}\gamma^3(-1+6\nu -
18\nu^2+8\nu^3)(\hat{w}_{ij} -\hat{x}_{ij})\;,\\ {\rm TII(2)}&=&{2 \over
2079}\gamma^3(8-67\nu + 134\nu^2-12\nu^3)(\hat{w}_{ij} -\hat{x}_{ij})
\;.\end{eqnarray}\end{subequations}$\!\!$
$T$-terms at 3PN:

\begin{subequations}\label{a9}\begin{eqnarray}
{\rm SI(17)}&=&{1 \over 9}\gamma^3\nu \left[(7-36\nu )\hat{w}_{ij}
+(-15+36\nu + 54\nu^2)\hat{x}_{ij}\right]\;,\\ {\rm SI(18)}&=&{1 \over
3}\gamma^3\nu \left[-3\hat{w}_{ij}+(-1-12\nu + 18\nu^2) \hat{x}_{ij}\right]\;,\\
{\rm SI(19C)}&=&\gamma^3\left[-{1 \over 6}(1-\nu )\hat{w}_{ij}
+{1\over 12}(2-7\nu + 70\nu^2+12\nu^3)\hat{x}_{ij}\right]\;,\\ {\rm
SI(21C)}&=&{1\over 12}\gamma^3\nu \left[(-2-8\nu )\hat{w}_{ij} +(-1+22\nu
+ 12\nu^2)\hat{x}_{ij}\right]\;,\\ {\rm SI(24)}&=&{1\over 18}\gamma^3\nu
\left[11\hat{w}_{ij}+(-15+36\nu + 54\nu^2) \hat{x}_{ij}\right]\;,\\ {\rm
SI(25)}&=&-{2 \over 9}\gamma^3\nu \left[-\hat{w}_{ij}+(-3-36\nu +
54\nu^2) \hat{x}_{ij}\right]\;,\\ {\rm SII(9)}&=&{2\over
9}\gamma^3(1-9\nu^2)(-\hat{w}_{ij}+\hat{x}_{ij})\;,\\ {\rm VI(14)}&=&-{2
\over 63}\gamma^3(20-71\nu + 36\nu^3)(\hat{w}_{ij}-\hat{x}_{ij})\;,\\
{\rm VI(15)}&=&-{2 \over 63}\gamma^3\nu (17-98\nu +
36\nu^2)(\hat{w}_{ij} -\hat{x}_{ij})\;,\\ {\rm VI(17)}&=&{1 \over
42}\gamma^3\nu (-43+94\nu + 36\nu^2)(\hat{w}_{ij} -\hat{x}_{ij})\;,\\
{\rm VI(18)}&=&{1 \over 42}\gamma^3(4-3\nu -
72\nu^2+36\nu^3)(\hat{w}_{ij} -\hat{x}_{ij})\;,\\ {\rm TI(5)}&=&{1 \over
189}\gamma^3(4-9\nu - 21\nu^2
+13\nu^3)(\hat{w}_{ij}-\hat{x}_{ij})
\;.\end{eqnarray}\end{subequations}$\!\!$
Cubic terms:

\begin{subequations}\label{a10}\begin{eqnarray}
{\rm SI(26+27}&+&{\rm 28+29+30+34+36)}\nonumber\\ &=& \gamma^3\nu
\left(\left({74\over 15}-{152\over 3}\nu
\right)\hat{x}_{ij}+\left({128\over 15}\ln r-{1024\over
225}\right)\hat{w}_{ij}\right)\;,\\ {\rm SII(13)}&=&{16\over
225}\gamma^3((45-75\nu )\ln r-9-25\nu ) (-\hat{x}_{ij}+\hat{w}_{ij})\;,\\
{\rm VI(22+23+24)}&=&{8\over 525}\gamma^3(60\ln r-7-265\nu + 275\nu^2)
(\hat{x}_{ij}-\hat{w}_{ij})
\;.\end{eqnarray}\end{subequations}$\!\!$
Non-compacts terms:

\begin{subequations}\label{a11}\begin{eqnarray}
{\rm SI(5NC)}&=&-{1 \over 2}\gamma^2(4+10\nu + \gamma (2-6\nu -
47\nu^2)) \hat{x}_{ij} \;,\\ {\rm SI(19NC)}&=&\left({1\over
6}+{71\over 36}\nu-{7\over 3}\nu^2\right)\gamma^3 \hat{x}_{ij}
+\left({2 \over 15}\nu \ln r-{1\over 6}-{77\over 450}\nu\right)
\gamma^3\hat{w}_{ij}\;,\\ {\rm SI(20)}&=&\left(\left({4
\over 15}+{4\over 3}\nu \right)\ln r-{77\over 225} -{25\over 6}\nu +
6\nu^2\right) \gamma^3 \hat{x}_{ij} \nonumber \\ &+&
\left(\left(-{4\over 15}+{4\over
15}\nu \right)\ln r + {77\over 225} -{77\over 225}\nu \right) \gamma^3
\hat{w}_{ij}\;,\\ {\rm SI(21NC)}&=&\left(\left({8\over
15}-{2\over 3}\nu \right)\ln r +{26\over 225}-{1\over 36}\nu
-\nu^2\right) \gamma^3 \hat{x}_{ij}  \nonumber \\ &+&
\left(\left(-{8\over 15}+{8\over 15}\nu \right)\ln r-{26\over
225} +{397\over 450}\nu \right) \gamma^3 \hat{w}_{ij}\;,\\ {\rm
SI(33NC)}&=&\left(-{164\over 15}\nu + {62\over
3}\nu^2\right)\gamma^3 \hat{x}_{ij} +\left(-{8\over 15}\nu \ln
r-{176\over 225}\nu \right)\gamma^3 \hat{w}_{ij}\;,\\ {\rm
SI(35NC)}&=&\left(\left(-{16\over 15}-{64\over 3}\nu \right)\ln
r +{1028\over 225} +{1148\over 45}\nu -{145\over
3}\nu^2\right)\gamma^3 \hat{x}_{ij} \nonumber \\ &+&
\left(\left({16\over 15}-{16\over 5}\nu \right)\ln r-{1028\over
225} +{568\over 75}\nu \right) \gamma^3 \hat{w}_{ij}\;,\\ {\rm
SI(37NC)}&=&\left(-{62\over 15}\nu - {17\over
3}\nu^2\right)\gamma^3 \hat{x}_{ij} +\left(-{4\over 15}\nu \ln
r+{512\over 225}\nu \right)\gamma^3 \hat{w}_{ij}\;,\\ {\rm
SI(38NC)}&=&\left({124\over 15}\nu + {34\over
3}\nu^2\right)\gamma^3 \hat{x}_{ij} +\left({8\over 15}\nu \ln
r+{1376\over 225}\nu \right)\gamma^3 \hat{w}_{ij}\;,\\ {\rm
SII(4NC)}&=& \left(\left({16\over 35}-{2\over 3}\nu \right)\ln
r+{139\over 1050} -{164\over 315}\nu + {11\over 7}\nu^2\right)\gamma^3
(\hat{w}_{ij} -\hat{x}_{ij})\;,\\ {\rm VI(25NC)}&=&\left( \left(-{48\over
35}+{8\over 15}\nu \right)\ln r-{968\over 1575} +{10792\over 1575}\nu
- {236\over 63}\nu^2\right)\gamma^3 (\hat{x}_{ij}-\hat{w}_{ij})\;,\\ {\rm
VI(26NC)}&+&{\rm VI(27NC)}+{\rm VI(28NC)}+{\rm VI(29NC)}=
\left(\left({32\over 105}-{8\over 15} \nu \right)\ln r
\right.\nonumber\\ &+&{2644\over 1575}-{488\over 175}\nu -
\left. {124\over 63}\nu^2\right) \gamma^3 (\hat{x}_{ij} -
\hat{w}_{ij})
\;.\end{eqnarray}\end{subequations}$\!\!$
Terms at 2.5PN:

\begin{equation}\label{a12}
{\rm SI(11+12)}={48\over 7}\gamma^{5/2}\nu x_{<i}w_{j>}
\;.\end{equation}
Divergence terms:

\begin{subequations}\label{a13}\begin{eqnarray}
{\rm SI(2)}&=&0\;,\\
{\rm SI(8)}&=&0\;,\\
{\rm SI(9)}&=&0\;,\\
{\rm SI(10)}&=&0\;,\\
{\rm SI(39)}&=&0\;,\\
{\rm SI(40)}&=&0\;,\\
{\rm SI(41)}&=&-{8\over 3}\gamma^3(-\hat{x}_{ij}+\hat{w}_{ij})\;,\\
{\rm SI(42)}&=&0\;,\\
{\rm SI(43)}&=&0\;,\\
{\rm SI(44)}&=&{1\over 3}\gamma^3(-\hat{x}_{ij}+\hat{w}_{ij})\;,\\
{\rm SI(45)}&=&\gamma^3(-\hat{x}_{ij}+\hat{w}_{ij})\;,\\
{\rm SI(46)}&=&{1\over 3}\gamma^3(\hat{x}_{ij}-\hat{w}_{ij})\;,\\
{\rm SI(47)}&=&0\;,\\
{\rm SI(48)}&=&-{2\over 3}\gamma^3(-\hat{x}_{ij}+\hat{w}_{ij})\;,\\
{\rm SI(49)}&=&16\gamma^3(\hat{x}_{ij}-\hat{w}_{ij})\;,\\
{\rm SI(50)}&=&{8\over 3}\gamma^3(\hat{x}_{ij}-\hat{w}_{ij})\;,\\
{\rm SII(8)}&=&0\;,\\
{\rm SII(10)}&=&0\;,\\
{\rm SII(12)}&=&{27\over 35}\gamma^3(\hat{x}_{ij}-\hat{w}_{ij})\;,\\
{\rm SII(14)}&=&{72\over 35}\gamma^3(-\hat{x}_{ij}+\hat{w}_{ij})\;,\\
{\rm SIII(3)}&=&0\;,\\
{\rm VI(6)}&=&0\;,\\
{\rm VI(30)}&=&0\;,\\
{\rm VI(31)}&=&0\;,\\
{\rm VI(32)}&=&0\;,\\
{\rm VI(33)}&=&{8\over 15}\gamma^3(\hat{x}_{ij}-\hat{w}_{ij})\;,\\
{\rm VI(34)}&=&0\;,\\
{\rm VI(35)}&=&-{4\over 15}\gamma^3(\hat{x}_{ij}-\hat{w}_{ij})\;,\\
{\rm VII(7)}&=&0\;,\\
{\rm TI(9)}&=&0
\;.\end{eqnarray}\end{subequations}$\!\!$

\subsection{The 2PN current quadrupole}

\noindent
Compact term at Newtonian order:

\begin{equation}\label{a14}
{\rm VI(1)}={1\over 8}\left[-8+4\gamma +\gamma^2(-15+88\nu +3\nu^2)\right]
\;.\end{equation}
Compact terms at 1PN:

\begin{subequations}\label{a15}\begin{eqnarray}
{\rm VI(2)}&=&\gamma\left[-2+2\nu +\gamma(1+\nu -4\nu^2)\right]\;,\\
{\rm VI(3)}&=&\gamma \nu \left[-2+\gamma(-1+4\nu )\right] \;,\\ {\rm
VII(1)}&=&{\gamma \over 28}\left[2-4\nu +\gamma(-7+16\nu
-3\nu^2)\right] \;,\\ {\rm TI(1)}&=&{\gamma \over 56}\left[2-4\nu
+\gamma(-7+16\nu - 3\nu^2)\right]
\;.\end{eqnarray}\end{subequations}$\!\!$
Compact terms at 2PN:

\begin{subequations}\label{a16}\begin{eqnarray}
{\rm VI(7)}&=&\gamma^2\nu (1-\nu ) \;,\\ 
{\rm VI(8)}&=&2\gamma^2(-1+2\nu ) \;,\\
{\rm VI(9)}&=&\gamma^2\nu (1+\nu ) \;,\\
{\rm VI(10C)}&=&-4\gamma^2\nu  \;,\\
{\rm VI(10NC)}&=&{3\over 2}\gamma^2\nu \;,\\ 
{\rm VI(11)}&=&2\gamma^2\nu (1-\nu )\;,\\ 
{\rm VI(12C)}&=&0\;,\\
{\rm VI(12NC)}&=&{\gamma^2\over 2}(1-6\nu )\;,\\
{\rm VI(13)}&=&2\gamma^2\nu (1-\nu )\;,\\
{\rm VII(2)}&=&{\gamma^2\over 7}(1-3\nu +\nu^2)\;,\\ 
{\rm VII(3)}&=&{\gamma^2\over 7}\nu(1-\nu)\;,\\
{\rm VIII(1)}&=&{\gamma^2\over 504}(1-3\nu )(-1+\nu )\;,\\ 
{\rm TI(3)}&=&{\gamma^2\over 7}(1-3\nu +\nu^2) \;,\\
{\rm TI(4)}&=&{\gamma^2\over 7}\nu(1-\nu) \;,\\
{\rm TII(1)}&=&{\gamma^2\over 504}(1-3\nu )(-1+\nu ) 
\;.\end{eqnarray}\end{subequations}$\!\!$
$Y$-terms at 1PN:

\begin{subequations}\label{a17}\begin{eqnarray}
{\rm VI(4)}&=&{\gamma\over 2}\left[-2+4\nu +\gamma(1-8\nu^2)\right] \;,\\
{\rm VI(5)}&=&-{3\over 4}\gamma \nu \left[2+\gamma(1-6\nu)\right]\;,\\
{\rm TI(2)}&=&0
\;.\end{eqnarray}\end{subequations}$\!\!$
$Y$-terms at 2PN:

\begin{subequations}\label{a18}\begin{eqnarray}
{\rm VI(16+20)}&+&{\rm VII(6)}={\gamma^2\over 3}(1-2\nu -3\nu^2)\;,\\ 
{\rm VI(19)}&=&{\gamma^2\over 3}(1-2\nu -3\nu^2)\;,\\ 
{\rm VI(21)}&=&-{2\over 3}\gamma^2\nu (5-6\nu )\;,\\
{\rm VI(25C)}&=&2\gamma^2(-1+2\nu) \;,\\
{\rm VI(26C)}&=&{2\over 3}\gamma^2\nu (1+3\nu )\;,\\
{\rm VI(27C)}&=&{\gamma^2\over 3}(1-2\nu) \;,\\
{\rm VI(28C)}&=&0\;,\\
{\rm VI(29C)}&=&{2\over 3}\gamma^2\nu  \;,\\
{\rm TI(6)}&=&{\gamma^2\over 21}(-2+4\nu -3\nu^2) \;,\\
{\rm TI(7)}&=&0\;,\\
{\rm TI(8)}&=&-{\gamma^2\over 21}\nu (7-3\nu ) 
\;.\end{eqnarray}\end{subequations}$\!\!$
$S$-terms at 2PN:

\begin{subequations}\label{a19}\begin{eqnarray}
{\rm VII(4)}&=&{1\over 42}\gamma^2(-2-3\nu )(1-2\nu )  \;,\\
{\rm VII(5)}&=&-{1\over 28}\gamma^2\nu (2+3\nu) \;,\\
{\rm TII(2)}&=&0
\;.\end{eqnarray}\end{subequations}$\!\!$
$T$-terms at 2PN:

\begin{subequations}\label{a20}\begin{eqnarray}
{\rm VI(14)}&=&{1\over 6}\gamma^2(1+4\nu -9\nu^2)  \;,\\
{\rm VI(15)}&=&{1\over 6}\gamma^2\nu (1-9\nu )  \;,\\
{\rm VI(17)}&=&{1\over 8}\gamma^2\nu (4+9\nu )  \;,\\
{\rm VI(18)}&=&{3\over 8}\gamma^2\nu (2+3\nu)  \;,\\
{\rm TI(5)}&=&{1\over 168}\gamma^2\nu (2+3\nu ) 
\;.\end{eqnarray}\end{subequations}$\!\!$
Cubic terms:

\begin{equation}\label{a21}
{\rm VI(22+23+24)}=2\gamma^2\left({1\over 3}-{9\over 4}\nu\right)
\;.\end{equation}
Non-compact terms:

\begin{subequations}\label{a22}\begin{eqnarray}
{\rm VI(25NC)}&=&\gamma^2\left(-{2\over 3}+{17\over 6}\nu
\right)\;,\\ {\rm
VI(26NC+27NC+28NC+29NC)}&=&\gamma^2\left({3\over 2}+{19\over
6}\nu\right)
\;.\end{eqnarray}\end{subequations}$\!\!$
Divergence terms:

\begin{subequations}\label{a23}\begin{eqnarray}
{\rm VI(6)}&=&0 \;,\\
{\rm VI(30)}&=&0\;,\\
{\rm VI(31)}&=&0\;,\\
{\rm VI(32)}&=&0\;,\\
{\rm VI(33)}&=&0\;,\\
{\rm VI(34)}&=&0\;,\\
{\rm VI(35)}&=&0\;,\\
{\rm VII(7)}&=&0\;,\\
{\rm TI(9)}&=&0
\;.\end{eqnarray}\end{subequations}$\!\!$

\subsection{The 2PN mass octupole}

\noindent
Compact term at Newtonian order:

\begin{equation}\label{a24}
{\rm SI(1)}={1\over 8}\left[-8+\gamma(-4+16\nu )+\gamma^2(13+24\nu +15\nu^2)\right]
       \hat{x}_{ijk}
\;.\end{equation}
Compact terms at 1PN:

\begin{subequations}\label{a25}\begin{eqnarray}
{\rm SII(1)}&=&{1\over 180}\gamma\left[(30-60\nu +\gamma(-75+120\nu
        +45\nu^2)) \hat{x}_{ijk}\right.\nonumber\\&+&\left. 3(-20+40\nu
        +\gamma(-10+60\nu -70\nu^2))x_{<i}w_{jk>}\right]\;,\\ {\rm
        VI(1)}&=&{1\over 90}\gamma\left[(30-60\nu +\gamma(-105+240\nu
        -45\nu^2))\hat{x}_{ijk} \right.\nonumber\\&+&\left. 3(-20+40\nu
        +\gamma(10-20\nu -10\nu^2))x_{<i}w_{jk>}\right]
\;.\end{eqnarray}\end{subequations}$\!\!$
Compact terms at 2PN:

\begin{subequations}\label{a26}\begin{eqnarray}
{\rm SI(3)}&=&-4\gamma^2(1-3\nu +\nu^2)\hat{x}_{ijk}\;,\\ {\rm
SIII(1)}&=&{1\over 3960}\gamma^2(1-3\nu )(1-\nu )(-105\hat{x}_{ijk}
+300x_{<i} w_{jk>})\;,\\ {\rm VI(2)}&=&{2\over 45}\gamma^2(1-3\nu
+\nu^2)(15\hat{x}_{ijk}-30x_{<i} w_{ij>})\;,\\ {\rm VI(3)}&=&{2\over
45}\gamma^2\nu (-1+\nu )(-15\hat{x}_{ijk} +30x_{<i} w_{jk>})\;,\\ {\rm
VII(1)}&=&{1\over 990}\gamma^2(1-3\nu )(1-\nu )(-105\hat{x}_{ijk}
+300x_{<i} w_{jk>})\;,\\ {\rm TI(1)}&=&{1\over 990}\gamma^2(1-3\nu
)(1-\nu )(-33\hat{x}_{ijk} +84x_{<i} w_{jk>})
\;.\end{eqnarray}\end{subequations}$\!\!$
$Y$-terms at 2PN:

\begin{subequations}\label{a27}\begin{eqnarray}
{\rm SI(4)}&=&-{2\over 15}\gamma^2\nu \left[(15-30\nu
)\hat{x}_{ijk}+60x_{<i}w_{jk>}\right]\;,\\ {\rm SI(5C)}&=&{1\over
30}\gamma^2\left[-(15-60\nu -60\nu^2)\hat{x}_{ijk} +3(-10+40\nu
)x_{<i}w_{jk>}\right]\;,\\ {\rm SI(6)}&=&-{1\over 30}\gamma^2\nu \left[(15-30\nu
)\hat{x}_{ijk} -30x_{<i}w_{jk>}\right]\;,\\ {\rm SI(7)}&=&-{2\over
15}\gamma^2\nu \left[(-15-30\nu )\hat{x}_{ijk}+30x_{<i} w_{jk>}\right]\;,\\ {\rm
SII(7)}&=&-{1\over 10}\gamma^2(1-2\nu )(-15\hat{x}_{ijk}+30x_{<i}
w_{jk})\;,\\ {\rm VI(4)}&=&-{1\over 450}\gamma^2(45-140\nu
+20\nu^2)(-15\hat{x}_{ijk} +30x_{<i} w_{jk})\;,\\ {\rm VI(5)}&=&{1\over
60}\gamma^2(2-9\nu +2\nu^2)(-15\hat{x}_{ijk} +30x_{<i}w_{jk>})\;,\\ {\rm
TI(2)}&=&{1\over 990}\gamma^2(8-21\nu +2\nu^2)(-15\hat{x}_{ijk}
+30x_{<i}w_{jk>})
\;.\end{eqnarray}\end{subequations}$\!\!$
Non-compact term:

\begin{equation}\label{a28}
{\rm SI(5NC)}=2\gamma^2(1+2\nu )\hat{x}_{ijk}
\;.\end{equation}

\end{document}